\documentclass[jcp,aps,showpacs,twocolumn,supergroupedaddress,epsfig,amsmath,amssymb,eqsecnum]{revtex4}
\usepackage{epsfig,amsmath,amsthm,amssymb,bm,epsf,graphics,psfrag,verbatim,subfigure,framed}
\usepackage[makeroom]{cancel}
\usepackage{mathtools}
\usepackage[usenames,dvipsnames]{color}
\usepackage{empheq}
\usepackage{bbm}
\usepackage{mathrsfs}
\usepackage{amsfonts}
\usepackage{color}
\usepackage{pbox}
\usepackage{xcolor}
\def\nv{\mathbf{n}}

\newcommand{\be}{\begin{equation}}
\newcommand{\ee}{\end{equation}}
\newcommand{\ba}{\begin{eqnarray}}
\newcommand{\ea}{\end{eqnarray}}
\newcommand{\bw}{\begin{widetext}}
\newcommand{\ew}{\end{widetext}}

\newcommand{\Ev}{{\bf E}}

\newcommand{\Pv}{{\bf{P}}}

\newcommand{\rv}{{\mathbf{r}}}

\newcommand{\kv}{{\bf k}}
\newcommand{\pv}{{\bf p}}

\newcommand{\dv}{{\bf d}}

\newcommand{\tk}{{\widetilde k}}

\newcommand{\Rv}{{\bf{R}}}

\begin{document}

\title{Spontaneous emission of a quantum emitter near a Chern insulator: 
\\
interplay of time reversal symmetry breaking and van Hove singularity}
\author{Bing-Sui Lu}
\email{binghermes@gmail.com, bslu@ntu.edu.sg}
\author{Khatee Zathul Arifa}
\author{Xing Ru Hong}
\affiliation{Division of Physics and Applied Physics, School of Physical and Mathematical Sciences, Nanyang Technological University, 21 Nanyang Link, 637371 Singapore.
}
\date{\today}

\begin{abstract}
We consider the generic problem of a two-level quantum emitter near a two-dimensional Chern insulator in the dipole approximation, and study how the frequency-dependent  response and electronic density of states of the insulator modifies the transition rate of the emitter between the ground and excited levels. 
To this end, we obtain the full real-frequency behavior of the conductivity tensor by performing a tight-binding calculation based on the Qi-Wu-Zhang model and using a Kubo formula, and derive the full electromagnetic Green tensor of the system, which breaks Onsager reciprocity. 
This enables us to find that for frequencies smaller than the maximum band gap, the system is sensitive to time reversal symmetry-breaking, whereas for much larger frequencies  the system becomes insensitive, with implications for the discrimination of the state of a circularly polarised dipole emitter. 
We also study the impact of a van Hove singularity on the surface-induced correction to the transition rate, finding that it can enhance its amplitude by a few orders of magnitude compared to the case where the conductivity is set to its static value.  
By considering configurations in which the dipole is circularly polarised or parallel with the surface of the Chern insulator, we find that the surface correction to the transition rate can exhibit a novel decay with sine integral-like oscillations. 
\end{abstract}

\maketitle

\section{Introduction} 
    
Chern insulators are two-dimensional electronic systems which exhibit the quantum anomalous Hall effect (QAHE) in the static limit, whereby the bulk interior is insulating but the edge of the system conducts a current~\cite{weng2015,liu2016}. The corresponding static Hall conductivity is known to be integer-quantized in units of $e^2/h$, this integer quantization being essentially related to a non-trivial Berry phase~\cite{cayssol2013,bernevig2013}. In principle, Chern insulators can be realised in a number of ways~\cite{weng2015}, {\em e.g.}, they can be materially approximated by ultrathin films of three-dimensional magnetic topological insulators~\cite{ren2016,zhang2016}. Similar to the quantum Hall effect, the QAHE emerges from the breaking of time-reversal symmetry (TRS); however, unlike the former effect, the latter effect arises without the aid of an external magnetic field, being entirely induced by magnetization. In the context of the Casimir effect~\cite{lifshitz1955}, studies have predicted the appearance of a repulsive Casimir force~\cite{grushin2011,pablo2014,hoye2018}, which also showed that the existence of the Hall conductivity is a necessary ingredient for Casimir repulsion to occur. 
In other physical contexts, for example, the domain of atom physics and spontaneous emission, there has also been growing awareness of how the lifetimes of excited atoms can be modified by being in the vicinity of topological materials. The problem was first considered in Ref.~\cite{song2014}, for a dipole aligned either perpendicular to or parallel with the surface of a three-dimensional magnetic topological insulator (MTI). The consideration was then extended in Ref.~\cite{fuchs2017} to a circularly polarised dipole configuration. Quantum interference from spontaneous emission has been studied for an atom in a cavity made out of semi-infinite three-dimensional MTI slabs~\cite{fang2015} and a cavity made out of multilayered slabs consisting of alternating layers of MTI and normal insulator~\cite{zeng2019}. 


In light of the above developments, we have carried out an investigation into the spontaneous emission behavior of a two-level quantum emitter near a two-dimensional Chern insulator. 
Chern insulators are theoretically simpler to analyse than MTIs, capturing the generic feature of the QAHE without the additional complications of a three-dimensional dielectric bulk. 
In spite of this simplification, we have not been aware of any study on the spontaneous emission of an emitter near a Chern insulator. 
Furthermore, in the above-referenced studies on the spontaneous emission behavior near three-dimensional MTIs, the practice has been to work within the framework of axion electrodynamics, and study the contribution of the topologically quantised, static value of the axion (the axion being essentially related to the Hall conductivity)~\cite{wilczek1987,qi2008,lu2018}. Such a framework inadvertently excludes the contribution of the longitudinal conductivity. On physical grounds, we expect that if the frequency of the ambient radiation is larger than the insulator's band gap, the insulator can absorb radiation to create electron-hole pairs which results in a nonzero longitudinal conductivity at that frequency~\cite{ziegler2013}. This consideration, as well as the physical requirement that the response function should vanish for an infinitely large frequency, behooves one to consider the full frequency dispersion of the conductivity \emph{tensor}. 
In the case of the Chern insulator, we can account for these requirements by starting from a lattice model (i.e., the Qi-Wu-Zhang model~\cite{qwz2006}), deriving the conductivity tensor from the Kubo formula, and identifying the Kubo conductivity tensor with the conductivity tensor of the Maxwell equations. Such an identification holds for transverse fields~\cite{eykholt1986}, which is the case we consider. 

Yet a further reason motivates our present investigation. It is known that van Hove singularities (VHS) in the electronic density of states (DOS) tend to be more pronounced for two-dimensional periodic systems than for three-dimensional ones~\cite{bassani1975,dressel2002,gonzalez-tudela2019}. Where the VHS appear as beaks and shoulders for three-dimensional systems, they are logarithmic divergences near saddle points of the electronic band structure for two-dimensional systems. Furthermore, as we see in the paper, at a certain value of the mass gap in a Chern insulator, there can be VHS which are effectively one-dimensional, resulting in a power-law divergence. It is thus of interest to investigate the impact of contributions near such VHS on spontaneous emission. 



 
To address the above-delineated physical questions, we calculate the modification to the transition rate of a quantum emitter (or atom) induced by the presence of a Chern insulator, based on the following assumptions: firstly, the quantum emitter is sufficiently small such that it can be approximated by a point dipole. 
Secondly, we assume that the quantum emitter is sufficiently far away such that its wavefunction does not overlap significantly with the orbital wavefunctions of the insulator. Thus, our calculations are valid for insulator-quantum emitter distances which fall within the nanometer to micron range. Our third assumption is that the temperature should be the same for both the quantum emitter and the insulator, being so low that it can be approximated by zero temperature. This means we are allowed to use linear response theory for small departures from thermal equilibrium to calculate the transition rate for the two-level quantum emitter and the conductivity tensor for the two-band insulator. 
On the basis of these assumptions, we derive the dyadic response function for a dipole emitter above a single-layered Chern insulator, and use it to study the behavior of the transition rate of the quantum emitter for various dipole configurations.  

\section{Dyadic response function} 

\begin{figure}[h]
\centering
  \includegraphics[width=0.37\textwidth]{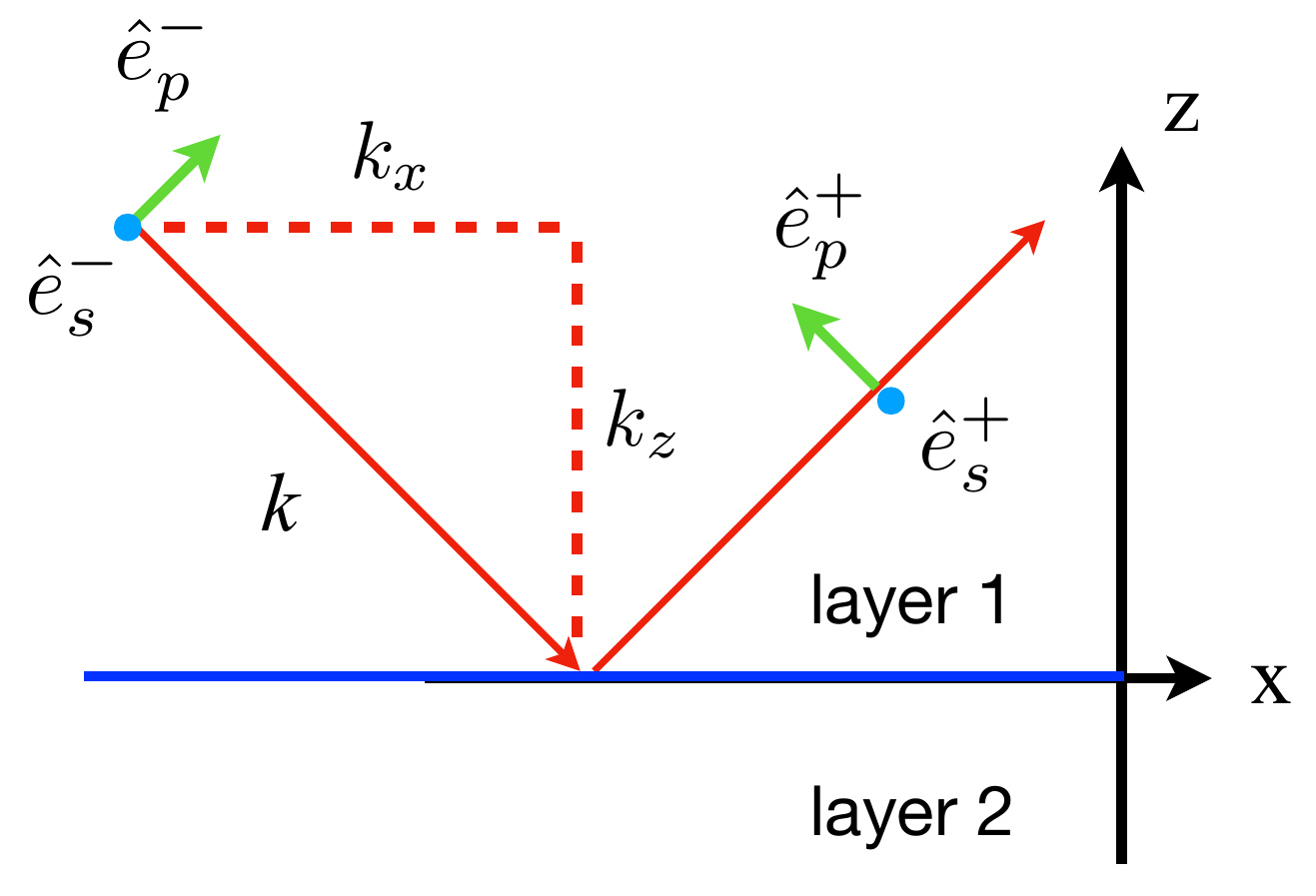}
  \caption{s and p polarisations, denoted respectively by $e_s^\pm$ (blue dot, directed along the negative y-direction) and $e_p^\pm$ (green arrow), for the case where the plane of incidence is the xz plane. For this case, the transverse wave vector lies entirely along the x direction, $k_y=0$, and $e_s^\pm = (0,-1,0)$ and $e_p^\pm = (1/k)(\mp k_z, 0, k_x)$. The $+$ ($-$) superscript refers to a wave propagating in the positive (negative) z-direction, $k = \omega/c$, and red arrows denote the propagation directions of incident and reflected waves.} 
  \label{polarisations}
\end{figure}
Consider a medium with a single point dipole source. The frequency-domain Maxwell equations are given by 
\begin{subequations}
\ba
&&\nabla\!\cdot\! \big( {\bf D} + 4\pi \Pv_d \big) = 0,
\\
&&\nabla\!\times{\bf H} = -i(\omega/c) \big( {\bf D} + 4\pi \Pv_{d} \big),
\\
&&\nabla\!\times\!{\bf E} = i(\omega/c) {\bf B},
\\
&&\nabla\!\cdot\!{\bf B} = 0
\ea
\end{subequations}
In the above, $\Pv_d(\rv,\omega) \equiv \pv(\omega) \delta(\rv-\rv_0)$, and describes an oscillating point dipole source of strength $\pv$ at position $\rv_0$. 
The quantity ${\bf D} \equiv \varepsilon {\bf E}$ denotes the displacement field. 
Taking the curl of the third Maxwell equation leads to
\ba
\nabla\times\nabla\times{\bf E} &=& i(\omega/c)\nabla\times{\bf B} 
= 
(\omega/c)^2 (\varepsilon {\bf E} + 4\pi {\bf P}_d). 
\ea
In the above, and we have made use of the equality ${\bf B} = {\bf H}$, this equality being only valid for non-magnetic materials, and this enabled us to make use of the second Maxwell equation on going from the first to the second equality.  
Rearranging terms, we obtain an inhomogeneous vector Helmholtz equation for ${\bf E}$:
\be
\label{eqE}
\left[ \nabla\times\nabla\times - \varepsilon(\omega/c)^2 \, {\mathbb I} \right] {\bf E}(\rv,\omega)
= 4\pi (\omega/c)^2 {\bf P}_d(\rv,\omega),
\ee
where ${\mathbb I}$ is the identity dyad. The equation above implies that the dipole ${\bf P}_d$ gives rise to the electric field ${\bf E}$. 
We can invert the vector Helmholtz equation to obtain 
\be
{\bf E}(\rv; \omega) 
= \mathbb{F}(\rv,\rv_0; \omega) \!\cdot\! \pv(\omega), 
\label{hellen1}
\ee
where ${\mathbb F}$ is the electromagnetic Green tensor, obeying 
\be
\label{green}
\left[ \nabla\times\nabla\times - \varepsilon(\omega/c)^2 \, {\mathbb I} \right] {\mathbb F}(\rv,\rv';\omega)
= 4\pi (\omega/c)^2 {\mathbb I} \, \delta(\rv - \rv'). 
\ee
In the Green tensor, it is useful to distinguish between \emph{source} and \emph{field} (or observation) points. The source (field) point is specified by the position $\rv_0$ ($\rv$), and refers to the location of the dipole source (field where measurement is made). 
Solving the boundary-value problem for ${\mathbb F}$ thus entails solving the boundary-value problem for ${\bf E}$. 

The Green tensor consists of bulk and scattering contributions: $\mathbb{F} = \mathbb{F}^{(0)} + \mathbb{F}^{(s)}$, where the bulk contribution $\mathbb{F}^{(0)}$ is the contribution in free space, and the scattering contribution $\mathbb{F}^{(s)}$ is the modification arising from the presence of boundary surfaces. If there is only a single dielectric interface, then we call the scattering contribution in the dielectric half-space with the source present the reflection Green tensor $\mathbb{F}^{R}$, whilst we call the scattering contribution in the dielectric half-space without the source the transmission Green tensor $\mathbb{F}^{T}$. 

For a single dielectric interface, the dyadic response function $\mathbb{G}$ (which is the quantity that the fluctuation-dissipation theorem relates to the field correlation function~\cite{LL5,fain1969}) is related to the Green tensor via 
\begin{subequations}
\ba
\label{GFE} 
G_{ab}^{(0)}(\rv,\rv'; \omega) &\!=\!& F_{ab}^{(0)}(\rv,\rv'; \omega) + 4\pi \delta_{ab} \delta(\rv-\rv'),
\\
G_{ab}^{(R)}(\rv,\rv'; \omega) &\!=\!& F_{ab}^{(R)}(\rv,\rv'; \omega). 
\ea
\label{GsFs}
\end{subequations}
The bulk Green tensor was obtained in Ref.~\cite{tomas1} (see also App.~\ref{app1}), and given in two-dimensional Fourier space by
\ba
&&{\mathbb F}^{(0)} (\kv_\parallel, z, z_0; \omega) 
\\
&\!=\!& 
-4\pi \delta(z-z_0) \, \hat{z} \hat{z} 
\nonumber\\
&&
+ \frac{2\pi i}{k_z} \left( \frac{\omega}{c} \right)^2
\sum_{\sigma = p, s}
\xi^\sigma
\Big[ 
\hat{e}_\sigma^+ (\kv_\parallel) \hat{e}_\sigma^- (-\kv_\parallel) 
e^{ik_z(z-z_0)} \Theta(z-z_0)
\nonumber\\
&&+
\hat{e}_\sigma^- (\kv_\parallel) \hat{e}_\sigma^+ (-\kv_\parallel) 
e^{-ik_z(z-z_0)} \Theta(z_0-z)
\Big]. 
\nonumber
\ea
Here $\kv_\parallel = (k_x, k_y)$, $k_z \equiv ((\omega/c)^2 - k_\parallel^2)^{1/2}$, $\Theta(z)$ is the Heaviside function which is equal to zero (unity) if $z < 0$ ($z > 0$). The symbol $\sigma=s,p$ refers to the s and p polarisations, $\xi^s = -1$ and $\xi^p = 1$, and polarisation vectors are given by $\hat{e}_p^\pm(\kv_\parallel) = (1/k)(\mp k_z \hat{\kv}_\parallel + k_\parallel \hat{z})$ and $\hat{e}_s^\pm(\kv_\parallel) = \hat{\kv}_\parallel \times \hat{z}$ (see Fig.~\ref{polarisations}). 

\subsection{scattering Green tensor}
\begin{figure}[h]
\centering
  \includegraphics[width=0.34\textwidth]{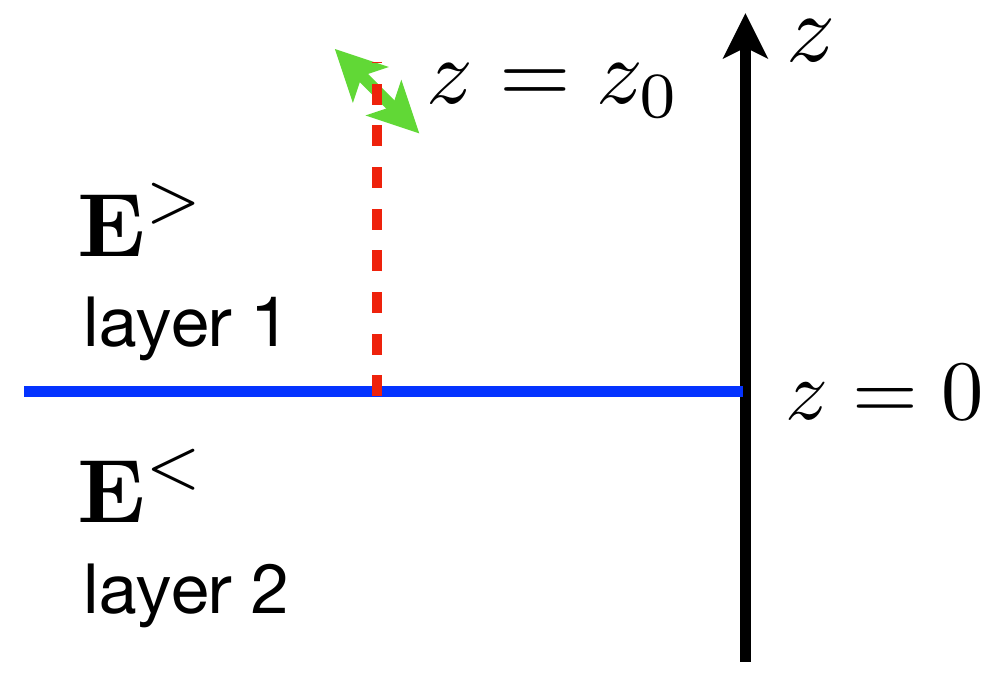}
  \caption{A dipole (green double-arrow) above Chern insulator layer (blue line), with layers 1 and 2 being the vacuum.} 
  \label{dipole}
\end{figure}
We now turn to derive the scattering Green tensor. We assume that the atom is at a height $z = z_0$ above the Chern insulator surface (which is at $z = 0$, cf. Fig.~\ref{dipole}), and write the total electric field as $\Ev^>$ ($\Ev^<$) in the $z > 0$ ($z < 0$) subspace: 
\begin{subequations}
\ba
\Ev^>(\kv_\parallel, z; \omega) &\!\!=\!\!& \Ev^{(0)}(\kv_\parallel, z; \omega) + \Ev^R (\kv_\parallel, z; \omega), 
\\
\Ev^<(\kv_\parallel, z; \omega) &\!\!=\!\!& \Ev^{T}(\kv_\parallel, z; \omega). 
\ea
\end{subequations}
In the above, we denote the scattering field in the $z>0$ ($z<0$) subspace by $\Ev^R$ ($\Ev^T$). The physical meaning of $\Ev^R$ ($\Ev^T$) is that it is the reflected (transmitted) wave, and so it only consists of wave propagating in the positive (negative) z-direction: 
\begin{subequations}
\ba
\label{ER}
{\bf E}^R(\kv_\parallel, z; \omega) &\!=\!& 
\sum_{\sigma=p, s} 
A_\sigma e^{i k_{z} z} \hat{e}_\sigma^+ (\kv_\parallel), 
\\
\label{ET}
{\bf E}^T(\kv_\parallel, z; \omega) &\!=\!& 
\sum_{\sigma=p, s} 
B_\sigma e^{-i k_{z} z} \hat{e}_\sigma^{-} (\kv_\parallel).  
\ea
\end{subequations}
Thus the total wave in the $z>0$ subspace is 
\ba
&&{\bf E}^>(\kv_\parallel, z; \omega) 
\\
&\!=\!& 
{\bf E}^{(0)}(\kv_\parallel, z; \omega) 
+ {\bf E}^R(\kv_\parallel, z; \omega)
\nonumber\\
&\!=\!&
-4\pi \delta(z-z_0) \, \hat{z} \hat{z} \cdot \pv(\omega)
\nonumber\\
&&+ 
\sum_{\sigma=p, s} 
\big[
A_\sigma^{(0)} e^{i k_{z} z} \Theta(z-z_0) \hat{e}_\sigma^+ (\kv_\parallel) 
\nonumber\\
&&+ 
B_\sigma^{(0)} e^{-i k_{z} z} \Theta(z_0-z) \hat{e}_\sigma^- (\kv_\parallel) 
\big]
+
\sum_{\sigma=p, s} 
A_\sigma e^{i k_{z} z} \hat{e}_\sigma^+ (\kv_\parallel)
\nonumber
\ea
Our task is to determine the values of these coefficients. 

At the interface, $z = 0$; correspondingly, we have 
\ba
&&{\bf E}^>(\kv_\parallel, 0; \omega) 
=
\sum_{\sigma=p, s} 
\big( 
B_\sigma^{(0)} \hat{e}_\sigma^- (\kv_\parallel) 
+
A_\sigma \hat{e}_\sigma^+ (\kv_\parallel)
\big), 
\nonumber\\
&&{\bf E}^T(\kv_\parallel, 0; \omega) 
=
\sum_{\sigma=p, s} 
B_\sigma \hat{e}_\sigma^{-} (\kv_\parallel).  
\label{346}
\ea
In each vacuum subspace, the s and p polarisations are eigensolutions to the Maxwell equations. On the other hand, the Hall current in the Chern insulator couples different polarisations. Consequently, the coefficients $A_\sigma$ and $B_\sigma$ each consist of one part which originates from the same polarisation $\sigma$ and another part which originates from the other polarisation $\sigma'$ of the incident wave:
\begin{subequations}
\label{347}
\ba
A_s &=& r_{ss} B_s^{(0)} + r_{ps} B_p^{(0)}, 
\\
A_p &=& r_{pp} B_p^{(0)} + r_{sp} B_s^{(0)}, 
\\
B_s &=& t_{ss} B_s^{(0)} + t_{ps} B_p^{(0)}, 
\\
B_p &=& t_{pp} B_p^{(0)} + t_{sp} B_s^{(0)}. 
\ea 
\end{subequations}
In the above, we have identified $B_s^{(0)}$ and $B_p^{(0)}$ as the amplitudes of the s- and p-polarised waves which are incident on the interface, and the coefficients $r_{ss}, r_{ps}, r_{sp}$, and $r_{pp}$ ($t_{ss}, t_{ps}, t_{sp}$, and $t_{pp}$) are the reflection (transmission) coefficients respectively for an incident s-polarised wave getting reflected (transmitted) as an s-polarised wave, an incident p-polarised wave getting reflected (transmitted) as an s-polarised wave, an incident s-polarised wave getting reflected (transmitted) as a p-polarised wave, and an incident p-polarised wave getting reflected (transmitted) as a p-polarised wave. 

To determine the values of the coefficients, a standard procedure is to consider the cases of incident s- and p-polarised waves separately, i.e., by considering cases for which $B_s^{(0)} = 0, B_p^{(0)} \neq 0$ and $B_p^{(0)} = 0, B_s^{(0)} \neq 0$, and solving the corresponding boundary-value problems for the tangential components of the E and H fields. 
The E and H fields can be obtained from the Green tensor. Moreover, because the Green tensor has both bulk and scattering contributions, the E and H fields are correspondingly also expressible in terms of bulk and scattering contributions, {{\it e.g.}}, ${\bf E} = {\bf E}^R + {\bf E}^T$. We have 
\begin{subequations}
\label{EG}
\ba
{\bf E}^{R}(\kv_\parallel, z; \omega) 
&\!\!=\!\!& 
{\mathbb F}^{R} (\kv_\parallel, z, z_0; \omega) \cdot \pv(\omega),
\\
{\bf E}^{T}(\kv_\parallel, z; \omega) 
&\!\!=\!\!& 
{\mathbb F}^{T} (\kv_\parallel, z, z_0; \omega) \cdot \pv(\omega).  
\ea
\end{subequations}
After some calculation (see App.~\ref{app1}), we find the following components of ${\mathbb{G}}^{R}(\rv_0,\rv_0;\omega)$:
\begin{subequations}
\label{dyadic}
\ba
\label{Gxx}
&&G_{xx}^R(\rv_0,\rv_0;\omega) = G_{yy}^R(\rv_0,\rv_0;\omega) 
\\
&\!\!=\!\!&
\frac{i}{2} \Big( \frac{\omega_{10}}{c} \Big)^3 \!\!
\int_0^\infty \!\!\! d\tk_\parallel (\tk_\parallel/\tk_z)
\big( (\omega/\omega_{10})^2 
 r_{ss} - \tk_z^2 r_{pp} 
\big)
e^{i \tk_z \eta},
\nonumber
\\
&&G_{xy}^R(\rv_0,\rv_0;\omega) = - G_{yx}^R(\rv_0,\rv_0;\omega) 
\label{Gxy}
\\
&\!\!=\!\!& 
\frac{i}{2} \Big( \frac{\omega_{10}}{c} \Big)^3 \!\!
\int_0^\infty \!\!\! d\tk_\parallel \tk_\parallel (\omega/\omega_{10}) (r_{ps} + r_{sp}) e^{i \tk_z \eta},
\nonumber\\
&&G_{zz}^R(\rv_0,\rv_0;\omega) = 
i \Big( \frac{\omega_{10}}{c} \Big)^3 \!\! 
\int_0^\infty \!\!\! d\tk_\parallel (\tk_\parallel^3/\tk_z) r_{pp} e^{i \tk_z \eta}. 
\nonumber\\
\label{Gzz}
\ea
\end{subequations}
Here, we defined $\widetilde{k}_z \equiv ck_z/\omega_{10}$, $\tk_\parallel \equiv ck_\parallel/\omega_{10}$, and $\eta \equiv 2 \omega_{10} z_0 / c$. 
We see that $G_{xy}^R(\rv_0,\rv_0;\omega) = - G_{yx}^R(\rv_0,\rv_0;\omega)$. This shows that the Chern insulator is nonreciprocal.

\subsection{Fresnel coefficients}

To obtain Fresnel coefficients for the Chern insulator, we use the fact that the surface conductivity only modifies the boundary conditions for $H_x$ and $H_y$, but otherwise leaves the bulk Maxwell equations unchanged. Thus, the s- and p-polarised modes are still the eigenmodes of electromagnetic waves in the region outside the Chern insulator. 
Boundary conditions (BCs) are obtained by integrating Maxwell equations from $z=0-$ to $z=0+$, which lead to 
\begin{subequations}
\label{BCs}
\ba
\label{BCsE}
&&E_x(0-) = E_x(0+), \,\, E_y(0-) = E_y(0+); 
\\
&&H_x(0+) - H_x(0-) = (4\pi/c) (\sigma_{xx} E_y(0) - \sigma_{xy} E_x(0)), 
\nonumber\\
&&H_y(0-) - H_y(0+) = (4\pi/c) (\sigma_{xx} E_x(0) + \sigma_{xy} E_y(0)). 
\nonumber\\
\label{BCsH}
\ea
\end{subequations}
In deriving the above BCs, we have made use of the relations $\sigma_{xx} = \sigma_{yy}$ and $\sigma_{xy} = - \sigma_{yx}$, which follow from spatial isotropy in the two-dimensional plane and the fact that the QAHE in the Chern insulator breaks time-reversal symmetry; the relations can also be deduced from the Kubo conductivity formula which we shall introduce in the next section. 
Solving the BCs leads to the following reflection coefficients~\cite{footnote1}: 
\begin{subequations}
\label{r-coeffs}
\ba
r_{ss} &\!\!=\!\!& 
-\frac{1}{\Delta} \big( \widetilde{\sigma}_{xx}^2 + \widetilde{\sigma}_{xy}^2 + \widetilde{k}_z^{-1} \widetilde{\sigma}_{xx} \big), 
\\
r_{ps} &\!\!=\!\!& r_{sp} = - \frac{\widetilde{\sigma}_{xy}}{\Delta}, 
\\
r_{pp} &\!\!=\!\!& 
\frac{1}{\Delta} \big( \widetilde{\sigma}_{xx}^2 + \widetilde{\sigma}_{xy}^2 + \widetilde{k}_z \widetilde{\sigma}_{xx} \big), 
\ea
\end{subequations}
where we defined $\widetilde{\sigma}_{\mu\nu} \equiv (2\pi/c) \sigma_{\mu\nu}$, and 
\be
\Delta \equiv 1 + \big( \widetilde{k}_z + {\widetilde{k}_z^{-1}} \big) \widetilde{\sigma}_{xx} + \widetilde{\sigma}_{xx}^2 + \widetilde{\sigma}_{xy}^2.
\label{Delta}
\ee
As the longitudinal conductivity $\sigma_{xx}$ is invariant under time reversal, whereas the Hall conductivity $\sigma_{xy}$ changes sign, the reflection coefficients $r_{ss}$ and $r_{pp}$ remain unchanged whereas $r_{ps}$ changes sign as we change the sign of $C$.  
In the nondispersive (static) limit, $\sigma_{xx}(0) = 0$ and $\sigma_{xy}(0) = Ce^2/h$ for a Chern insulator with Chern number $C$. The Fresnel coefficients then become 
\begin{subequations}
\label{r-static}
\ba
r_{ss} &\!\!=\!\!& - r_{pp} = -\frac{(C \alpha)^2}{1 + (C \alpha)^2}, 
\\
r_{sp} &\!\!=\!\!& r_{ps} = -\frac{C \alpha}{1 + (C \alpha)^2}, 
\ea
\end{subequations}
where $\alpha \equiv e^2/(\hbar c)$ is the fine-structure constant. 
Here, the sign of $C$ is positive (negative) if the magnetization or orientation vector of the Hall current on the Chern insulator is in the positive (negative) $\hat{z}$ direction, i.e., opposite to (in the same direction as) the normal direction of the incident wave. 

\subsection{Kubo conductivity tensor}

The conductivity tensor can be calculated from the Kubo formula, given by~\cite{czycholl2017} 
\ba
\sigma_{\mu\nu}(\omega) &=& 
-\frac{i}{\hbar}\lim_{\epsilon\rightarrow 0} 
\int_{BZ} \!\frac{d^2k}{(2\pi)^2} \sum_{\ell,\ell'=0,1} 
\frac{\langle E_\ell | j_\mu | E_{\ell'} \rangle \langle E_{\ell'} | j_\nu | E_{\ell} \rangle}{E_\ell - E_{\ell'} + \hbar \omega + i\epsilon} 
\nonumber\\
&&\times 
\frac{f(E_\ell) - f(E_{\ell'})}{E_\ell - E_{\ell'}}. 
\label{kubo}
\ea
The integration is over wavevectors belonging to the Brillouin zone (which we denote by $BZ$), $f(E_\ell) = (\exp(\beta (E_\ell - \mu)) + 1)^{-1}$ is the Fermi-Dirac distribution (with chemical potential $\mu$) and $j_\mu$ is the current operator. 
As we are interested in the case where the Fermi level is in the band gap, we assume for simplicity that the chemical potential is zero. 
The symbol $| E_\ell \rangle$ denotes the eigenket of $H$ with eigenvalue $E_\ell$ ($\ell$ being the band index, and $\ell = 0$ ($\ell = 1$) denoting the valence (conduction) band. 
Using the representation of the Dirac delta-function $\delta(x) = (1/\pi) \lim_{\epsilon \rightarrow 0} \epsilon/(\epsilon^2 + x^2)$, we can write
\ba
\label{kubo1}
&&\sigma_{\mu\nu}(\omega)
\\
&\!=\!& 
-\frac{1}{\hbar}
\int_{BZ} \!\!\frac{d^2k}{(2\pi)^2} 
\sum_{\ell,\ell'} 
\langle E_\ell | j_\mu | E_{\ell'} \rangle \langle E_{\ell'} | j_\nu | E_{\ell} \rangle
\frac{f(E_\ell) - f(E_{\ell'})}{E_\ell - E_{\ell'}} 
\nonumber\\
&&\times 
\left( \frac{i}{E_\ell - E_{\ell'} + \hbar\omega}
+ \pi \delta(E_\ell - E_{\ell'} + \hbar\omega)
\right)
\nonumber
\ea
For a model of the Chern insulator, we adopt the Qi-Wu-Zhang (QWZ) model~\cite{qwz2006}, which is defined on a square lattice with a lattice constant $a$, and described by~\cite{asboth2016}
\be
\label{H2band}
H = {\bf{d}}(\kv) \cdot {\bm{\sigma}} = d_x(\kv) \sigma_x + d_y(\kv) \sigma_y + d_z(\kv) \sigma_z, 
\ee
where $\sigma_x, \sigma_y, \sigma_z$ are the Pauli matrices, and 
\ba
d_x(\kv) &\!=\!& t \sin k_x a, 
\,\,
d_y(\kv) = t \sin k_y a,
\nonumber\\
d_z(\kv) &\!=\!& t (\cos k_x a + \cos k_y a) + u. 
\label{qwz}
\ea
The symbol $t$ denotes the hopping parameter, and $u$ denotes the band gap. 
It is known~\cite{qwz2006,asboth2016} that the Chern number is $1$ ($-1$) for $0 < u/t < 2$ ($-2 < u/t < 0$), and zero for $|u|/t > 2$. 
We depict the band structure in Fig.~\ref{bandstructureC1} for $u/t = 1$. 
The corresponding energy eigenvalues are 
\be
E_{\pm}(\kv) = \pm d(\kv) = \pm \sqrt{d_x^2(\kv) + d_y^2(\kv) + d_z^2(\kv)}.
\ee
We can express the corresponding eigenkets in the Bloch sphere representation~\cite{cayssol2013}, viz.,  
\ba
| \kv, + \rangle &\!\equiv\!& 
\begin{pmatrix}
\cos(\theta(\kv)/2) e^{-i\phi(\kv)}
\\
\sin(\theta(\kv)/2)
\end{pmatrix} 
e^{i\kv\cdot\rv}
\nonumber\\ 
&\!=\!& 
\frac{1}{2\sin(\theta(\kv)/2)}
\begin{pmatrix}
\sin \theta(\kv) e^{-i\phi(\kv)}
\\
1 - \cos \theta(\kv)
\end{pmatrix} 
e^{i\kv\cdot\rv},
\nonumber\\
| \kv, - \rangle &\!\equiv\!& 
\begin{pmatrix}
\sin(\theta(\kv)/2) e^{-i\phi(\kv)}
\\
- \cos(\theta(\kv)/2)
\end{pmatrix} 
e^{i\kv\cdot\rv}
\nonumber\\ 
&\!=\!& 
\frac{1}{2\cos(\theta(\kv)/2)}
\begin{pmatrix}
\sin \theta(\kv) e^{-i\phi(\kv)}
\\
- 1 - \cos \theta(\kv)
\end{pmatrix} 
e^{i\kv\cdot\rv}, 
\label{eigenkets}
\ea
where $d_x/d = \cos\phi\sin\theta$, $d_y/d = \sin\phi\sin\theta$, $d_z/d = \cos\theta$, or equivalently, $\cos\phi = d_x/(d_x^2 + d_y^2)^{1/2}$, $\sin\phi = d_y/(d_x^2 + d_y^2)^{1/2}$, $\cos\theta = d_z/d$, and $\sin\theta = (d_x^2 + d_y^2)^{1/2}/d$. 
For the purpose of numerical evaluation, it may be easier to work with the eigenkets expressed using $\dv = (d_x, d_y, d_z)$:
\begin{subequations}
\ba
| \kv, + \rangle &\!\equiv\!& \sqrt{\frac{d(\kv)-d_z(\kv)}{2d(\kv)}} 
\begin{pmatrix}
\frac{d_x(\kv) - id_y(\kv)}{d(\kv)-d_z(\kv)}
\\
1
\end{pmatrix} 
e^{i\kv\cdot\rv}, 
\\
| \kv, - \rangle &\!\equiv\!& \sqrt{\frac{d(\kv)+d_z(\kv)}{2d(\kv)}} 
\begin{pmatrix}
\frac{d_x(\kv) - id_y(\kv)}{d(\kv)+d_z(\kv)}
\\
- 1
\end{pmatrix} 
e^{i\kv\cdot\rv}. 
\ea
\end{subequations}
Next, we turn to the current operator. In the Heisenberg picture, this is given by 
$
j_\mu = 
- e(\partial H/\partial k_\mu).
$
In the eigenket basis, the current matrix elements are given by
\begin{subequations}
\label{jxjy}
\ba
j_x &\!\!=\!\!& 
- e 
\begin{pmatrix}
\frac{\partial d_z}{\partial k_x} & \frac{\partial d_x}{\partial k_x} - i \frac{\partial d_y}{\partial k_x}
\\
\frac{\partial d_x}{\partial k_x} + i \frac{\partial d_y}{\partial k_x} & - \frac{\partial d_z}{\partial k_x}
\end{pmatrix}, 
\\
j_y &\!\!=\!\!& 
- e 
\begin{pmatrix}
\frac{\partial d_z}{\partial k_y} & \frac{\partial d_x}{\partial k_y} - i \frac{\partial d_y}{\partial k_y}
\\
\frac{\partial d_x}{\partial k_y} + i \frac{\partial d_y}{\partial k_y} & - \frac{\partial d_z}{\partial k_y}
\end{pmatrix}.
\ea
\begin{figure}[h]
\centering
  \includegraphics[width=0.4\textwidth]{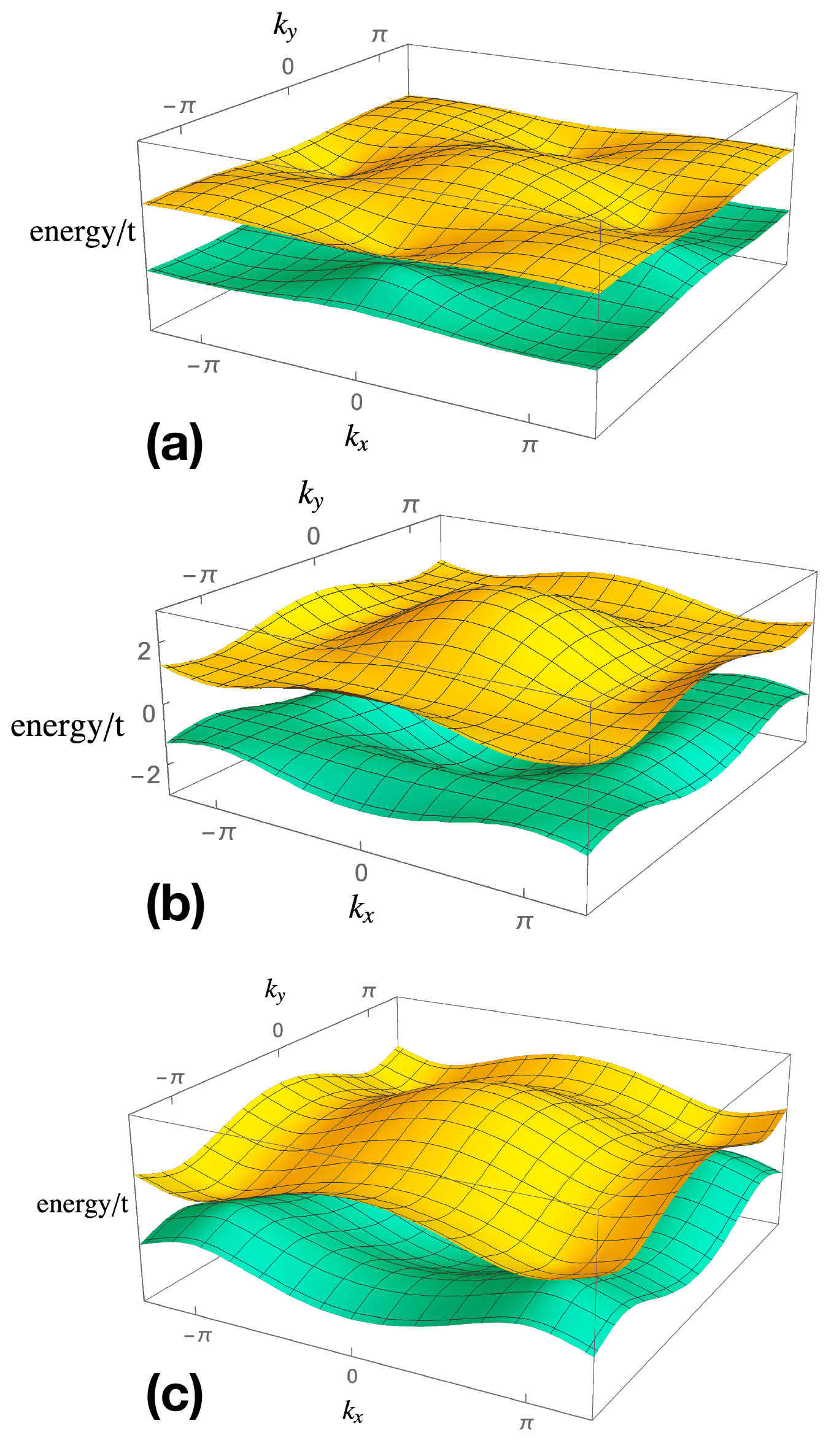}
  \caption{Band structures of Chern insulators with $C = 1$ modelised by Eq.~(\ref{qwz}): (a)~$u/t=1/4$; (b)~$u/t=1$; (c)~$u/t=7/4$ (where $u$ and $t$ are respectively the band gap and hopping parameter for the Qi-Wu-Zhang model). Points $\kv^{{\rm T}} = (\pm \pi/a, 0), (0, \pm \pi/a)$ are two-dimensional minima for $0 < u/t < 1$, and saddle points for $1 < u/t < 2$. For the special case $u = t$, they are effectively one-dimensional minima as one of the eigenvalues of the Hessian matrix vanishes.} 
  \label{bandstructureC1}
\end{figure}
\end{subequations}
\begin{figure*}[t]
\begin{center}
\begin{minipage}[b]{0.45\textwidth}
\includegraphics[width=\textwidth]{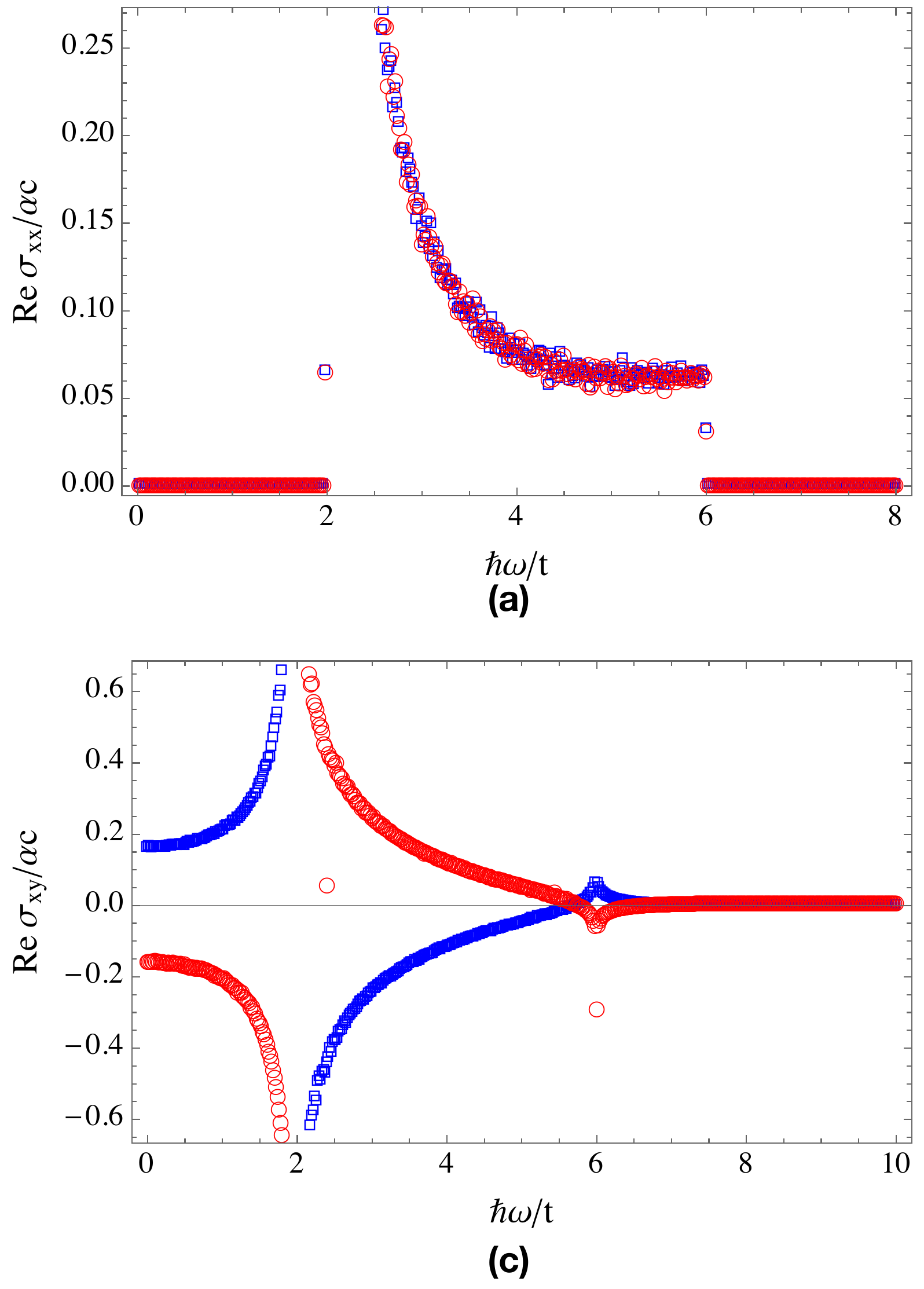} 
\end{minipage} 
\begin{minipage}[b]{0.45\textwidth}
\includegraphics[width=\textwidth]{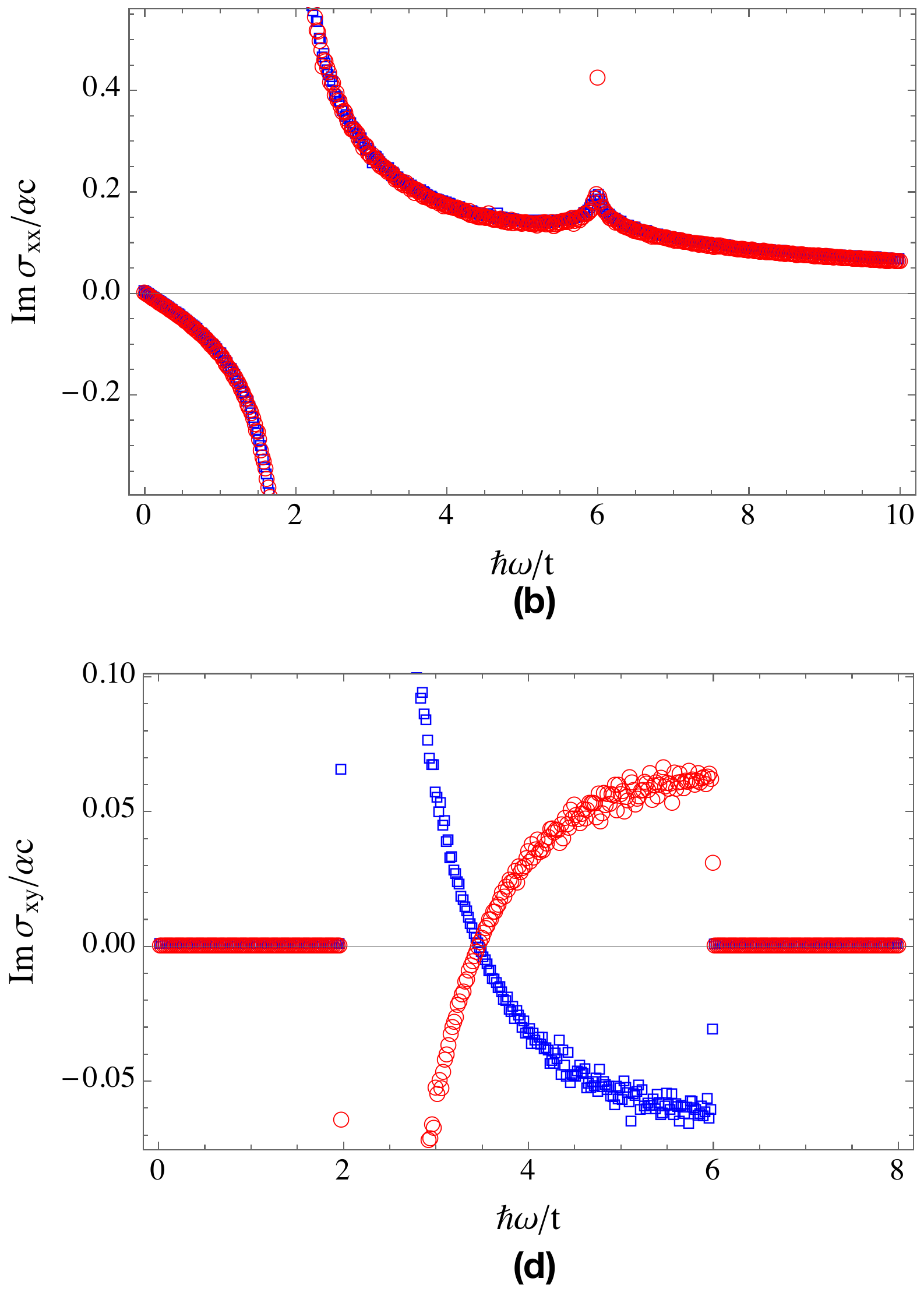} 
\end{minipage}
\end{center}
\caption{(color online) Real-frequency dispersion behavior, shown by curves with blue squares (red circles), of the conductivity tensor of a $C = 1$ ($C = -1$) Chern insulator for $u/t = 1$ ($u/t = -1$) in Eq.~(\ref{qwz}): (a)~${{\rm Re}} \, \sigma_{xx}(\omega)/(\alpha c)$, (b)~${{\rm Im}} \, \sigma_{xx}(\omega)/(\alpha c)$, (c)~${{\rm Re}} \, \sigma_{xy}(\omega)/(\alpha c)$, and (d)~${{\rm Im}} \, \sigma_{xy}(\omega)/(\alpha c)$ as functions of $\hbar \omega/t$ (horizontal axis).}
 \label{conduct-realC1}
 \end{figure*}
The Kubo formula Eq.~(\ref{kubo1}) involves an integration over $k_x$ and $k_y$, which causes the terms odd in $k_x$ and/or $k_y$ to vanish (see App.~\ref{app:current} for details). 
The Fermi function $f(E_\ell)$ and the terms $E_\ell - E_{\ell'}$ are even in $k_x$ and in $k_y$ (as these are functions of $d$, which is even in $k_x$ and in $k_y$). 
We find (see App.~\ref{app:current}) that the term ${{\rm Re}}\, [\langle + | j_x | - \rangle \langle - | j_y | + \rangle]$ is odd in both $k_x$ and $k_y$, so this vanishes under the integration. 
Thus the only surviving contributions to the conductivity tensor are given by 
\begin{subequations}
\ba
&&\sigma_{xx}(\omega) 
\\
&=& -\frac{1}{\hbar}
\int_{BZ} \!\frac{d^2\kv}{(2\pi)^2} 
{{\rm Re}} \, [\langle + | j_x | - \rangle \langle - | j_x | + \rangle]
\frac{f(d) - f(-d)}{2d}
\nonumber\\
&&\times
\bigg(
\pi \big(
\delta(\hbar\omega + 2d) + \delta(\hbar\omega - 2d)
\big)
+ 
\frac{2 i \hbar\omega}{\hbar^2\omega^2 - 4d^2} 
\bigg),
\nonumber\\
&&\sigma_{xy}(\omega) 
\\
&=& \frac{1}{\hbar}
\int_{BZ} \!\frac{d^2\kv}{(2\pi)^2} \, 
{{\rm Im}} \, [\langle + | j_x | - \rangle \langle - | j_y | + \rangle]
\frac{f(d) - f(-d)}{2d}
\nonumber\\
&&\times
\bigg(
\frac{4d}{4d^2-\hbar^2\omega^2} 
+
i \pi \big( 
\delta(\hbar\omega - 2d) - \delta(\hbar\omega + 2d)
\big)
\bigg). 
\nonumber
\ea
\label{cond-tensor}
\end{subequations}
In Fig.~\ref{conduct-realC1}, we show the behaviors of the longitudinal and Hall conductivities as functions of frequency for the cases $C = 1$ (blue, square) and $C = -1$ (red, circle). 
We have implemented the Brillouin zone integration numerically using quasi and adaptive Monte Carlo methods, modeling the Dirac delta function $\delta(x)$ by the heat kernel $\sqrt{\beta/(2\pi)} \exp (-\beta x^2/2)$, where $\beta$ is the inverse temperature (in units of $1/t$). For our calculations we have taken $\beta = 2 \times 10^4$. As we see in Fig.~\ref{conduct-realC1} (and also later in Figs.~\ref{dipole-para-intermediate}, \ref{dipole-parallel}a, \ref{dipole-circ-intermediate-Cone}a, \ref{dipole-circ-intermediate-Cminusone}a, \ref{dipole-circular}a), there is scatter which arises from sampling integration points for the Dirac delta and Fermi functions, whose values change rapidly over a very narrow interval. 
The scatter is more pronounced for non-analytic spectral regions such as van Hove singularities. 
In the plots, we have rescaled $\dv$ and $\hbar \omega$ in units of energy $t$, and $k_x$ and $k_y$ in units of $1/a$ in the conductivity tensor, so that the quantity $\sigma_{\mu\nu}(\omega)/(\alpha c)$ is a dimensionless function of a dimensionless frequency $\hbar\omega/t$. 
We see that changing the sign of $C$ does not affect the longitudinal conductivity, whereas the Hall conductivity changes sign. This is because changing the sign of $C$ can be regarded as a time reversal operation, and the longitudinal (Hall) conductivity is insensitive (sensitive) to time reversal. 

To understand the salient features of Fig.~\ref{conduct-realC1}, we first note that the expressions for $\sigma_{xx}'$ and $\sigma_{xy}''$ (where $\sigma_{\mu\nu}'$ and $\sigma_{\mu\nu}''$ denote the real and imaginary parts of $\sigma_{\mu\nu}$ respectively) in Eq.~(\ref{cond-tensor}) involve a Dirac delta function which is non-zero only if $\hbar\omega = 2d(\kv)$. For $0 < |u|/t < 2$, $2t - |u| \leq d(\kv) \leq 2t + |u|$, which implies that the frequencies at which $\sigma_{xx}', \sigma_{xy}'' \neq 0$ occur in the range $2(2t - |u|) \leq \hbar \omega \leq 2(2t + |u|)$, which for our considered case of $u/t = 1$ gives $2 \leq \hbar\omega/t \leq 6$. 
Within this frequency range, nonzero dissipative current fluctuations can appear as the insulator can absorb radiation of energy $2d(\kv)$ to promote an electron (with a Bloch wavevector $\kv$) from the valence band to the conduction band. 
Secondly, the wavevector integral over the Dirac delta function in Eqs.~(\ref{cond-tensor}) represents an electronic density of states (DOS), which is known to exhibit van Hove singularities (VHS) whenever $|{\bm{\nabla}}_\kv d(\kv)|$ vanishes~\cite{dressel2002}. These singularities are more pronounced in two-dimensional periodic systems than in three-dimensional ones, with the electronic DOS diverging logarithmically near a zero of $|{\bm{\nabla}}_\kv d(\kv)|$ which is also a saddle point (i.e., one of the eigenvalues of the corresponding Hessian matrix is positive and the other negative)~\cite{bassani1975}. 
For $1 < |u|/t < 2$, such saddle points appear at $\kv = (\pm\pi/a, 0), (0, \pm\pi/a)$, with the corresponding frequency being $\omega = 2|u|/\hbar$ (cf. Fig.~\ref{bandstructureC1}c). 
On the other hand, for $|u|/t = 1$, one of the eigenvalues of the Hessian matrix vanishes, and the points $\kv = (\pm\pi/a, 0), (0, \pm\pi/a)$ effectively become one-dimensional minima (cf. Fig.~\ref{bandstructureC1}b). The VHS is even more pronounced at such points, diverging with $|\hbar \omega - 2t|^{-1/2}$ (cf. App.~\ref{app:VHS}). 
As we see later in Sec.~IV, this can lead to a dramatic van Hove singularity-assisted enhancement of the surface correction to the transition rate for an emitter emitting a photon at the corresponding frequency.  

For frequencies $\hbar\omega/t \gg 2(2t + |u|)$, $\sigma_{xx}''$ decays as $\omega^{-1}$ whereas $\sigma_{xy}'$ decays as $\omega^{-2}$. Thus, for this frequency regime only $\sigma_{xx}''$ effectively contributes to the surface-induced correction in the emitter's transition rate. In this regime, we can therefore neglect $r_{ps}$, $r_{sp}$, $G_{xy}^R$, and $G_{yx}^R$, and we expect the transition rate for a circularly polarised dipole to be unable to discriminate between a $C = 1$ and a $C = -1$ Chern insulator surface. 

The nondispersive limit contrasts with the aforementioned frequency regime, in that both $\sigma_{xx}$ and $\sigma_{xy}''$ now vanish whereas $\sigma_{xy}' \neq 0$, and the Chern insulator exhibits the QAHE. The dispersive features of the conductivity tensor thus allows us to distinguish three qualitatively distinct ranges of frequency, and study the behavior of the transition rate behavior in each range. These ranges are: (i)~the nondispersive limit of the conductivity tensor (i.e., approximating the conductivity tensor by its static value), (ii)~low ($\hbar\omega/t < 2(2t - |u|)$) to intermediate $(2(2t - |u|) < \hbar\omega < 2(2t+|u|)$) frequencies, and (iii)~high frequencies ($\hbar\omega/t > 2(2t+|u|)$).

\section{Transition rate}

The transition rate can be obtained from the dyadic response function of the atom~\cite{wylie-sipe1,wylie-sipe2}, which we denote by $\mathbb{G}$. This relates the change in the expectation value of the displacement field at time $t$ induced by the appearance of a dipole $\pv$ at time $t'$, viz., 
\be
D_a(t)|_{\mu} = D_a(t)|_{\mu=0} + \int_{-\infty}^\infty \!\!\! dt' \, G_{ab}(t-t') \, p_b(t'). 
\label{greenD}
\ee
Here, $a,b = x,y,z$ labels the Cartesian coordinates. 
The response function can be expressed as a sum of two contributions, the first being the expression for an atom in free space (called the bulk contribution), whilst the second (called the scattering contribution) is the correction introduced by the presence of an insulating layer. Denoting the former contribution by $\mathbb{G}^{0}$ and the latter contribution by $\mathbb{G}^{R}$, the transition rate $R_{10}$ of a two-level atom from an excited state $|1\rangle$ to the ground state $|0\rangle$ at $T=0$ near the Chern insulator is given by 
\be
\frac{R_{10}}{R_{10}^{(0)}}
=
1 - \frac{3 i c^3 \mu_a^{01} \mu_b^{10}}{4\omega^3 |{\bm{\mu}}^{10}|^2} 
\big( 
G_{ab}^R(\rv_0, \rv_0; \omega_{10}) - G_{ba}^{R*}(\rv_0, \rv_0; \omega_{10}) 
\big). 
\label{R10}
\ee
We shall call $R_{10}/R_{10}^{(0)}$ the normalised transition rate. 
The above result can be obtained from the fluctuation-dissipation theorem. A version which is valid for reciprocal media, though not for nonreciprocal media, was employed in Ref.~\cite{wylie-sipe1}. For the Chern insulator, which is electromagnetically nonreciprocal, we have to adopt the more general form of the fluctuation-dissipation formula which involves the anti-Hermitian part of the dyadic response function rather than the imaginary part~\cite{LL5}. 
The symbol ${\bm{\mu}}^{10} \equiv \langle 1 | {\bm{\mu}} | 0 \rangle$ denotes the dipole transition matrix element from an initial state $| 0 \rangle$ to a final state $| 1 \rangle$, ${\bm{\mu}} = -e\, \hat{q}\, \nv$ is the electric dipole operator of the atom, with $\nv$ specifying the orientation of the dipole moment and $\hat{q}$ is the position operator, $\omega_{10} \equiv \omega_1 - \omega_0 \equiv  (E_1 - E_0)/\hbar$. 
If we adopt an oscillator model for the atom, we have that the dipole transition matrix element is
$\mu_a^{10} = \mu_a^{01*}
= \mu \, n_a$, 
where $\mu \equiv -e ( \hbar/(2m\omega_{10}) )^{1/2}$. 
The quantity $R_{10}^{(0)}$ denotes the transition rate for an atom in free space, which has the value $R_{10}^{(0)} = 4\omega_{10}^3 |{\bm{\mu}}^{10}|^2/(3 \hbar c^3)$~\cite{sakurai1967,grynberg2010,milonni2019}. 

As we are interested in how the presence of the Chern insulator modifies the transition rate, we focus on effects arising from the scattering contribution to the dyadic response function, i.e., $\mathbb{G}^R$. 
To this end, we consider the following four configurations: 
(i)~a dipole aligned perpendicular to the surface of a $C = 1$ Chern insulator; 
(ii)~a dipole aligned parallel with the surface of a $C = 1$ Chern insulator; 
(iii)~a right circularly polarised dipole with its quantisation axis perpendicular to the surface of a $C = 1$ Chern insulator; and 
(iv)~a right circularly polarised dipole with its quantisation axis perpendicular to the surface of a $C = -1$ Chern insulator (or equivalently, a left circularly polarised dipole with its quantisation axis perpendicular to a $C = 1$ Chern insulator). 
In particular, the circular dipole polarisation, now leads to a surface correction to the transition rate which is sensitive to time reversal, as the Chern insulator is nonreciprocal and the circular dipole polarisation is non-invariant under time reversal. Furthermore, as both the circular dipole polarisation and the Chern number are sensitive to time reversal, we should expect that the transition rate behaviors for configurations (iii) and (iv) would be different. Our expectation is borne out by the results of Sec.~\ref{sec:dipole-circ}. 


\section{Results and discussion} 

As described in Sec. II, we can distinguish three qualitatively distinct frequency ranges. In this Section, we first study the transition rate behavior in the nondispersive limit, then compare the behaviors in the low to intermediate frequency regime (where we consider $\hbar\omega_{10}/t = 1, 1.9, 2.1$, and $3$), where the feature of interest is the dramatic enhancement of the surface-induced correction to the transition rate as the energy approaches an effectively one-dimensional van Hove singularity (i.e., $\hbar\omega_{10} = 2t$ for $u = t$). 
Lastly, we look at the behavior in the high frequency regime (where we consider $\hbar\omega_{10}/t = 10, 100$). If the energy level spacing of the emitter is $\hbar \omega_{10} = 2\, {{\rm eV}}$ and $u = t$, then the mass gaps of the Chern insulator corresponding to $\hbar\omega/t = 1, 10, 100$ would respectively be $u = 2\, {{\rm eV}}$, $u = 0.2\, {{\rm eV}}$, and $u = 0.02\, {{\rm eV}}$.

In the low to intermediate frequency ranges, both real and imaginary parts of the longitudinal and Hall conductivities can be nonzero. 
Therefore in this regime (and also the nondispersive limit), we expect the transition rate for the circularly polarised dipole to be able to discriminate between $C = 1$ and $C = -1$ as long as the Hall conductivity is nonzero. 
As we shall see, differences in the conductivity behavior between the low and high frequency ranges can result in qualitative differences of power-law decay behavior for the transition rates. 

\subsection{dipole aligned perpendicular to surface} 
\begin{figure}[h]
\centering
  \includegraphics[width=0.46\textwidth]{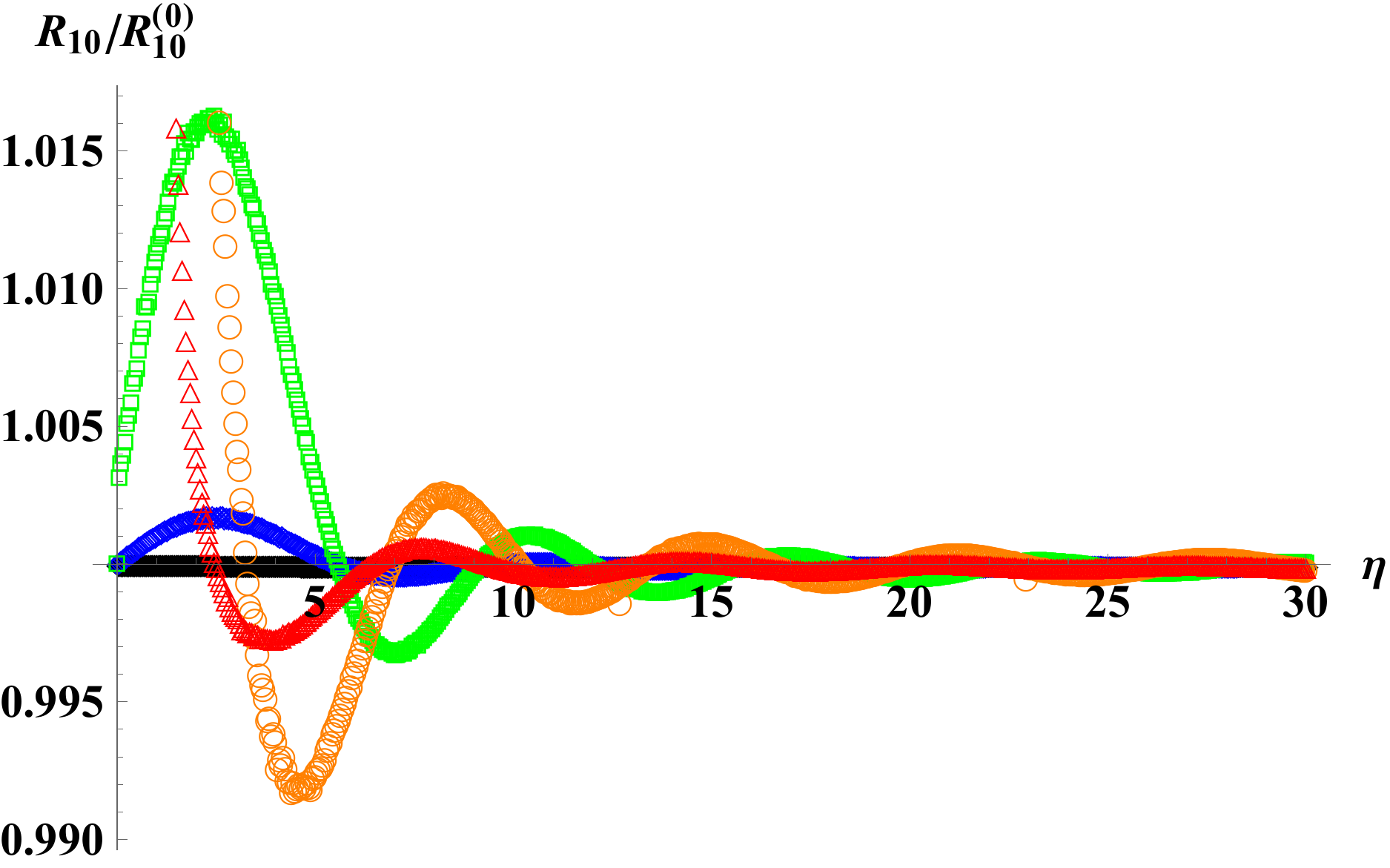}
\caption{Behavior of $R_{10}/R_{10}^{(0)}$ (vertical axis) as a function of $\eta = 2z_0 c/\omega_{10}$ (horizontal axis) for a dipole aligned perpendicular to the surface of a $C = 1$ Chern insulator specified by $u/t = 1$ in Eq.~(\ref{qwz}):~nondispersive case (black, filled diamond); dispersive case with $\hbar \omega/t = 1$ (blue, diamond), $\hbar \omega/t = 1.9$ (green, square), $\hbar \omega/t = 2.1$ (orange, circle), and $\hbar \omega/t = 3$ (red, triangle).}
 \label{dipole-vert-intermediate}
 \end{figure} 
\begin{figure}[h]
\centering
  \includegraphics[width=0.46\textwidth]{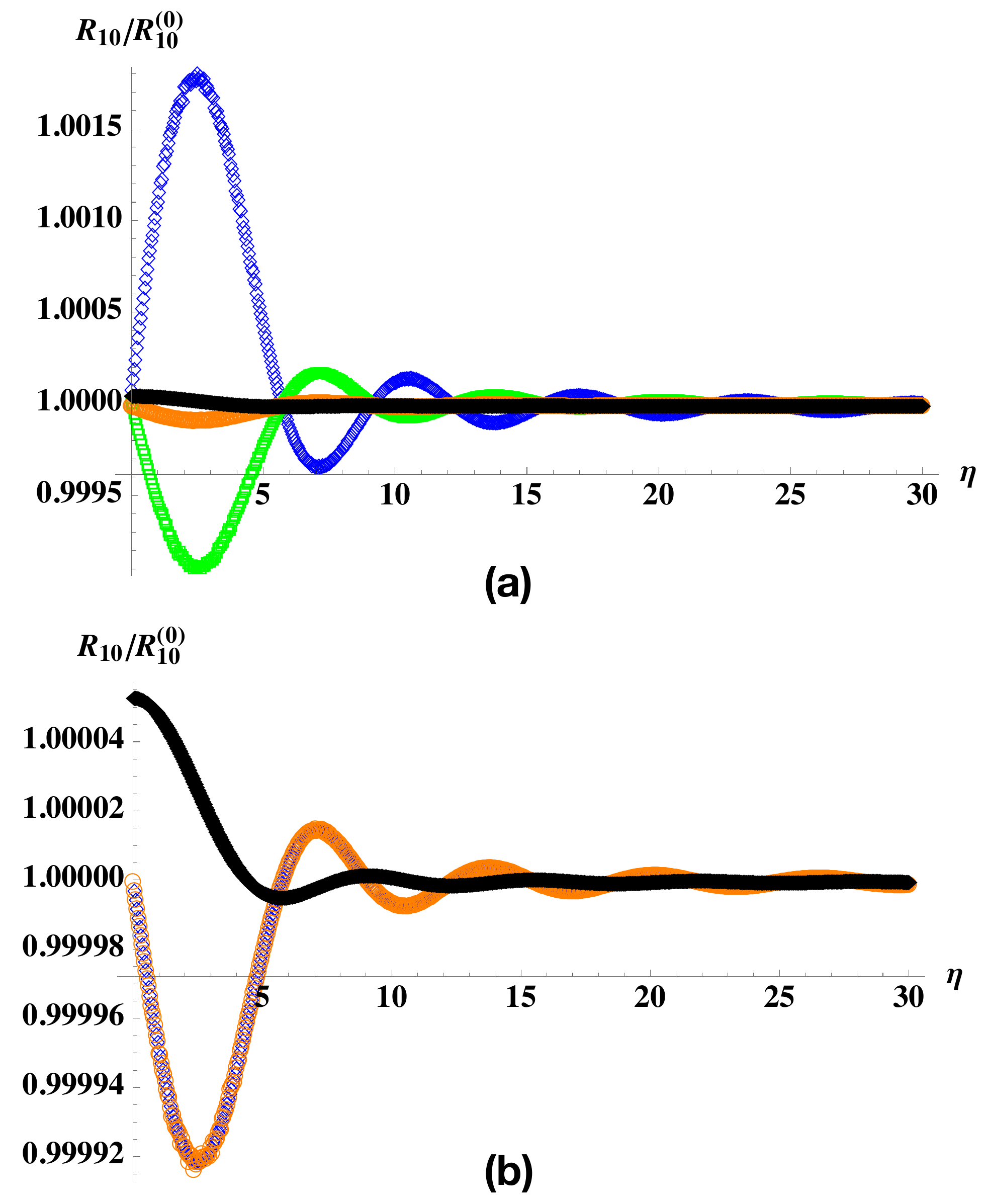}
  \caption{Behavior of $R_{10}/R_{10}^{(0)}$ (vertical axis) as a function of $\eta = 2z_0 \omega_{10}/c$ (horizontal axis) for a dipole oriented perpendicularly to the surface of a $C = 1$ Chern insulator with $u/t = 1$: (a)~Behaviors for the case of nondispersive conductivity (black, filled diamond) as well as dispersive conductivity with $t = \hbar\omega_{10}$ (blue, diamond), $t = 0.1 \hbar\omega_{10}$ (green, square), and $t = 0.01 \hbar\omega_{10}$ (orange, circle). The behavior is unchanged for the parameter choice $u/t = - 1$ which corresponds to $C = -1$.
  (b)~Nondispersive conductivity case (black, filled diamond), and comparison of the exact plot for $t = 0.01 \hbar\omega_{10}$ (orange, circle) with its intermediate asymptotic approximation Eq.~(\ref{asympperp}) (blue, diamond). The match between the latter two plots is excellent.} 
  \label{dipole-vertical}
\end{figure}
First, we consider a dipole oriented along the $z$-axis so that it lies perpendicular to the Chern insulator surface, i.e., $\nv^{{\rm T}} = (0,0,1)$. Applying this to Eq.~(\ref{R10}) gives
\be
\label{dipperp}
\frac{R_{10}}{R_{10}^{(0)}} = 1 + \frac{3 c^3}{2 \omega_{10}^3} {{\rm Im}}\, G_{zz}^R(\rv_0,\rv_0; \omega_{10}),   
\ee
where $G_{zz}^R$ is given by Eq.~(\ref{Gzz}). 

If we assume that the conductivity tensor is nondispersive, we can use Eqs.~(\ref{r-static}) and (\ref{Gzz}) to find the normalised transition rate in the retarded regime: 
\be
\label{tara}
\frac{R_{10, \, {{\rm nondisp}}}}{R_{10}^{(0)}} = 1 + \frac{3(C \alpha)^2}{1 + (C \alpha)^2} \frac{\sin \eta - \eta \cos \eta}{\eta^3}. 
\ee
In the far-field regime ($\eta \gg 1$), we can approximate this to leading order by 
\be
\frac{R_{10, \, {{\rm nondisp}}}}{R_{10}^{(0)}} \approx 1 - \frac{3(C \alpha)^2 \cos\eta}{(1 + (C \alpha)^2) \eta^2}.
\ee
In the near-field / nonretarded regime (i.e., $\eta \ll 1$), the normalised transition rate can be approximated by 
\be
\frac{R_{10, \, {{\rm nondisp}}}}{R_{10}^{(0)}} \approx \frac{1 + 2(C \alpha)^2}{1 + (C \alpha)^2} - \frac{(C \alpha)^2 \eta^2}{10 (1 + (C \alpha)^2)}. 
\ee
Next, let's consider the effects of frequency dispersion in the conductivity tensor. 
Figure~\ref{dipole-vert-intermediate} shows the behavior of the transition rate for a perpendicular dipole in the low and intermediate frequency regimes. 
A key feature that we see is a dramatic enhancement in the amplitude of the surface correction to the transition rate as the frequency approaches $2t/\hbar$, a value which is associated with the effectively one-dimensional VHS points $\kv^{{\rm T}} = (\pm \pi/a, 0), (0, \pm \pi/a)$ for $u = t$. 
For example, the amplitude of the surface correction to the transition rate for $\hbar\omega/t = 1.9$ can be 8 times greater than that for $\hbar\omega/t = 1$, and 76 times greater than in the nondispersive limit (cf. Fig.~\ref{dipole-vert-intermediate}). 
As we show in App.~\ref{app:VHS}, both real and imaginary parts of the longitudinal conductivity diverge as $|\hbar\omega/t - 2|^{-1/2}$, and the Fresnel coefficients approximate those of a perfect conductor. For $1 < u/t < 2$, the VHS in the DOS at $\kv^{{\rm T}} = (\pm \pi/a, 0), (0, \pm \pi/a)$ are of saddle-point type with logarithmic divergence~\cite{bassani1975}, and ${{\rm Im}} \, \sigma_{xx}(\omega)$ evinces a finite discontinuity instead of a power-law singularity at $\hbar\omega/t = 2$. 

We also observe that for a Chern insulator with $C = 1$, the normalised transition rates for frequencies in the low frequency range oscillate in phase (cf. the curves with black filled diamonds, blue diamonds, and green squares in Fig.~\ref{dipole-vert-intermediate}). As the frequency approaches $2t/\hbar$ from below, the conductivity begins to diverge and the normalised transition rate can be approximated by (see App.~\ref{app:asymp-vhs})
\be
\frac{R_{10}}{R_{10}^{(0)}} \approx 1 + 3 \left( \frac{\sin \eta - \eta \cos \eta}{\eta^3} \right), 
\ee
which has the same phase as in Eq.~(\ref{tara}).   

On the other hand, a phase shift occurs for frequencies within the intermediate frequency range. At high frequencies (cf. Fig.~\ref{dipole-vertical}) the normalised transition rates (cf. the curves with green squares and orange circles) become anti-phasal to the normalised transition rate from the low frequency range (cf. the curve with blue diamonds) and the nondispersive limit. 
In the high frequency regime ($\hbar \omega/t > 6$), the phases of the transition amplitude oscillations are approximately the same, owing to the fact that only $\sigma_{xx}''$ effectively contributes to the conductivity tensor. This is reflected in the behavior of the transition rates for $\hbar\omega/t = 10, 100$ in Fig.~\ref{dipole-vertical}. 
From the figure, we see that the amplitude tends to decrease with decreasing $t/(\hbar\omega_{10})$, being larger than the amplitude in the nondispersive limit for $t/(\hbar\omega_{10}) > 0.01$, and with the amplitudes becoming of comparable magnitude at $t/(\hbar\omega_{10}) = 0.01$ 
(the quantity $t$ is about 20 meV if we consider the transition between the 1s and 2p levels of a Na atom, which is typically smaller than the surface band gap induced by QAHE in presently known materials~\cite{zhang2016}). 
Although the magnitudes are comparable, the shapes of the oscillatory decay in the dispersive case is distinct from that in the nondispersive case, with the normalised transition rate in the nondispersive case starting from a value larger than unity at $\eta = 0$ and undergoing a parabolic decrease as $\eta$ increases (cf. the curve with black diamonds), whereas in the dispersive case it starts at unity for $\eta = 0$ (cf. the curve with orange circles). 

For high frequencies ($\hbar\omega/t > 2(2t+|u|)$) and at intermediate emitter-surface separations, the normalised transition rate can be approximated by (see App. C2 for derivational details) 
\be
\frac{R_{10}}{R_{10}^{(0)}} \approx 
1 + \frac{3 \widetilde{\sigma}_{xx}''}{\eta^4} 
\big(  
(\eta^2 - 3) \sin \eta + 3 \eta \cos \eta
\big).
\label{asympperp}
\ee
In Fig.~\ref{dipole-vertical}b, we compare the exact normalised transition rate behavior with the above, intermediate asymptotic, approximation for the case of $t = 0.01 \hbar\omega_{10}$, observing that the fit is excellent.  

For large emitter-surface separations (i.e., $\eta \gg (\widetilde{\sigma}_{xx}'')^{-1}$), we find (see App.~\ref{sec:app2}) that the normalised transition rate decays to unity as 
\be
\frac{R_{10}}{R_{10}^{(0)}} \approx 1 + \frac{3}{2} \frac{\widetilde{\sigma}_{xx}''}{\eta} \cos\eta,
\ee
which is a cosinusoidal decay with $1/\eta$ dependence. 
Thus, for a perpendicular dipole, the decay is longer-ranged in the dispersive scenario than in the nondispersive scenario (the latter decaying with inverse square separation).

Finally, we note that the behavior of the transition rate for the perpendicular dipole would remain the same if we reverse the sign of $u/t$ (which would describe a $C = -1$ insulator). This is because such a transition rate depends only on $r_{pp}$ which is insensitive to a sign change in $C$.

\subsection{dipole aligned parallel with surface} 
\begin{figure}[h]
\centering
  \includegraphics[width=0.46\textwidth]{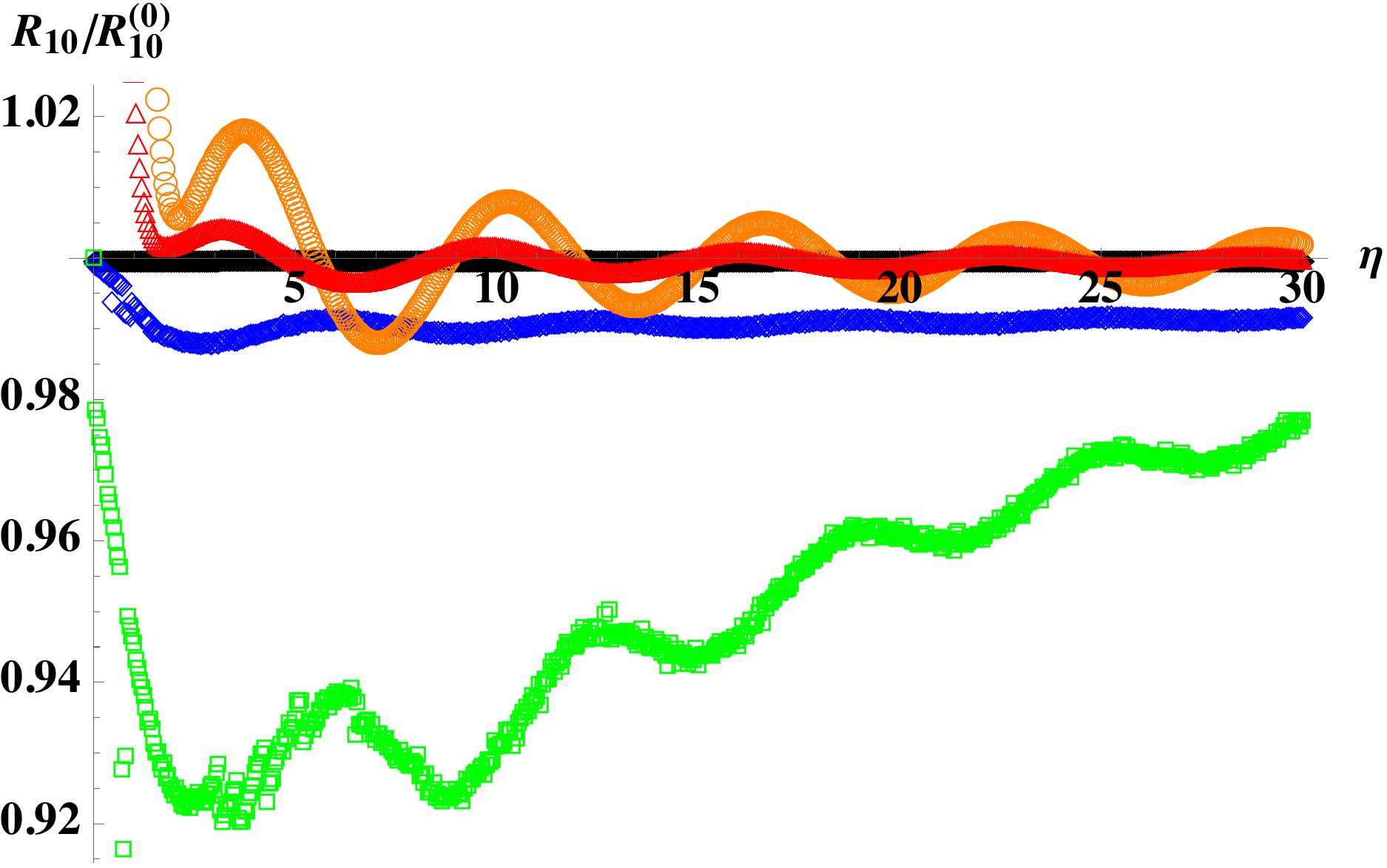}
\caption{Behavior of $R_{10}/R_{10}^{(0)}$ (vertical axis) as a function of $\eta = 2z_0 c/\omega_{10}$ (horizontal axis) for a dipole aligned parallel with the surface of a $C = 1$ Chern insulator with $u/t = 1$: nondispersive (black, filled diamond), $\hbar \omega/t = 1$ (blue, diamond), $\hbar \omega/t = 1.9$ (green, square), $\hbar \omega/t = 2.1$ (orange, circle), and $\hbar \omega/t = 3$ (red, triangle).}
 \label{dipole-para-intermediate}
 \end{figure}
\begin{figure}[h]
\centering
  \includegraphics[width=0.46\textwidth]{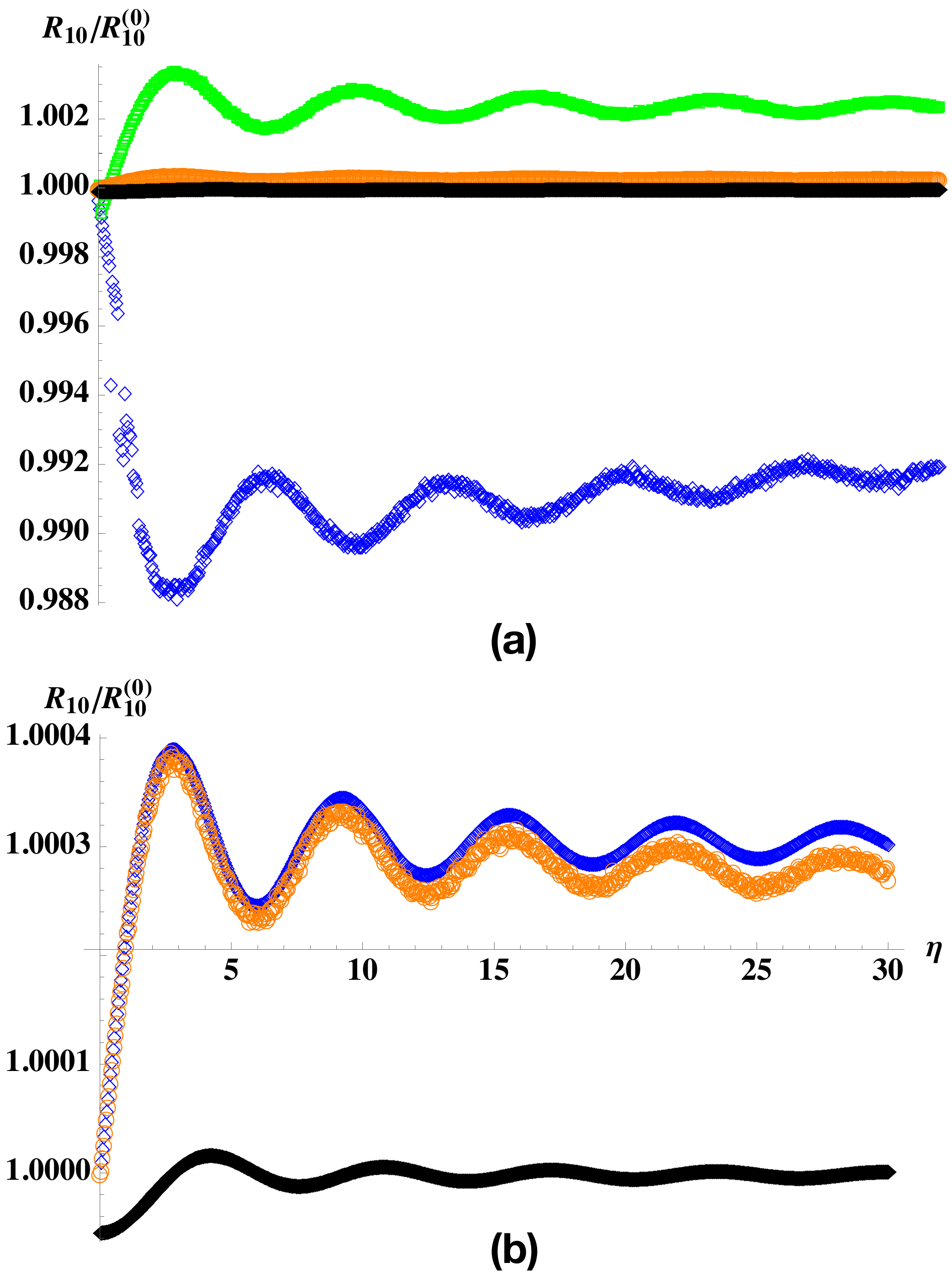}
\caption{Behavior of $R_{10}/R_{10}^{(0)}$ (vertical axis) as a function of $\eta = 2z_0 c/\omega_{10}$ (horizontal axis) for a dipole aligned parallel with the surface of a $C = 1$ Chern insulator with $u/t = 1$: (a)~Behaviors corresponding to the case of nondispersive conductivity (black, filled diamond) as well as dispersive conductivity with $t = \hbar\omega_{10}$ (blue, diamond), $t = 0.1 \hbar\omega_{10}$ (green, square), and $t = 0.01 \hbar\omega_{10}$ (orange, circle). (b)~An enlarged view of the same behavior of $R_{10}/R_{10}^{(0)}$ for the case of a non-dispersive conductivity tensor (black, filled diamond) and dispersive conductivity case with $t = 0.01 \hbar\omega_{10}$ (orange, circle). We compare with the intermediate asymptotic approximation Eq.~(\ref{asymppara}) to the case $t = 0.01 \hbar\omega_{10}$ (blue, diamond), noting its relatively good match with the exact transition rate behavior for $\eta < 15$ (we provide an explanation in Sec.~IV B).}
 \label{dipole-parallel}
 \end{figure}
Next, consider a dipole which is parallel to the surface of the Chern insulator, with $\nv^{{\rm T}} = (1,0,0)$. This leads to 
\be
    \label{dippar}
\frac{R_{10}}{R_{10}^{(0)}} = 1+ \frac{3 c^3}{2 \omega_{10}^3} {{\rm Im}}\, G_{xx}^R(\rv_0,\rv_0; \omega_{10}),   
\ee
where $G_{xx}^R$ is given by Eq.~(\ref{Gxx}). 
For the case where the conductivity tensor is nondispersive, the use of Eqs.~(\ref{r-static}) and (\ref{Gxx}) leads to the following expression for the normalised transition rate in the retarded regime: 
\be
\label{tara1}
\frac{R_{10, \, {{\rm nondisp}}}}{R_{10}^{(0)}} = 1 - \frac{3(C \alpha)^2}{2 \big( 1 + (C \alpha)^2 \big)} \frac{\eta \cos \eta + (\eta^2 - 1) \sin \eta}{\eta^3}. 
\ee
In the far-field regime, we can approximate the above by 
\be
\frac{R_{10, \, {{\rm nondisp}}}}{R_{10}^{(0)}} \approx 
1 - \frac{3(C\alpha)^2}{2(1 + (C\alpha)^2)}\frac{\sin \eta}{\eta}, 
\ee
whilst in the near-field / nonretarded regime the normalised transition rate can be approximated by 
\be
\frac{R_{10, \, {{\rm nondisp}}}}{R_{10}^{(0)}} \approx \frac{1}{1+(C \alpha)^2} + \frac{(C \alpha)^2 \eta^2}{5 \big( 1 + (C \alpha)^2 \big)}. 
\ee
As before, we can compare the normalised transition rate behavior for the nondispersive and dispersive conductivity cases. 
In the low to intermediate frequency regimes, the behavior is shown in Fig.~\ref{dipole-para-intermediate} for $u/t=1$. 
As in the case of the perpendicularly aligned dipole, for the parallel aligned dipole the amplitude of oscillation of the surface-induced correction to the transition rate also becomes greatly enhanced as the frequency approaches the VHS-associated value, $\omega = 2t/\hbar$. 
For example, for $\eta = 2.5$, the amplitude for $\hbar\omega/t = 1.9$ ($\hbar\omega/t = 2.1$) is approximately 13000 (1400) times greater than that of the nondispersive case. 

As the frequency approaches $2t/\hbar$ from below, the phase of oscillatory decay coincides with that in the nondispersive limit. 
As we see from App.~\ref{app:asymp-vhs}, the normalised transition rate can be approximated by  
\be
\frac{R_{10}}{R_{10}^{(0)}} \approx 1 - \frac{3}{2} \left( \frac{\eta \cos \eta + (\eta^2 - 1) \sin \eta}{\eta^3} \right), 
\ee
which has the same phase as in Eq.~(\ref{tara1}).   

In the high frequency regime and for large emitter-surface separations ($\eta \gg (\widetilde{\sigma}_{xx}'')^{-1}$), we find (see App.~C) that the normalised transition rate decays to unity as 
\be
\frac{R_{10}}{R_{10}^{(0)}} \approx 1 - \frac{3}{4\eta} \sin\eta,
\ee
i.e., it is a sinusoidal decay with $1/\eta$ dependence. 

Compared to both the nondispersive case of the parallel aligned dipole and the dispersive case of the perpendicular dipole, the shape of oscillatory decay for the dispersive case of the parallel aligned dipole is different. 
Where the normalised transition rate undergoes an oscillatory decay about unity in the first two cases (cf. Figs.~\ref{dipole-vertical}b and the black diamond curve in \ref{dipole-parallel}b),  
the shape of oscillations of the transition rate in the latter case is reminiscent of a sine integral function (cf. the orange circle curve in Fig.~\ref{dipole-parallel}b). This shape persists into the high frequency regime, as we see from Fig.~\ref{dipole-parallel}. 
The sine integral-like shape can be traced to the contribution involving $r_{ss}$ in ${{\rm Im}}\,G_{xx}^R$. 
For high frequencies ($\hbar\omega/t > 2(2t+|u|)$) and at intermediate emitter-surface separations, the normalised transition rate can be approximated by (see App. C2 for derivational details) 
\be
\frac{R_{10}}{R_{10}^{(0)}} \approx 
1 + \frac{3 \widetilde{\sigma}_{xx}''}{4} \bigg( {{\rm Si}}(\eta) + \frac{3(\eta^2-2)\sin\eta - \eta(\eta^2-6) \cos\eta}{\eta^4} \bigg). 
\label{asymppara}
\ee
In Fig.~\ref{dipole-parallel}b, we compare the above, intermediate asymptotic, approximation Eq.~(\ref{asymppara}) with the exact normalised transition rate behavior for the case $t = 0.01 \hbar\omega_{10}$. We see that there is a relatively good match for $\eta < 15$. For $\eta > 15$ the match becomes poorer; this is to be expected as the intermediate asymptotic approximation tends towards $1 + 3\pi \widetilde{\sigma}_{xx}''/8$ for $\eta \rightarrow \infty$, whereas the exact normalised transition rate would approach 1.

As with the previously considered dipole configuration, the oscillatory amplitudes tend to decrease with decreasing values of $t/(\hbar\omega_{10})$. Similar to the case of the perpendicularly aligned dipole configuration, the transition rate is unchanged under a change of sign of $C$. This is because the transition rate depends only on $r_{pp}$ and $r_{ss}$ which are insensitive to a sign change in $C$. 


\subsection{circularly polarised dipole}
\begin{figure}[h]
\centering
  \includegraphics[width=0.46\textwidth]{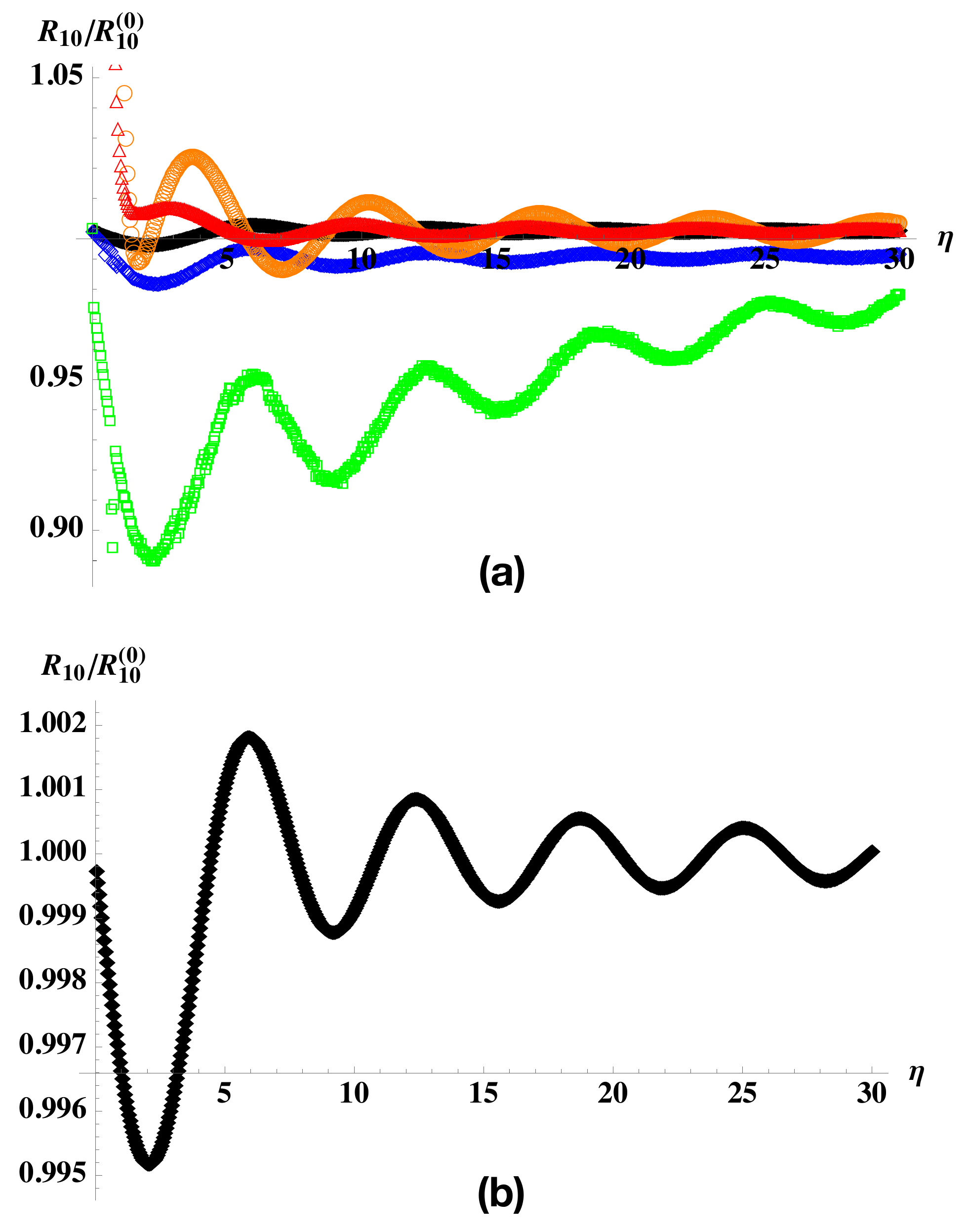}
\caption{Behavior of $R_{10}/R_{10}^{(0)}$ (vertical axis) as a function of $\eta = 2z_0 c/\omega_{10}$ (horizontal axis) for a right circularly polarised dipole near the surface of a $C = -1$ Chern insulator with $u/t = -1$: (a)~nondispersive (black, filled diamond), $\hbar \omega/t = 1$ (blue, diamond), $\hbar \omega/t = 1.9$ (green, square), $\hbar \omega/t = 2.1$ (orange, circle), and $\hbar \omega/t = 3$ (red, triangle); (b)~an enlarged view of the nondispersive case (black, filled diamond).}
 \label{dipole-circ-intermediate-Cone}
 \end{figure}
\begin{figure}[h]
\centering
  \includegraphics[width=0.46\textwidth]{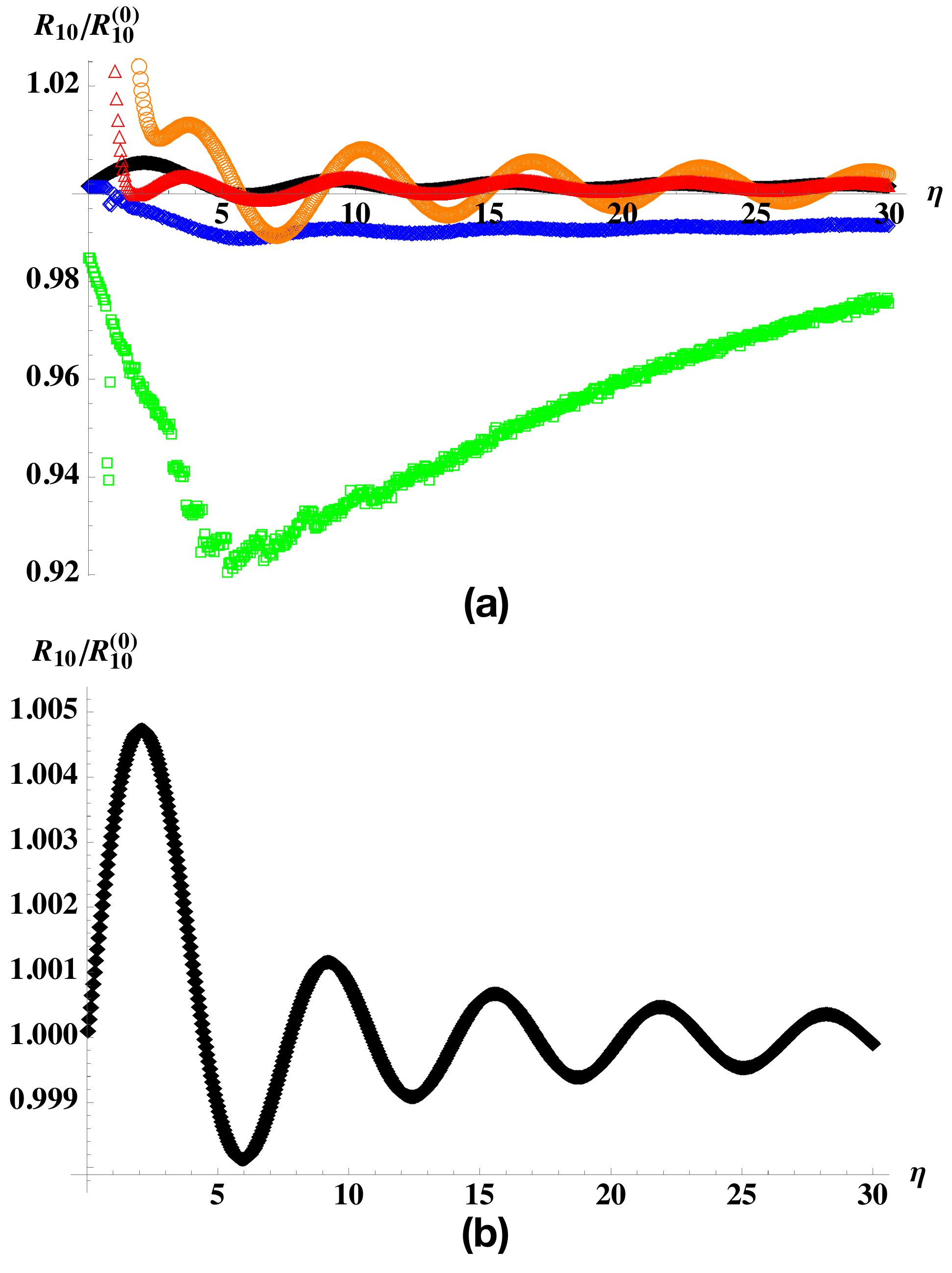}
\caption{Behavior of $R_{10}/R_{10}^{(0)}$ (vertical axis) as a function of $\eta = 2z_0 c/\omega_{10}$ (horizontal axis) for a right circularly polarised dipole near the surface of a $C = 1$ Chern insulator with $u/t = 1$: (a)~nondispersive (black, filled diamond), $\hbar \omega/t = 1$ (blue, diamond), $\hbar \omega/t = 1.9$ (green, square), $\hbar \omega/t = 2.1$ (orange, circle), and $\hbar \omega/t = 3$ (red, triangle); (b)~an enlarged view of the nondispersive case (black, filled diamond).}
 \label{dipole-circ-intermediate-Cminusone}
 \end{figure}

\label{sec:dipole-circ}
\begin{figure}[h]
\centering
  \includegraphics[width=0.46\textwidth]{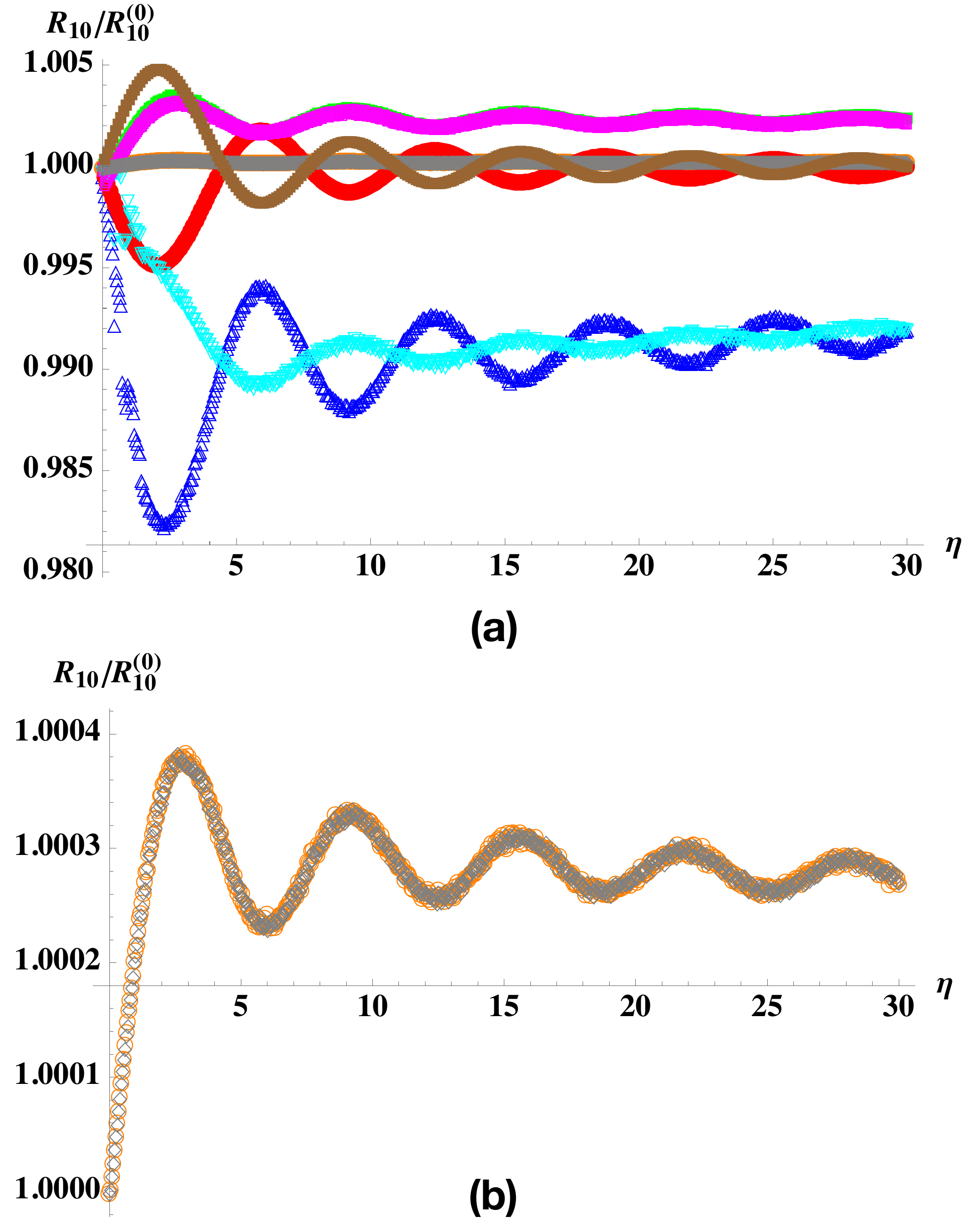}
  \caption{Behavior of $R_{10}/R_{10}^{(0)}$ (vertical axis) as a function of $\eta = 2z_0 c/\omega_{10}$ (horizontal axis) for a right-circularly polarised dipole whose quantisation axis is perpendicular to the surface of a $C = 1$ Chern insulator with $u/t = 1$ in Eq.~(\ref{qwz}), as well as a similarly polarised dipole next to a $C = -1$ insulator with $u/t = -1$. (a)~Behaviors corresponding to the case of nondispersive conductivity for $C = -1$ (red, filled circle) and $C = 1$ (brown, filled square) as well as dispersive conductivity for $C = -1$ with $t = \hbar\omega_{10}$ (blue, up-triangle), $t = 0.1 \hbar\omega_{10}$ (green, square), and $t = 0.01 \hbar\omega_{10}$ (orange, circle), and for $C = 1$ with $t = \hbar\omega_{10}$ (cyan, down-triangle), $t = 0.1 \hbar\omega_{10}$ (magenta, rectangle), and $t = 0.01 \hbar\omega_{10}$ (gray, diamond). (b)~Close-up of the behavior for $t = 0.01 \hbar\omega_{10}$ with $C = -1$ (orange, circle) and $C = 1$ (gray, diamond).}
  \label{dipole-circular}
\end{figure}
Finally, we consider a right-circularly or $\sigma_+$ polarised dipole configuration with $\nv^{{\rm T}} = (1,i,0)/\sqrt{2}$. 
The corresponding transition rate is given by 
\be
    \label{dipcirc}
\frac{R_{10}}{R_{10}^{(0)}} = 1+ \frac{3 c^3}{2 \omega_{10}^3} 
\big( {{\rm Im}}\, G_{xx}^R(\rv_0,\rv_0; \omega_{10}) + {{\rm Re}}\, G_{xy}^R(\rv_0,\rv_0; \omega_{10}) \big),   
\ee
where $G_{xy}^R$ is given by Eq.~(\ref{Gxy}). 

For the case of nondispersive conductivity, the normalised transition rate for a circularly polarised dipole with $C = 1$ oscillates antiphasally to that for the same dipole with $C = -1$, as we see from 
Figs.~9 and 10. 
We can use Eqs.~(\ref{r-static}) and (\ref{Gxy}) to find the normalised transition rate in the retarded regime: 
\ba
\frac{R_{10, \, {{\rm nondisp}}}}{R_{10}^{(0)}} 
&\!\!=\!\!& 1 - 
\frac{3 (C\alpha)^2 \big( \eta \cos \eta + (\eta^2 - 1) \sin\eta \big)}{2(1+(C\alpha)^2) \eta^3}
\nonumber\\
&&\mp \frac{3 \, C\alpha \big( \eta \cos \eta - \sin \eta \big)}{2(1+(C\alpha)^2) \eta^2}
, 
\ea
where the $+$ ($-$) sign in front of the third term applies to $C=1$ ($C=-1$). 
In the far-field regime, we can approximate the above to leading order by 
\ba
\frac{R_{10, \, {{\rm nondisp}}}}{R_{10}^{(0)}} 
&\!\!\approx\!\!& 
1 \mp \frac{3 C\alpha}{2(1+(C\alpha)^2)} \frac{\cos \eta}{\eta} 
\nonumber\\
&&- \frac{3(C\alpha)^2}{2(1+(C\alpha)^2)} \frac{\sin \eta}{\eta}, 
\ea
where the $+$ ($-$) sign in front of the second term applies to $C=1$ ($C=-1$). 
In the near-field / nonretarded regime, the normalised transition rate becomes 
\be
\frac{R_{10, \, {{\rm nondisp}}}}{R_{10}^{(0)}} 
\approx 
\frac{1}{1 + (C \alpha)^2} 
\pm \frac{C \alpha \eta}{2(1+(C\alpha)^2)}
+ \frac{(C \alpha)^2 \eta^2}{5 ( 1 + (C \alpha)^2 )}, 
\ee
where the $-$ ($+$) sign in front of the second term applies to $C=1$ ($C=-1$).

Next, we consider the effect of conductivity dispersion. 
The behavior of the transition rate in the low to intermediate frequency regimes is shown in Fig.~\ref{dipole-circ-intermediate-Cone} and Fig.~\ref{dipole-circ-intermediate-Cminusone} for $C = -1$ (with $u = -t$) and $C=1$ (with $u = t$), respectively. There is again an enhancement in the oscillation amplitude near the VHS-associated frequency of $\omega = 2t/\hbar$, however this is less drastic than for the perpendicularly and parallel aligned dipole configurations. 
We also see that the transition rate for the same circularly polarised dipole with $C = 1$ has a different appearance from the one with $C = -1$. 
Such a difference can enable one to distinguish the sign of the Chern number for a Chern insulator near a dipole of a given circular polarisation state. 

As the left circularly (or $\sigma_-$) polarised dipole is a time-reversed version of the right circularly (or $\sigma_+$) polarised one, and $C = 1$ is a time-reversed version of $C = -1$, the normalised transition rate for a right (left) circularly polarised dipole near a $C = 1$ insulator would coincide with that for a left (right) circularly polarised dipole near a $C = -1$ insulator. We can see this explicitly too, by applying Eq.~(\ref{R10}) to a dipole moment with a normalised orientation vector $\nv^{{\rm T}} = (n_x, n_y, 0)$: 
\be
\frac{R_{10}}{R_{10}^{(0)}} 
= 1 + \frac{3}{2}\Big( \frac{c}{\omega_{10}} \Big)^3 
\big( 
{{\rm Im}} \, G_{xx}^R 
- i \, {{\rm Re}} \, G_{xy}^R (n_x^\ast n_y - n_x n_y^\ast)
\big). 
\label{R10nxny}
\ee
The term $(n_x^\ast n_y - n_x n_y^\ast)$ changes sign on replacing a left circularly-polarised dipole with a right circularly-polarised one, but this sign change is annulled if we also change the sign of the Hall conductivity (to which $G_{xy}^R$ is proportional). 
This implies that we can also use the transition rate in the low to intermediate frequency ranges to distinguish which circular polarisation state the dipole emitter is in, for a Chern insulator of a given Chern number. 
If the medium is \emph{reciprocal}, the term proportional to $G_{xy}^R$ in Eq.~(\ref{R10nxny}) vanishes, and the resulting normalised transition rate for the circularly polarised dipole is the same as that for a dipole aligned parallel to the surface. Physically, the two configurations are now equivalent because of rotational symmetry in the plane of the (normal) insulator~\cite{footnote2}. 

In the high frequency regime, the behavior of the transition rate for $u/t = 1$ at intermediate emitter-surface separations is shown in Fig.~\ref{dipole-circular}. 
In contrast to the zero to intermediate frequency regimes, the transition rate of the circularly polarised dipole with $C = 1$ now oscillates \emph{in phase} with that of the same dipole with $C = -1$ (see Fig.~\ref{dipole-circular}b). This is again related to the sole effective contribution of $\sigma_{xx}''$ to the conductivity tensor in the high frequency regime.  
The oscillation amplitudes of the normalised transition rate decrease with $t/(\hbar\omega_{10})$. 

In the high frequency regime, we find (see App.~C) that the transition rate decays at $\eta \gg (\widetilde{\sigma}_{xx}'')^{-1}$ as 
\be
\frac{R_{10}}{R_{10}^{(0)}} \approx 1 - \frac{3}{4\eta} \sin\eta,
\ee
i.e., it is a sinusoidal decay with $1/\eta$ dependence. 

\section{Conclusion} 

We have investigated the transition rate behavior for a two-level quantum emitter near a two-dimensional Chern insulator, working in the framework of linear response theory and representing the Chern insulator by the Qi-Wu-Zhang model. 
To summarise: we have derived the full electromagnetic Green tensor for the dipole emitter near the Chern insulator, including both the reflection and transmission contributions. We have also identified three qualitatively distinct regimes of behavior: (i)~the nondispersive limit of the conductivity tensor, (ii)~the low to intermediate frequency range (i.e., frequencies smaller than the maximum energy gap between the valence and conduction bands of the Chern insulator), and (iii)~the high frequency range (i.e., frequencies larger than the maximum energy gap). For example, physics in the high frequency range is dominated by the contribution of the imaginary part of the longitudinal conductivity, which is invariant under time reversal, whereas in the low to intermediate frequency ranges, the response of the insulator is sensitive to the breaking of time reversal symmetry. Thus, we have found that if we change the sign of the Chern number of the insulator near a dipole emitter of a given circular polarisation state, the surface-induced correction to the transition rate behaves differently in the low to intermediate frequency ranges, but the surface correction is essentially unchanged in the high frequency range. This enables one to distinguish the sign of the Chern number for a given circular polarisation, and equivalently, also enables us to distinguish whether the emitter is in a left or right circularly polarised state for a Chern insulator of a given Chern number.  

The decay behavior of the surface-induced correction to the transition rate at large emitter-surface separations can also follow distinct power laws depending on the frequency regime. For a dipole which is aligned perpendicular to the insulator's surface and the system is being probed in the high frequency regime, the surface correction in the far-field limit undergoes a long-ranged decay with $\eta^{-1}$ scaling, compared to a shorter-ranged decay with $\eta^{-2}$ scaling if the conductivity tensor is assumed to be nondispersive. 

The presence of van Hove singularities in the electronic density of states can have a significant impact on the amplitude and shape of the surface correction to the transition rate. In the context of the Qi-Wu-Zhang model, we have found that when the mass gap parameter is tuned equal to the hopping parameter, two-dimensional saddle point-type van Hove singularities can become effectively one-dimensional, giving rise to a power-law divergence rather than a logarithmic one. We found that as the emitter's energy level spacing approaches the energy associated with such VHS, the oscillation amplitude of the surface correction to the transition rate can be enhanced by a few orders of magnitude relative to that for nondispersive conductivity.  
Accounting for frequency dispersion in the conductivity tensor has also led us to recognise qualitative differences in the shape of oscillatory decay of the surface-induced correction depending on the dipole configuration. For a dipole perpendicularly aligned to the surface and also the parallel and circularly aligned dipole configurations in the nondispersive conductivity limit, the surface-induced correction to the transition rate undergoes an algebraic decay with sinusoidal oscillations; on the other hand, when conductivity dispersion is taken into account, the surface-induced correction for the parallel and circularly aligned dipole configurations decays with sine integral-like oscillations. 

By tuning the ratios $\hbar\omega/t$ and $u/t$, one can access and probe the transition rate behavior corresponding to different frequency regimes. Such tuning may be effected by suitably engineered quantum emitters and/or Chern insulators. At present, the largest achieved value for the band gap in Chern insulators appears to be $0.226$ eV~\cite{xue2018}, whilst for the sodium atom the energy difference between the 1s and 2p levels is around 2 eV, which would put the transition behavior into the high frequency regime. If materials with larger band gaps and/or quantum emitters with smaller absorption energies are found, it will be possible to access the transition rate behavior in the intermediate and low frequency regimes as well.
Our results suggest that in order to obtain a more complete description of the spontaneous emission of a quantum emitter near a Chern insulator, it is imperative to account for the finite-frequency, non-topological contributions, in addition to the zero-frequency, topological one. The methodology used in this paper can be straightforwardly extended to Chern insulators of higher Chern numbers, e.g., $C = 2$ (for this one can use a model proposed in Ref.~\cite{grushin2012}).

\section{Acknowledgments}

The authors would like to thank Martial Ducloy and David Wilkowski for their feedback on the manuscript. BSL also thanks them for introducing him to the fields of atomic physics and quantum electrodynamics. The authors acknowledge support from the Academic Research Fund (Tier 1) of the Ministry of Education (Singapore) under the grant reference no. RG 160/19 and a start-up grant under the reference no. M4082095.110 from Nanyang Technological University.

\appendix

\section{electromagnetic Green tensor} 
\label{app1}

\subsection{bulk Green tensor}

We want to derive the field solution corresponding to a dipole source placed above a dielectric interface. First, let's obtain the \emph{bulk} field solution, which describes waves emitted by the source. As later on we will also address wave reflection from the planar dielectric interface, it is useful to obtain the bulk solution in terms of plane waves propagating along the direction normal to the interface (which we can define to be the z-axis). The amplitude of the plane waves (which are emitted by the dipole source) can then be related to the amplitudes of the incident waves which impinge on the interface. 

\paragraph{full Fourier-space decomposition.} As the bulk solution is the solution without the consideration of boundaries, it is also the solution to a system which is spatially \emph{homogeneous}, i.e., looks the same everywhere. (On the other hand, the presence of boundaries makes the system spatially \emph{inhomogeneous}, spatial inhomogeneity meaning that translation symmetry is broken in certain directions, in this case, the $z$-direction. When translation symmetry is broken we cannot apply Fourier integral transformation along the broken-symmetry direction, this is because the Fourier integral is the sum of all wavevectors, whereas symmetry breaking boundaries cause certain wavevectors to be allowed, whilst disallowing certain other wavevectors.) Because of spatial homogeneity we can express the bulk solution ${\bf E}^{(0)}(\rv,\omega)$ as a three-dimensional Fourier integral, i.e., 
\be
{\bf E}^{(0)}(\rv,\omega) = \int\!\! \frac{d^3\kv}{(2\pi)^3}\, e^{i\kv\cdot\rv}\, {\bf E}^{(0)}(\kv,\omega),
\label{hellen2}
\ee
where the three-dimensional wavevector $\kv = (k_x, k_y, k_z)$. 
For the purpose of finding the Green function ${\mathbb G}(\rv, \rv'; \omega)$, which is the field response at $\rv$ to the presence of a unit dipole $\pv$ at $\rv_0$, we substitute the above into Eq.~(\ref{eqE}) with ${\bf P}_d(\rv ,\omega) = {\bf p}(\omega) \, \delta(\rv - \rv_0)$ to obtain
\ba
&&\big[ -k_a k_b + \big( k^2 - (\omega/c)^2 \big) \delta_{ab} \big] {\bf E}^{(0)}(\kv,\omega) 
\nonumber\\
&=& 
4\pi (\omega/c)^2 p_a(\omega) e^{-i\kv\cdot\rv_0}, 
\label{senc}
\ea
where we have set $\varepsilon = 1$ for the vacuum. 
We have to invert the equation to find the solution for ${\bf E}^{(0)}$. To invert it, we resolve the Fourier-space Helmholtz operator into transverse and longitudinal parts, by making use of transverse and longitudinal projection operators, ${\mathcal P}^{({\rm T})}$ and ${\mathcal P}^{({\rm L})}$
(here ``longitudinal" (``transverse") refers to the component parallel (perpendicular) to the direction of $\kv$). 
The Fourier-space projection operators are given by 
\begin{subequations}
\ba
\delta_{ab} &\!=\!& {\mathcal P}_{ab}^{({\rm L})} + {\mathcal P}_{ab}^{({\rm T})}, 
\\
{\mathcal P}_{ab}^{({\rm L})} &\!=\!& \frac{k_a k_b}{k^2}, 
\\
{\mathcal P}_{ab}^{({\rm T})} &\!=\!& \delta_{ab} - \frac{k_a k_b}{k^2}
\ea 
\end{subequations}
Let us also decompose $\pv$ into longitudinal and transverse parts, i.e., $\pv = \pv^{({\rm L})} + \pv^{({\rm T})}$, where $\pv^{({\rm L})} \equiv {\mathcal P}^{({\rm L})}\cdot \pv$ and $\pv^{({\rm T})} \equiv {\mathcal P}^{({\rm T})}\cdot \pv$. Equation~(\ref{senc}) becomes
\ba
&&\big[ -(\omega/c)^2 {\mathcal P}_{ab}^{({\rm L})}  +(k^2 - (\omega/c)^2) {\mathcal P}_{ab}^{({\rm T})} \big] E_b^{(0)}(\kv, \omega) 
\nonumber\\
&=& 4\pi (\omega/c)^2 \big( p_a^{({\rm L})} + p_a^{({\rm T})}   \big) e^{-i\kv\cdot\rv_0}
\ea
Equating the longitudinal parts on the LHS and RHS, and doing the same for the transverse parts, we obtain
\begin{subequations}
\ba
E_a^{({\rm 0L})}(\kv, \omega) &\!\equiv\!& 
{\mathcal P}_{ab}^{({\rm L})} E_b^{(0)}(\kv, \omega) 
\\
&=&
 -4\pi \left( \frac{k_a k_b}{k^2} \right) p_b(\omega) \, e^{-i\kv\cdot\rv_0}, 
\nonumber\\
E_a^{({\rm 0T})}(\kv, \omega) &\!\equiv\!& 
{\mathcal P}_{ab}^{({\rm T})} E_b^{(0)}(\kv, \omega) 
\\
&=&
\frac{4\pi (\omega/c)^2}{k^2 - (\omega/c)^2} \left( \delta_{ab} - \frac{k_a k_b}{k^2} \right) p_b(\omega) \, e^{-i\kv\cdot\rv_0}. 
\nonumber
\ea
\end{subequations}
Combining longitudinal and transverse parts, we obtain
\ba
\label{irina}
&&E_a^{(0)}(\kv, \omega) = E_a^{({\rm 0L})}(\kv, \omega) + E_a^{({\rm 0T})}(\kv, \omega) 
\\
&=&
4\pi (\omega/c)^2 e^{-i\kv\cdot\rv_0} 
\nonumber\\
&&\times 
\left( \frac{\delta_{ab}}{k^2 - (\omega/c)^2} - \frac{k_a k_b}{(\omega/c)^2 \big( k^2 - (\omega/c)^2 \big)} \right) p_b(\omega) 
\nonumber
\ea
Noting that the inverse Fourier transform of $4\pi \exp(i \kv\cdot\rv_0) /(k^2 - (\omega/c)^2)$ is $\exp(i (\omega/c) R)/R$ (which can be obtained using residue calculus), we take the inverse Fourier transform of Eq.~(\ref{irina}) and obtain
\be
E_a^{(0)}(\rv, \omega) = \big( (\omega/c)^2 \delta_{ab} + \partial_a \partial_b \big) 
\left( \frac{p_b(\omega) \, e^{i(\omega/c)R}}{R} \right), 
\label{angel}
\ee
where $\Rv \equiv \rv - \rv_0$, and $\partial_a \equiv \partial/\partial R_a$ acts on $R$. This is just the field induced by dipole in a homogeneous medium. 

To make contact with Ref.~\cite{wylie-sipe2}, we can use Eqs.~(\ref{hellen1}) and (\ref{hellen2}) to rewrite Eq.~(\ref{irina}) in the form: 
\be
E_a^{(0)}(\kv, \omega) =
F_{ab}^{(0)} (\kv, \omega) p_b(\omega) e^{-i\kv\cdot\rv_0}, 
\ee
where $\mathbb{F}^{(0)}(\kv, \omega)$ is the bulk Green tensor in \emph{full} Fourier space, given by 
\be
F_{ab}^{(0)} (\kv, \omega) 
\equiv
4\pi \frac{k_a k_b - (\omega/c)^2 \delta_{ab}}{(\omega/c)^2 - k^2}. 
\ee
We deduce the dyadic response function from Eq.~(\ref{GFE}): 
\be
G_{ab}^{(0)} (\kv, \omega) = 
4\pi \frac{k_a k_b - (\omega/c)^2 \delta_{ab}}{(\omega/c)^2 - k^2} + 4\pi \delta_{ab}. 
\ee
This agrees with Eq.~(3.6) of Ref.~\cite{wylie-sipe2}. 

\paragraph{partial Fourier-space decomposition.} As we shall treat the problem of finding the Green function for a system with a planar interface (which breaks the translation symmetry of space in one direction), it would be expedient for us to now obtain the bulk Green tensor in two-dimensional (instead of three-dimensional) wave-vector space, retaining the dependence on the Cartesian coordinate $z$. Again we relate the bulk electric field to the dipole source via the Green tensor, but now in two-dimensional wave-vector space: 
\be
{\bf E}^{(0)}(\kv_\parallel, z; \omega) = 
{\mathbb F}^{(0)} (\kv_\parallel, z, z_0; \omega) \cdot \pv(\omega).
\ee
Here we have taken the Fourier transform of the electric field in the x and y directions, which we define parallel to the plane of the interface. We can choose the s and p polarisations to be basis vectors for the Green tensor. 

As the translation symmetry is broken in the $z$-direction but preserved in the $x$- and $y$-directions, we can express the bulk solution $E_a^{(0)}$ in terms of a two-dimensional Fourier integral over the wave vector $\kv_\rho = (k_x, k_z)$. To do this, we make use of the \emph{Weyl expansion}: 
\be
\frac{e^{i(\omega/c)|\rv- \rv_0|}}{|\rv - \rv_0|} = \int\!\!\frac{d^2 \kv_\rho}{(2\pi)^2} 
\frac{2\pi i}{k_z} e^{ik_z|z - z_0|} e^{i\kv_\parallel\cdot(\rv_\parallel - \rv_{0,\parallel})}, 
\ee
where $\kv_\parallel \equiv (k_x, k_y, 0)$, $\rv_\parallel \equiv (x, y, 0)$ and $k_z \equiv ( (\omega/c)^2 - k_\parallel^2 )^{1/2}$. 
Next, we express Eq.~(\ref{angel}) as a Fourier transform over the wavevector $\kv_\parallel$, and make use of the fact that $\{ \hat{\kv}_\parallel, \hat{z}, \hat{\kv}_\parallel \times \hat{z} \}$ form an orthonormal basis, so the identity dyad ${\mathbb I} = \hat{\kv}_\parallel \hat{\kv}_\parallel + \hat{z} \hat{z} + (\hat{\kv}_\parallel \times \hat{z}) (\hat{\kv}_\parallel \times \hat{z})$: 
\ba
\label{bigo}
&&{\bf E}^{(0)}(\rv, \omega) 
\\
&=& \pv(\omega) \!\cdot\! 
\int\!\!\frac{d^2\kv_\rho}{(2\pi)^2} 
\frac{2\pi i}{k_z} 
\big[  
\nabla \nabla
e^{ik_z|z - z_0|} e^{i\kv_\parallel\cdot(\rv_\parallel - \rv_{0,\parallel})}
\nonumber\\
&&+ 
(\omega/c)^2 
\big( \hat{\kv}_\parallel \hat{\kv}_\parallel + \hat{z} \hat{z} + (\hat{\kv}_\parallel \times \hat{z}) (\hat{\kv}_\parallel \times \hat{z}) \big)
\nonumber\\
&&\quad\times 
e^{ik_z|z - z_0|} e^{i\kv_\parallel\cdot(\rv_\parallel - \rv_{0,\parallel})}
\big], 
\nonumber
\ea
where $\nabla \equiv (\partial_x, \partial_y, \partial_z)$. 
We evaluate the first term: 
\ba
&&\nabla \nabla 
e^{ik_z|z - z_0|} e^{i\kv_\parallel\cdot(\rv_\parallel - \rv_{0,\parallel})} 
\\
&=& 
\nabla \big[ i k_\parallel \hat{\kv}_\parallel + i k_z \,{\rm Sgn}\, (z-z_0) \hat{z} \big] 
e^{ik_z|z - z_0|} e^{i\kv_\parallel\cdot(\rv_\parallel - \rv_{0,\parallel})} 
\nonumber\\
&=&
\Big\{  
2i k_z \delta(z-z_0) \hat{z} \hat{z} 
- 
\big[ k_\parallel \hat{\kv}_\parallel + k_z \,{\rm Sgn}\, (z-z_0) \hat{z} \big]
\nonumber\\
&&
\big[ k_\parallel \hat{\kv}_\parallel + k_z \,{\rm Sgn}\, (z-z_0) \hat{z} \big]
\Big\} 
e^{ik_z|z - z_0|} e^{i\kv_\parallel\cdot(\rv_\parallel - \rv_{0,\parallel})} 
\nonumber
\ea
On going to the last step, we made use of the fact that the sign function, ${\rm Sgn} \, (z-z_0) = \Theta(z-z_0) - \Theta(z_0-z)$ (where $\Theta(x)$ is the Heaviside function, defined to be zero for $x<0$ and 1 for $x>0$), which implies $\partial_z \, {\rm Sgn} \, (z-z_0) = 2\delta(z-z_0)$, where $\delta(x)$ is the Dirac delta-function (here we noted that the derivative of a Dirac delta-function gives the Heaviside function). Equation~(\ref{bigo}) becomes
\ba
\label{selo}
&&{\bf E}^{(0)}(\rv, \omega) 
\\
&\!=\!& \pv(\omega) \!\cdot\! \int\!\!\frac{d^2\kv_\parallel}{(2\pi)^2} 
\frac{2\pi i}{k_z} 
\Big\{
2i k_z \delta(z-z_0) \hat{z} \hat{z}  
\nonumber\\
&&+ (\omega/c)^2 \big[ \hat{\kv}_\parallel \hat{\kv}_\parallel + \hat{z} \hat{z} + (\hat{\kv}_\parallel \times \hat{z}) (\hat{\kv}_\parallel \times \hat{z}) \big]
\nonumber\\
&&- 
\big[ k_\parallel \hat{\kv}_\parallel + k_z \,{\rm Sgn}\, (z-z_0) \hat{z} \big]
\big[ k_\parallel \hat{\kv}_\parallel + k_z \,{\rm Sgn}\, (z-z_0) \hat{z} \big]
\Big\} 
\nonumber\\
&&\times 
e^{ik_z|z - z_0|} e^{i\kv_\parallel\cdot(\rv_\parallel - \rv_{0,\parallel})}, 
\nonumber
\ea
Consider the following terms: 
\ba
&&(\omega/c)^2 \big[ \hat{\kv}_\parallel \hat{\kv}_\parallel + \hat{z} \hat{z} \big] 
\\
&&- 
\big[ k_\parallel \hat{\kv}_\parallel + k_z \,{\rm Sgn}\, (z-z_0) \hat{z} \big]
\big[ k_\parallel \hat{\kv}_\parallel + k_z \,{\rm Sgn}\, (z-z_0) \hat{z} \big]
\nonumber\\
&=&
(\omega/c)^2 \big[ \hat{\kv}_\parallel \hat{\kv}_\parallel + \hat{z} \hat{z} \big] 
- k_\parallel^2 \hat{\kv}_\parallel \hat{\kv}_\parallel 
- k_z^2 \hat{z} \hat{z} 
\nonumber\\
&&- k_\parallel k_z (\hat{\kv}_\parallel \hat{z} + \hat{z} \hat{\kv}_\parallel) \, {\rm Sgn} \, (z-z_0)
\nonumber\\
&=& 
k_z^2 \hat{\kv}_\parallel \hat{\kv}_\parallel + k_\parallel^2 \hat{z} \hat{z} 
- k_\parallel k_z (\hat{\kv}_\parallel \hat{z} + \hat{z} \hat{\kv}_\parallel) \, {\rm Sgn} \, (z-z_0)
\nonumber\\
&=& 
[ -k_z \,{\rm Sgn} \, (z-z_0) \hat{\kv}_\parallel + k_\parallel \hat{z} ]
[ -k_z \,{\rm Sgn} \, (z-z_0) \hat{\kv}_\parallel + k_\parallel \hat{z} ]
\nonumber
\ea
Plug this result back into Eq.~(\ref{selo}) to obtain
\ba
&&{\bf E}^{(0)}(\rv, \omega) 
\\
&\!=\!& \pv(\omega) \!\cdot\! \int\!\!\frac{d^2\kv_\parallel}{(2\pi)^2} 
\frac{2\pi i}{k_z} 
\Big\{
2i k_z \delta(z-z_0) \hat{z} \hat{z}  
\nonumber\\
&&+ 
(\omega/c)^2 (\hat{\kv}_\parallel \times \hat{z}) (\hat{\kv}_\parallel \times \hat{z}) 
+ 
\big[ -k_z \,{\rm Sgn} \, (z-z_0) \hat{\kv}_\parallel + k_\parallel \hat{z} \big]
\nonumber\\
&&\times 
\big[ -k_z \,{\rm Sgn} \, (z-z_0) \hat{\kv}_\parallel + k_\parallel \hat{z} \big] 
\Big\} \, 
e^{ik_z|z - z_0|} e^{i\kv_\parallel\cdot(\rv_\parallel - \rv_{0,\parallel})}, 
\nonumber
\ea
If we express ${\bf E}^{(0)}(\rv, \omega)$ in terms of a Fourier transform over $\kv_\parallel$, i.e., 
\be
{\bf E}^{(0)}(\rv, \omega) = 
\int\!\!\frac{d^2\kv_\parallel}{(2\pi)^2} 
e^{i\kv_\parallel\cdot(\rv_\parallel - \rv_{0,\parallel})}
{\bf E}^{(0)}(\kv_\parallel, z; \omega), 
\ee
we have that 
\ba
\label{phayao}
&&{\bf E}^{(0)}(\kv_\parallel, z; \omega) 
\\
&\!\!=\!\!& 
-4\pi \delta(z-z_0) \, \hat{z} \hat{z} \cdot \pv(\omega)
+ \frac{2\pi i}{k_z} 
\Big\{ 
 (\omega/c)^2 (\hat{\kv}_\parallel \times \hat{z}) (\hat{\kv}_\parallel \times \hat{z}) 
\nonumber\\
&&+
[ -k_z \,{\rm Sgn} \, (z-z_0) \hat{\kv}_\parallel + k_\parallel \hat{z} ]
[ -k_z \,{\rm Sgn} \, (z-z_0) \hat{\kv}_\parallel + k_\parallel \hat{z} ] 
\Big\} 
\nonumber\\
&&\cdot 
\pv(\omega) 
\, 
e^{ik_z|z - z_0|} 
\nonumber
\ea
We can express this result in terms of polarisation vectors $\hat{e}_p^\pm(\kv_\parallel) = (1/k)(\mp k_z \hat{\kv}_\parallel + k_\parallel \hat{z})$ and $\hat{e}_s^\pm(\kv_\parallel) = \hat{\kv}_\parallel \times \hat{z}$. To do so, we make use of the identities
\ba
&&1 = \Theta(z-z_0) + \Theta(z_0-z), 
\\
&&{\rm Sgn} (z-z_0) = \Theta(z-z_0) - \Theta(z_0-z), 
\\
&&\Theta(z-z_0)\Theta(z-z_0) = \Theta(z-z_0),
\\
&&\Theta(z-z_0)\Theta(z_0-z) = 0, 
\\
&&\Theta(z_0-z) \Theta(z_0-z) = \Theta(z_0-z). 
\ea
We can thus write the expression the curly braces times the phase factor in Eq.~(\ref{phayao}) as 
\ba
&& \big\{ 
(\omega/c)^2 (\Theta(z-z_0) + \Theta(z_0-z)) 
(\hat{\kv}_\parallel \times \hat{z}) (\hat{\kv}_\parallel \times \hat{z}) 
\nonumber\\
&&+
[ -k_z ( \Theta(z-z_0) - \Theta(z_0-z) ) \hat{\kv}_\parallel + k_\parallel \hat{z} ]
\nonumber\\
&&[ -k_z ( \Theta(z-z_0) - \Theta(z_0-z) ) \hat{\kv}_\parallel + k_\parallel \hat{z} ] 
\big\} e^{i k_z |z - z_0|}
\nonumber\\
&=& 
[
(\omega/c)^2 
(\hat{\kv}_\parallel \times \hat{z}) 
(\hat{\kv}_\parallel \times \hat{z})  
\nonumber\\
&&+ 
(- k_z \hat{\kv}_\parallel + k_\parallel \hat{z}) 
(- k_z \hat{\kv}_\parallel + k_\parallel \hat{z})
]
e^{i k_z (z - z_0)} \Theta(z - z_0) 
\nonumber\\
&&+
[
(\omega/c)^2 (\hat{\kv}_\parallel \times \hat{z}) (\hat{\kv}_\parallel \times \hat{z})  
\nonumber\\
&&+ 
(k_z \hat{\kv}_\parallel + k_\parallel \hat{z}) 
(k_z \hat{\kv}_\parallel + k_\parallel \hat{z})
]
e^{-i k_z (z - z_0)} \Theta(z_0 - z) 
\nonumber\\
&=& 
[
(\omega/c)^2 
\hat{e}_s^+(\kv_\parallel) \hat{e}_s^+(\kv_\parallel)
+ k^2 \hat{e}_p^+(\kv_\parallel) \hat{e}_p^+(\kv_\parallel)
]
\nonumber\\
&&\times
e^{i k_z (z - z_0)} \Theta(z - z_0) 
\nonumber\\
&&+
[
(\omega/c)^2 \hat{e}_s^-(\kv_\parallel) \hat{e}_s^-(\kv_\parallel)
+ 
k^2 \hat{e}_p^-(\kv_\parallel) \hat{e}_p^-(\kv_\parallel) ]
\nonumber\\
&&
\times 
e^{-i k_z (z - z_0)} \Theta(z_0 - z) 
\ea
Defining the symbols $\xi^s = -1$ and $\xi^p = 1$, and making use of the reflection properties $\hat{e}_s^\pm(\kv_\parallel) = - \hat{e}_s^\mp(-\kv_\parallel)$, $\hat{e}_p^\pm(\kv_\parallel) = \hat{e}_p^\mp(-\kv_\parallel)$, we can write the above equation as 
\ba
&&[
(\omega/c)^2 \xi^s 
\hat{e}_s^+(\kv_\parallel) \hat{e}_s^-(-\kv_\parallel)
+ k^2 \xi^p \hat{e}_p^+(\kv_\parallel) \hat{e}_p^-(-\kv_\parallel)
]
\nonumber\\
&&\times
e^{i k_z (z - z_0)} \Theta(z - z_0) 
\nonumber\\
&&+
[
(\omega/c)^2 \xi^s \hat{e}_s^-(\kv_\parallel) \hat{e}_s^+(-\kv_\parallel)
+ 
k^2 \xi^p \hat{e}_p^-(\kv_\parallel) \hat{e}_p^+(-\kv_\parallel) ]
\nonumber\\
&&
\times 
e^{-i k_z (z - z_0)} \Theta(z_0 - z) 
\nonumber\\
&=& 
(\omega/c)^2 
[\xi^s \hat{e}_s^+(\kv_\parallel) \hat{e}_s^-(-\kv_\parallel)
+ \xi^p \hat{e}_p^+(\kv_\parallel) \hat{e}_p^-(-\kv_\parallel)
]
\nonumber\\
&&\times
e^{i k_z (z - z_0)} \Theta(z - z_0) 
\nonumber\\
&&+
(\omega/c)^2 
[\xi^s \hat{e}_s^-(\kv_\parallel) \hat{e}_s^+(-\kv_\parallel)
+ 
\xi^p \hat{e}_p^-(\kv_\parallel) \hat{e}_p^+(-\kv_\parallel) ]
\nonumber\\
&&
\times 
e^{-i k_z (z - z_0)} \Theta(z_0 - z) 
\ea
In the above, we have also noted that $k^2 = (\omega/c)^2$ in the vacuum. 
Plugging into Eq.~(\ref{phayao}), we obtain
\ba
&&{\bf E}^{(0)}(\kv_\parallel, z; \omega) 
\\
&=& 
-4\pi \delta(z-z_0) \, \hat{z} \hat{z} \cdot \pv(\omega)
+ \frac{2\pi i}{k_z} 
\nonumber\\
&&\times 
\left( \frac{\omega}{c} \right)^2
\sum_{\sigma = p, s}
\xi^\sigma
\Big[ 
\hat{e}_\sigma^+ (\kv_\parallel) \hat{e}_\sigma^- (-\kv_\parallel) 
e^{ik_z(z-z_0)} \Theta(z-z_0)
\nonumber\\
&&+
\hat{e}_\sigma^- (\kv_\parallel) \hat{e}_\sigma^+ (-\kv_\parallel) 
e^{-ik_z(z-z_0)} \Theta(z_0-z)
\Big] 
\cdot 
\pv(\omega)
\nonumber
\ea
We can rewrite the above expression as
\ba
\label{uppo}
&&{\bf E}^{(0)}(\kv_\parallel, z; \omega) 
\\
&=& 
-4\pi \delta(z-z_0) \, \hat{z} \hat{z} \cdot \pv(\omega)
\nonumber\\
&&+ 
\sum_{\sigma=p, s} 
\big[
A_\sigma^{(0)} e^{i k_z z} \Theta(z-z_0) \hat{e}_\sigma^+ (\kv_\parallel) 
\nonumber\\
&&+ 
B_\sigma^{(0)} e^{-i k_z z} \Theta(z_0-z) \hat{e}_\sigma^- (\kv_\parallel) 
\big], 
\nonumber
\ea
where the coefficients $A_\sigma^{(0)}$ and $B_\sigma^{(0)}$ are given by 
\begin{subequations}
\label{AB-coeffs}
\ba
A_\sigma^{(0)} 
&\!=\!& 
\frac{2\pi i}{k_z} \left( \frac{\omega}{c} \right)^2
\xi^\sigma
e^{-ik_z z_0}
\hat{e}_\sigma^- (-\kv_\parallel) \cdot \pv(\omega); 
\\
B_\sigma^{(0)} 
&\!=\!& 
\frac{2\pi i}{k_z} \left( \frac{\omega}{c} \right)^2
\xi^\sigma
e^{ik_z z_0}
\hat{e}_\sigma^+ (-\kv_\parallel) \cdot \pv(\omega).
\ea
\end{subequations}
Equation~(\ref{uppo}) describes an upward-propagating wave to above the source ($z > z_0$), and a downward-propagating wave to below the source ($z < z_0$). 
If we also express the bulk Green function in terms of a Fourier integral over $\kv_\parallel$, i.e., 
\be
{\mathbb F}^{(0)}(\rv, \rv_0; \omega) = \int\!\!\frac{d^2\kv_\parallel}{(2\pi)^2} 
{\mathbb F}^{(0)} (\kv_\parallel, z, z_0; \omega) \, 
e^{i\kv_\parallel\cdot(\rv_\parallel - \rv_{0,\parallel})},
\ee
then we have that 
\be
\label{E0G0}
{\bf E}^{(0)}(\kv_\parallel, z; \omega) =  
{\mathbb F}^{(0)} (\kv_\parallel, z, z_0; \omega) \cdot \pv(\omega),
\ee
where the \emph{bulk Green tensor} is given by 
\ba
&&{\mathbb F}^{(0)} (\kv_\parallel, z, z_0; \omega) 
\\
&\!=\!& 
-4\pi \delta(z-z_0) \, \hat{z} \hat{z} 
\nonumber\\
&&
+ \frac{2\pi i}{k_z} \left( \frac{\omega}{c} \right)^2
\sum_{\sigma = p, s}
\xi^\sigma
\Big[ 
\hat{e}_\sigma^+ (\kv_\parallel) \hat{e}_\sigma^- (-\kv_\parallel) 
e^{ik_z(z-z_0)} \Theta(z-z_0)
\nonumber\\
&&+
\hat{e}_\sigma^- (\kv_\parallel) \hat{e}_\sigma^+ (-\kv_\parallel) 
e^{-ik_z(z-z_0)} \Theta(z_0-z)
\Big] 
\nonumber
\ea

\subsection{reflection Green tensor}
To find ${\mathbb F}^{R}$, we express $\Ev^R$ in terms of the reflection coefficients, starting from Eq.~(\ref{ER}): 
\ba
&&\Ev^R(\kv_\parallel, z; \omega) 
\\
&\!\!=\!\!& 
\big( A_s \hat{e}_s^+ (\kv_\parallel) + A_p \hat{e}_p^+ (\kv_\parallel) \big) e^{i k_{z} z}
\nonumber\\
&\!\!=\!\!& 
\big( (r_{ss} B_s^{(0)} + r_{ps} B_p^{(0)}) \hat{e}_s^+ (\kv_\parallel) 
\nonumber\\
&&+ (r_{pp} B_p^{(0)} + r_{sp} B_s^{(0)}) \hat{e}_p^+ (\kv_\parallel) \big) e^{i k_{z} z}
\nonumber\\
&\!\!=\!\!& 
(2\pi i/k_{z}) (\omega/c)^2 e^{i k_{z} (z+z_0)} \pv(\omega) \cdot
\nonumber\\
&&
\big( r_{ss} \xi^s \hat{e}_s^+ (-\kv_\parallel) \hat{e}_s^+ (\kv_\parallel)  
+ r_{ps} \xi^p \hat{e}_p^+ (-\kv_\parallel) \hat{e}_s^+ (\kv_\parallel) 
\nonumber\\
&&+ r_{sp} \xi^s \hat{e}_s^+ (-\kv_\parallel) \hat{e}_p^+ (\kv_\parallel)
+ r_{pp} \xi^p \hat{e}_p^+ (-\kv_\parallel) \hat{e}_p^+ (\kv_\parallel) \big) 
\nonumber
\ea
We can thus identify the reflection Green tensor: 
\ba
\label{GR}
&&{\mathbb F}^{R} (\kv_\parallel, z, z_0; \omega) 
= {\mathbb G}^{R} (\kv_\parallel, z, z_0; \omega)
\\
&\!=\!& 
(2\pi i/k_{z}) (\omega/c)^2 e^{i k_{z} (z+z_0)} 
\nonumber\\
&&
\big( r_{ss} \xi^s \hat{e}_s^+ (\kv_\parallel) \hat{e}_s^+ (-\kv_\parallel) 
+ r_{ps} \xi^p \hat{e}_s^+ (\kv_\parallel) \hat{e}_p^+ (-\kv_\parallel)
\nonumber\\
&&+ r_{sp} \xi^s \hat{e}_p^+ (\kv_\parallel) \hat{e}_s^+ (-\kv_\parallel)
+ r_{pp} \xi^p \hat{e}_p^+ (\kv_\parallel) \hat{e}_p^+ (-\kv_\parallel) \big) 
\nonumber\\
&\!=\!& 
(2\pi i/k_{z}) (\omega/c)^2 e^{i k_{z} (z+z_0)} 
\nonumber\\
&&
\big( r_{ss} \hat{e}_s^+ (\kv_\parallel)  \hat{e}_s^- (\kv_\parallel) 
+ r_{ps} \hat{e}_s^+ (\kv_\parallel) \hat{e}_p^- (\kv_\parallel)
\nonumber\\
&&+ r_{sp} \hat{e}_p^+ (\kv_\parallel) \hat{e}_s^- (\kv_\parallel)
+ r_{pp} \hat{e}_p^+ (\kv_\parallel) \hat{e}_p^- (\kv_\parallel) \big) 
\nonumber
\ea
The first equality obtains because of Eq.~(\ref{GsFs}), and on going to the second identity, we made use of 
\be
\hat{e}_p^{\pm}(\kv_\parallel) = \hat{e}_p^{\mp}(-\kv_\parallel), 
\quad
\hat{e}_s^{\pm}(\kv_\parallel) = - \hat{e}_s^{\mp}(-\kv_\parallel), 
\ee
To obtain the components of the Green tensor, we make use of the identities
\ba
\hat{x} \!\cdot\! \hat{e}_s^{\pm}(\kv_\parallel) &\!=\!& 
\frac{k_y}{k_\parallel}, 
\,\,\, 
\hat{y} \!\cdot\! \hat{e}_s^{\pm}(\kv_\parallel) =
- \frac{k_x}{k_\parallel}, 
\nonumber\\
\hat{x} \!\cdot\! \hat{e}_p^{+}(\kv_\parallel) &\!=\!& 
- \frac{c k_x k_{z}}{\omega k_\parallel}, 
\,\,\, 
\hat{y} \!\cdot\! \hat{e}_p^{+}(\kv_\parallel) =
- \frac{c k_y k_{z}}{\omega k_\parallel}, 
\nonumber\\
\hat{x} \!\cdot\! \hat{e}_p^{-}(\kv_\parallel) &\!=\!& 
 \frac{c k_x k_{z}}{\omega k_\parallel}, 
\,\,\, 
\hat{y} \!\cdot\! \hat{e}_p^{-}(\kv_\parallel) =
 \frac{c k_y k_{z}}{\omega k_\parallel}, 
\nonumber\\
\hat{z} \!\cdot\! \hat{e}_p^{\pm}(\kv_\parallel) &\!=\!& 
\frac{c k_\parallel}{\omega}, 
\,\,\,
\hat{z} \!\cdot\! \hat{e}_s^{\pm}(\kv_\parallel) = 0. 
\ea
In what follows, we will neglect to write the argument of $\hat{e}_\sigma^\pm (\kv_\parallel)$ with the understanding that the argument is positive, i.e., $\hat{e}_\sigma^\pm  \equiv \hat{e}_\sigma^\pm (\kv_\parallel)$, and let us write ${\mathbb{G}}^{R} (\kv_\parallel, z_0; \omega) \equiv {\mathbb{G}}^{R} (\kv_\parallel, z_0, z_0; \omega).$ 
The xx component of the scattering Green tensor with $z = z_0$ is then 
\ba
&&G_{xx}^{R} (\kv_\parallel, z_0; \omega) = \hat{x} \!\cdot\! {\mathbb{G}}^{R} (\kv_\parallel, z_0; \omega) \!\cdot\! \hat{x}
\\
&\!=\!& 
(2\pi i/k_{z}) (\omega/c)^2 e^{2i k_{z} z_0} 
\nonumber\\
&&
\big( r_{ss} (\hat{x} \!\cdot\! \hat{e}_s^-) (\hat{x} \!\cdot\! \hat{e}_s^+) 
+ r_{ps} ( \hat{x} \!\cdot\! \hat{e}_p^-) (\hat{x} \!\cdot\! \hat{e}_s^+) 
\nonumber\\
&&+ r_{sp} ( \hat{x} \!\cdot\! \hat{e}_s^-) (\hat{x} \!\cdot\! \hat{e}_p^+) 
+ r_{pp} (\hat{x} \!\cdot\! \hat{e}_p^-) (\hat{x} \!\cdot\! \hat{e}_p^+) \big) 
\nonumber\\
&\!=\!& 
(2\pi i/k_{z}) (\omega/c)^2 e^{2i k_{z} z_0} 
\nonumber\\
&&\times 
\bigg( 
r_{ss} \Big( \frac{k_y}{k_\parallel} \Big)^2 
+ (r_{ps} - r_{sp}) \frac{c k_x k_y k_z}{\omega k_\parallel^2}
- r_{pp} \Big( \frac{c k_x k_z}{\omega k_\parallel} \Big)^2
\bigg)
\nonumber
\ea
The yy component of the scattering Green tensor with $z = z_0$ is given by 
\ba
&&G_{yy}^{R} (\kv_\parallel, z_0; \omega) = \hat{y} \!\cdot\! {\mathbb{G}}^{R} (\kv_\parallel, z_0; \omega) \!\cdot\! \hat{y}
\\
&\!=\!& 
(2\pi i/k_{z}) (\omega/c)^2 e^{2i k_{z} z_0} 
\nonumber\\
&&
\big( r_{ss} (\hat{y} \!\cdot\! \hat{e}_s^-) (\hat{y} \!\cdot\! \hat{e}_s^+) 
+ r_{ps} ( \hat{y} \!\cdot\! \hat{e}_p^-) (\hat{y} \!\cdot\! \hat{e}_s^+) 
\nonumber\\
&&+ r_{sp} ( \hat{y} \!\cdot\! \hat{e}_s^-) (\hat{y} \!\cdot\! \hat{e}_p^+) 
+ r_{pp} (\hat{y} \!\cdot\! \hat{e}_p^-) (\hat{y} \!\cdot\! \hat{e}_p^+) \big) 
\nonumber\\
&\!=\!& 
(2\pi i/k_{z}) (\omega/c)^2 e^{2i k_{z} z_0} 
\nonumber\\
&&\times 
\bigg( 
r_{ss} \Big( \frac{k_x}{k_\parallel} \Big)^2 
- (r_{ps} - r_{sp}) \frac{c k_x k_y k_z}{\omega k_\parallel^2}
- r_{pp} \Big( \frac{c k_y k_z}{\omega k_\parallel} \Big)^2
\bigg)
\nonumber
\ea
The xy component of the scattering Green tensor with $z = z_0$ is given by 
\ba
&&G_{xy}^{R} (\kv_\parallel, z_0; \omega) 
= \hat{x} \!\cdot\! {\mathbb{G}}^{R} (\kv_\parallel, z_0; \omega) \!\cdot\! \hat{y}
\\
&\!=\!& 
(2\pi i/k_{z}) (\omega/c)^2 e^{2i k_{z} z_0} 
\nonumber\\
&&
\big( r_{ss} (\hat{x} \!\cdot\! \hat{e}_s^+) (\hat{y} \!\cdot\! \hat{e}_s^-)
+ r_{ps} (\hat{x} \!\cdot\! \hat{e}_s^+) ( \hat{y} \!\cdot\! \hat{e}_p^-)
\nonumber\\
&&+ r_{sp} (\hat{x} \!\cdot\! \hat{e}_p^+) ( \hat{y} \!\cdot\! \hat{e}_s^-)
+ r_{pp} (\hat{x} \!\cdot\! \hat{e}_p^+) (\hat{y} \!\cdot\! \hat{e}_p^-) \big) 
\nonumber\\
&\!=\!& 
(2\pi i/k_{z}) (\omega/c)^2 e^{2i k_{z} z_0} 
\nonumber\\
&&\times 
\bigg( 
- r_{ss} \frac{k_x k_y}{k_\parallel^2} + r_{ps} \frac{c k_y^2 k_z}{\omega k_\parallel^2}
+r_{sp} \frac{c k_x^2 k_z}{\omega k_\parallel^2} - r_{pp} \frac{c^2 k_x k_y k_z^2}{\omega^2 k_\parallel^2}
\bigg)
\nonumber
\ea
The yx component of the scattering Green tensor with $z = z_0$ is given by 
\ba
&&G_{yx}^{R} (\kv_\parallel, z_0; \omega) 
= \hat{y} \!\cdot\! {\mathbb{G}}^{R} (\kv_\parallel, z_0; \omega) \!\cdot\! \hat{x}
\\
&\!=\!& 
(2\pi i/k_{z}) (\omega/c)^2 e^{2i k_{z} z_0} 
\nonumber\\
&&
\big( r_{ss} (\hat{y} \!\cdot\! \hat{e}_s^+) (\hat{x} \!\cdot\! \hat{e}_s^-)
+ r_{ps} (\hat{y} \!\cdot\! \hat{e}_s^+) ( \hat{x} \!\cdot\! \hat{e}_p^-)
\nonumber\\
&&+ r_{sp} (\hat{y} \!\cdot\! \hat{e}_p^+) ( \hat{x} \!\cdot\! \hat{e}_s^-)
+ r_{pp} (\hat{y} \!\cdot\! \hat{e}_p^+) (\hat{x} \!\cdot\! \hat{e}_p^-) \big) 
\nonumber\\
&\!=\!& 
- (2\pi i/k_{z}) (\omega/c)^2 e^{2i k_{z} z_0} 
\nonumber\\
&&\times 
\bigg( 
r_{ss} \frac{k_x k_y}{k_\parallel^2} + r_{ps} \frac{c k_x^2 k_z}{\omega k_\parallel^2}
+ r_{sp} \frac{c k_y^2 k_z}{\omega k_\parallel^2} + r_{pp} \frac{c^2 k_x k_y k_z^2}{\omega^2 k_\parallel^2}
\bigg)
\nonumber
\ea
The zz component of the scattering Green tensor with $z = z_0$ is given by 
\ba
&&G_{zz}^{R} (\kv_\parallel, z_0; \omega) 
= \hat{z} \!\cdot\! {\mathbb{G}}^{R} (\kv_\parallel, z_0; \omega) \!\cdot\! \hat{z}
\\
&\!=\!& 
(2\pi i/k_{z}) (\omega/c)^2 e^{2i k_{z} z_0} 
\nonumber\\
&&
\big( r_{ss} (\hat{z} \!\cdot\! \hat{e}_s^-) (\hat{z} \!\cdot\! \hat{e}_s^+) 
+ r_{ps} ( \hat{z} \!\cdot\! \hat{e}_p^-) (\hat{z} \!\cdot\! \hat{e}_s^+) 
\nonumber\\
&&+ r_{sp} ( \hat{z} \!\cdot\! \hat{e}_s^-) (\hat{z} \!\cdot\! \hat{e}_p^+) 
+ r_{pp} (\hat{z} \!\cdot\! \hat{e}_p^-) (\hat{z} \!\cdot\! \hat{e}_p^+) \big) 
\nonumber\\
&\!=\!&  (2\pi i \, k_\parallel^2/k_{z}) e^{2i k_{z} z_0} 
r_{pp}.
\nonumber
\ea
Writing $\int d^2\kv_\parallel = \int_{-\infty}^\infty dk_x \int_{-\infty}^\infty k_y = \int_0^\infty dk_\parallel k_\parallel \int_0^{2\pi} d\phi$ with $k_x = k_\parallel \cos\phi$ and $k_y = k_\parallel \sin\phi$, and observing that $k_z$ and the reflection coefficients are even in $k_x$ and in $k_y$, and functions odd in $k_x$ and/or $k_y$ vanish under integration over $k_x$ and $k_y$, we obtain after performing an inverse Fourier transform the result in Eqs.~(\ref{dyadic}). 

\subsection{transmission Green tensor} 

For completeness, we present our results for the transmission coefficients below:  
\ba
t_{ss} &\!=\!& \frac{1}{\Delta} \big( 1 + \widetilde{k}_z \widetilde{\sigma}_{xx} \big), 
\nonumber\\
t_{sp} &\!=\!& - t_{ps} = \frac{\widetilde{\sigma}_{xy}}{\Delta}, 
\nonumber\\
t_{pp} &\!=\!& \frac{1}{\Delta} \big( 1 + \widetilde{k}_z^{-1} \widetilde{\sigma}_{xx} \big), 
\label{transmission}
\ea 
where the denominator $\Delta$ is defined by Eq.~(\ref{Delta}). 
Similar to how we derived the reflection Green tensor, to find the transmission Green tensor ${\mathbb F}^{T}$, we express $\Ev^T$ in terms of the reflection coefficients using Eq.~(\ref{ET}): 
\ba
&&\Ev^T(\kv_\parallel, z; \omega) 
\\
&\!\!=\!\!& 
\big( B_s \hat{e}_s^{-} (\kv_\parallel) + B_p \hat{e}_p^{-} (\kv_\parallel) \big) e^{-i k_{z} z}
\nonumber\\
&\!\!=\!\!& 
\big( (t_{ss} B_s^{(0)} + t_{ps} B_p^{(0)}) \hat{e}_s^{-} (\kv_\parallel) 
\nonumber\\
&&+ (t_{pp} B_p^{(0)} + t_{sp} B_s^{(0)}) \hat{e}_p^{-} (\kv_\parallel) \big) e^{-i k_{z} z}\nonumber\\
&\!\!=\!\!& 
(2\pi i/k_{z}) (\omega/c)^2 e^{-i k_{z} (z - z_0)} \pv(\omega) \cdot
\nonumber\\
&&
\big( t_{ss} \xi^s \hat{e}_s^+ (-\kv_\parallel) \hat{e}_s^{-} (\kv_\parallel)  
+ t_{ps} \xi^p \hat{e}_p^+ (-\kv_\parallel) \hat{e}_s^{-} (\kv_\parallel) 
\nonumber\\
&&+ t_{sp} \xi^s \hat{e}_s^+ (-\kv_\parallel) \hat{e}_p^{-} (\kv_\parallel)
+ t_{pp} \xi^p \hat{e}_p^+ (-\kv_\parallel) \hat{e}_p^{-} (\kv_\parallel) \big) 
\nonumber
\ea
We can thus identify the transmission Green tensor:
\ba
\label{GT}
&&{\mathbb F}^{T} (\kv_\parallel, z, z_0; \omega) 
= {\mathbb G}^{T} (\kv_\parallel, z, z_0; \omega)
\\
&\!=\!& 
(2\pi i/k_{z}) (\omega/c)^2 e^{-i k_{z} (z - z_0)} 
\nonumber\\
&&
\big( t_{ss} \xi^s \hat{e}_s^{-} (\kv_\parallel) \hat{e}_s^+ (-\kv_\parallel)
+ t_{ps} \xi^p \hat{e}_s^{-} (\kv_\parallel) \hat{e}_p^+ (-\kv_\parallel)
\nonumber\\
&&+ t_{sp} \xi^s \hat{e}_p^{-} (\kv_\parallel) \hat{e}_s^+ (-\kv_\parallel) 
+ t_{pp} \xi^p \hat{e}_p^{-} (\kv_\parallel) \hat{e}_p^+ (-\kv_\parallel) \big) 
\nonumber\\
&\!=\!& 
(2\pi i/k_{z}) (\omega/c)^2 e^{-i k_{z} (z - z_0)} 
\nonumber\\
&&
\big( t_{ss} \hat{e}_s^- (\kv_\parallel) \hat{e}_s^{-} (\kv_\parallel) 
+ t_{ps} \hat{e}_s^- (\kv_\parallel) \hat{e}_p^{-} (\kv_\parallel) 
\nonumber\\
&&+ t_{sp} \hat{e}_p^- (\kv_\parallel) \hat{e}_s^{-} (\kv_\parallel) 
+ t_{pp} \hat{e}_p^- (\kv_\parallel) \hat{e}_p^{-} (\kv_\parallel) \big) 
\nonumber
\ea

\section{current matrix elements} 
\label{app:current}
Here we show our results for the elements of the current matrix in the Qi-Wu-Zhang model of the Chern insulator for general values of $u$: 
\begin{widetext}
\begin{subequations}
\label{current1}
\ba
&&{{\rm Re}}\, [\langle + | j_x | - \rangle \langle - | j_y | + \rangle] 
= {{\rm Re}}\, [\langle - | j_x | + \rangle \langle + | j_y | - \rangle] 
\nonumber\\
&\!\!=\!\!& 
\frac{(tae)^2 \sin k_x a \sin k_y a}{32 \, d^2 d_\parallel^2} 
\Big(
t^2 \big( \cos 2 k_x a + \cos 2 k_y a \big) - 2 (d_\parallel^2 + t^2)
\Big)
\nonumber\\
&&\qquad\times 
\Big(
t^2 \big( \cos 2 k_x a + \cos 2 k_y a \big) - 4 (d_\parallel^2 + t^2) 
- 8 t d_z \cos k_y a 
\Big)
\nonumber\\
&&+ 
\frac{(tae)^2 \sin 2 k_x a \sin k_y a}{16 \, d^2 d_\parallel^2} 
\Big(
4 t d_z (d_\parallel^2 + t^2) 
- t^2 
\big(
8 d_\parallel^2 \cos k_y a 
+ t d_z ( \cos 2 k_x a + 2 \cos 2 k_y a)
\big)
\Big)
,
\\
&&{{\rm Im}}\, [\langle + | j_x | - \rangle \langle - | j_y | + \rangle] 
= - {{\rm Im}}\, [\langle - | j_x | + \rangle \langle + | j_y | - \rangle] 
\nonumber\\
&\!\!=\!\!& 
\frac{t^3 (ae)^2}{4\,d d_\parallel^2} 
\Big(
\big( \cos k_x a \sin^2 k_y a + \cos k_y a \sin^2 k_x a \big) 
\big( 2(d_\parallel^2 + t^2) - t^2 (\cos 2k_x a + \cos 2k_y a) \big)
\nonumber\\
&&\qquad\qquad+ 
4td_z \cos k_x a \cos k_y a \big( \sin^2 k_x a + \sin^2 k_y a \big)
\Big), 
\\
&&{{\rm Re}}\, [\langle + | j_x | - \rangle \langle - | j_x | + \rangle] 
\nonumber\\
&\!\!=\!\!& 
\frac{(tae)^2}{16\,d^2 d_\parallel^2} 
\Big(
\sin^2 k_x a
\big(
t^2 (\cos 2k_x a + \cos 2k_y a) - 2(d_\parallel^2 + t^2) 
\big)
\big(
t^2 (\cos 2k_x a + \cos 2k_y a) - 2(d_\parallel^2 + t^2) - 8 t d_z \cos k_x a
\big)
\nonumber\\
&&\qquad\qquad+ 4 t^2 d_z^2 \sin^2 2k_x a
+ 16 t^2 d^2 \cos^2 k_x a \sin^2 k_y a
\Big),   
\\
&&{{\rm Im}}\, [\langle + | j_x | - \rangle \langle - | j_x | + \rangle] 
= 0.
\ea
\end{subequations}
In the above, we have defined $d_\parallel^2 \equiv d_x^2 + d_y^2$, where $d_x$ and $d_y$ are given by Eq.~(\ref{qwz}). 

\end{widetext}


\section{van Hove singularities at $\kv^{{\rm T}} = (\pm \pi/a,0), (0,\pm \pi/a)$ for $|u| = t$} 
\label{app:VHS}
The case of van Hove singularities (VHS) for two-dimensional saddle points has been discussed in Ref.~\cite{bassani1975}, which shows that the electronic DOS exhibits logarithmic divergence. For the special case $|u| = t$, the points $\kv^{{\rm T}} = (\pm \pi/a,0), (0,\pm \pi/a)$ are no longer two-dimensional saddle points, instead becoming one-dimensional minima as one of the eigenvalues of the corresponding Hessian matrix vanishes. To see this, we perform a Taylor expansion on $d(\kv)$ around a VHS point $(\kv^0)^{{\rm T}} = (k_x^0, k_y^0)$: 
\be
d(\kv) \approx d_0 + ({\bm{\nabla}}_\kv d(\kv))_{\kv_0} \!\cdot\! \Delta \kv + \frac{1}{2} \Delta \kv^{{\rm T}} \!\cdot\! \mathcal{H} \!\cdot\! \Delta \kv, 
\ee
where $d_0 \equiv d(\kv^0)$, $\Delta \kv^{{\rm T}} \equiv (\Delta k_x, \Delta k_y)$, $\Delta k_{x/y} \equiv (k_{x/y} - k_{x/y}^0)^2$, and $\mathcal{H}$ is the Hessian matrix, defined by 
\be
\mathcal{H} \equiv 
\left. 
\begin{pmatrix}
\frac{\partial^2 d(\kv)}{\partial k_x^2} & \frac{\partial^2 d(\kv)}{\partial k_x \partial k_y} 
\\
\frac{\partial^2 d(\kv)}{\partial k_y \partial k_x} & \frac{\partial^2 d(\kv)}{\partial k_y^2}
\end{pmatrix}
\right\vert_{\kv_0}. 
\ee
At a VHS point, ${\bm{\nabla}}_\kv d(\kv))_{\kv_0} = 0$, and we can approximate the argument of the Dirac delta function in the DOS by  
\be
\hbar \omega - 2d(\kv) \approx \hbar \omega - 2d_0 + Q_x^2 + Q_y^2,  
\ee
where 
\ba
Q_x^2 &=& \frac{m_x}{2} \Delta k_x^2 \equiv \frac{(ta)^2}{d_0} \cos k_x^0 a (\cos k_y^0 a + \frac{u}{t}) \Delta k_x^2, 
\nonumber\\
Q_y^2 &=& \frac{m_y}{2} \Delta k_y^2 \equiv \frac{(ta)^2}{d_0} \cos k_y^0 a (\cos k_x^0 a + \frac{u}{t}) \Delta k_y^2. 
\nonumber
\ea 
Here $m_x$ and $m_y$ are the diagonal elements of the Hessian matrix (which has vanishing off-diagonal elements). 

To be specific, let's consider the VHS point $(\kv^0)^{{\rm T}} = (0, -\pi/a)$ (similar conclusions will hold for $(\kv^0)^{{\rm T}} = (0, \pi/a), (\pm \pi/a, 0)$). At this point, $d_0 = u = t$, $Q_x^2 = 0$, and $Q_y^2 = - 2ta^2 \Delta k_y^2 \equiv - q_y^2$, so one of the eigenvalues of the Hessian matrix vanishes, implying that the electronic DOS has become effectively \emph{one-dimensional.}  We also have that $d\Delta k_y = (2ta^2)^{-1/2} dq_y$, and $\int_{-\pi/a}^{\pi/a} dk_y = \int_{0}^{2\pi/a} d\Delta k_y = (2ta^2)^{-1/2} \int_0^{(2ta^2)^{1/2}2\pi/a} dq_y$. The integral over $k_x$ evaluates simply to $2\pi/a$, and we can therefore rewrite the electronic DOS as
\ba
&&\int_{BZ} \!\!\! dk_x dk_y \, \delta(\hbar\omega - 2d_0 + \frac{m_y}{a} \Delta k_y^2) 
\nonumber\\
&=& \frac{2\pi}{a(2ta^2)^{1/2}} 
\int_0^{(2ta^2)^{1/2}2\pi/a} 
\!\!\!\!\!\!\!\!\!\!\!\!\!\!\!\!\!\!\!\!\!\!\!\!\!\!\!\!
dq_y \, \delta(\hbar\omega - 2d_0 - q_y^2). 
\nonumber
\ea
The DOS vanishes if $\hbar\omega < 2t$ (as the argument of the delta function becomes negative). For $\hbar\omega > 2t$, the argument vanishes if $q_y^2 = \hbar\omega - 2t$, i.e., if $q_y = (\hbar\omega - 2t)^{1/2}$ (we reject the negative root as this is outside the integration range of $q_y$). 
Using the fact that $\int_a^b g(x) \delta(f(x)) dx = \sum_{x_0} g(x_0)/|\partial f/\partial x|_{x_0}$ (where $x_0$ is a zero of $f(x)$ in the interval $(a, b)$), we find that the electronic DOS becomes 
\be
\int_{BZ} \!\!\! dk_x dk_y \, \delta(\hbar\omega - 2t - q_y^2) = \frac{\pi}{a \sqrt{2ta^2}} \frac{1}{(\hbar \omega - 2t)^{1/2}}. 
\ee
Near the VHS points $\kv^{{\rm T}} = (0, \pm \pi/a), (\pm \pi/a, 0)$ and for $u = t$, we can approximate ${{\rm Im}}\, [\langle + | j_x | - \rangle \langle - | j_y | + \rangle] \approx -(tae)^2$ and ${{\rm Re}}\, [\langle + | j_x | - \rangle \langle - | j_x | + \rangle] \approx (tae)^2$ (cf. Eqs.~(\ref{current1})). We can also replace $(f(d)-f(-d))/d$ by $1/t$ as  the temperature tends to zero, whereupon we obtain (from Eq.~(\ref{cond-tensor}))
\be
\label{divrexx}
\frac{{{\rm Re}} \, \sigma_{xx}(\omega)}{\alpha c} 
\approx 
\frac{1}{8\sqrt{2}\sqrt{\widetilde{\omega} -2}}  \quad (\text{$\widetilde{\omega} > 2$})
\ee
where we have defined $\widetilde{\omega} \equiv \hbar\omega/t$. 
The above function exhibits the $(\hbar \omega - 2t)^{-1/2}$ divergence associated with VHS in the DOS of one-dimensional periodic systems, and accounts for the peak in ${{\rm Re}} \, \sigma_{xx}$ (see Fig.~4a) as $\hbar\omega/t \rightarrow 2 + \epsilon$. 
This leads to a similar divergence in ${{\rm Im}} \, \sigma_{xx}$ as $\hbar\omega/t \rightarrow 2-\epsilon$ in Fig.~4b, as we now see. 
A Kramers-Kr\"{o}nig relation leads to the following expression for ${{\rm Im}} \, \sigma_{xx}(\omega)$, valid for contributions from the neighborhood of the effectively one-dimensional VHS points $\kv^{{\rm T}} = (0, \pm \pi/a), (\pm \pi/a, 0)$: 
\ba
\label{divimxx}
&&\frac{{{\rm Im}} \, \sigma_{xx}(\omega)}{\alpha c} 
\\
&\approx& 
-\frac{\widetilde{\omega}}{4\sqrt{2}\pi} \mathcal{P} \! \int_2^\infty \!\!\! \frac{d\widetilde{\omega}'}{((\widetilde{\omega}')^2 - \widetilde{\omega}^2)\sqrt{\widetilde{\omega}'-2}} 
\nonumber\\ 
&=& 
\frac{1}{8\sqrt{2}} \left( \frac{1}{\sqrt{2+\widetilde{\omega}}} - \frac{1}{\sqrt{2-\widetilde{\omega}}} \right) \quad (\text{$\widetilde{\omega} < 2$})
\nonumber
\ea
This diverges to negative infinity as $\hbar\omega/t \rightarrow 2 - \epsilon$, consistent with the behavior shown in Fig.~4b. 

Similarly, we find 
\ba
\label{divimxy}
&&\frac{{{\rm Im}} \, \sigma_{xy}(\omega)}{\alpha c} 
\approx 
\frac{1}{8\sqrt{2}\sqrt{\widetilde{\omega} -2}}  \quad (\text{$\widetilde{\omega} > 2$});
\\
&&\frac{{{\rm Re}} \, \sigma_{xy}(\omega)}{\alpha c} 
\approx 
\frac{1}{4\sqrt{2}\pi} \mathcal{P} \! \int_2^\infty \!\!\! \frac{\widetilde{\omega} \, d\widetilde{\omega}'}{((\widetilde{\omega}')^2 - \widetilde{\omega}^2)\sqrt{\widetilde{\omega}'-2}} 
\nonumber\\
&=& 
\frac{1}{4\sqrt{2}\pi} 
\sqrt{\frac{1}{4 - \widetilde{\omega}^2} + \frac{1}{2\sqrt{4 - \widetilde{\omega}^2}}} 
 \quad (\text{$\widetilde{\omega} < 2$})
\ea

\section{asymptotia}
\label{sec:app2}

To study the asymptotic behavior with respect to the emitter-surface separation distance, we scale out the dependence on $\eta$ in the exponential factors in Eqs.~(\ref{dyadic}) by defining a new dimensionless variable $t \equiv \tk_z \eta$ for the range $0 \leq \tk_\parallel < 1$ (recalling that $\tk_z = (1 - \tk_\parallel^2)^{1/2}$), whence $\tk_\parallel d\tk_\parallel = - \tk_z d\tk_z = - t \, dt/\eta^2$. For this range, $0 \leq \tk_z < 1$ and thus $0 \leq t < \eta$. 
On the other hand, for the range $1 \leq \tk_\parallel < \infty$, we have $0 \leq \tk_z < i\infty$. If we define $\tk_z \equiv i \ell$, then $0 \leq \ell < \infty$, and we can define $t \equiv \ell \eta$, whereupon $\tk_\parallel d\tk_\parallel = \ell \, d\ell = t \, dt/\eta^2$. 
After rescaling and using Eqs.~(\ref{r-coeffs}), Eqs.~(\ref{dyadic}) become  
\ba
&&\Big(\frac{c}{\omega_{10}}\Big)^3 
G_{xx}^R(\rv_0,\rv_0;\omega_{10}) 
\nonumber\\
&\!\!=\!\!& 
- \frac{i}{2\eta} \int_0^\eta \!\!dt \frac{(1 + (\frac{t}{\eta})^2)(\widetilde{\sigma}_{xx}^2 + \widetilde{\sigma}_{xy}^2) + (\frac{\eta}{t} + (\frac{t}{\eta})^3)\widetilde{\sigma}_{xx}}{1 + \widetilde{\sigma}_{xx}^2 + \widetilde{\sigma}_{xy}^2 + (\frac{\eta}{t} + \frac{t}{\eta})\widetilde{\sigma}_{xx}} e^{it}
\nonumber\\
&&- 
\frac{1}{2\eta} \int_0^\infty \!\!dt \frac{(1 - (\frac{t}{\eta})^2)(\widetilde{\sigma}_{xx}^2 + \widetilde{\sigma}_{xy}^2) - i ((\frac{t}{\eta})^3 + \frac{\eta}{t})\widetilde{\sigma}_{xx}}{1 + \widetilde{\sigma}_{xx}^2 + \widetilde{\sigma}_{xy}^2 + i (\frac{t}{\eta} - \frac{\eta}{t})\widetilde{\sigma}_{xx}} e^{-t}, 
\nonumber\\
&&\Big(\frac{c}{\omega_{10}}\Big)^3 
 G_{yx}^R(\rv_0,\rv_0;\omega_{10}) = 
\Big(\frac{c}{\omega_{10}}\Big)^3 
- G_{xy}^R(\rv_0,\rv_0;\omega_{10}) 
\nonumber\\
&\!\!=\!\!& 
\frac{i}{\eta^2} \int_0^\eta \!\!dt \, t \frac{\widetilde{\sigma}_{xy}}{1 + \widetilde{\sigma}_{xx}^2 + \widetilde{\sigma}_{xy}^2 + (\frac{\eta}{t} + \frac{t}{\eta})\widetilde{\sigma}_{xx}} e^{it}
\nonumber\\
&&+ 
\frac{i}{\eta^2} \int_0^\infty \!\!dt \, t \frac{\widetilde{\sigma}_{xy}}{1 + \widetilde{\sigma}_{xx}^2 + \widetilde{\sigma}_{xy}^2 + i (\frac{t}{\eta} - \frac{\eta}{t})\widetilde{\sigma}_{xx}} e^{-t}, 
\nonumber\\
&&\Big(\frac{c}{\omega_{10}}\Big)^3 
G_{zz}^R(\rv_0,\rv_0;\omega_{10}) 
\nonumber\\
&\!\!=\!\!& 
\frac{i}{\eta} \int_0^\eta \!\!dt \, 
\bigg( 1 - \frac{t^2}{\eta^2} \bigg)
 \frac{\widetilde{\sigma}_{xx}^2 + \widetilde{\sigma}_{xy}^2 + \frac{t}{\eta} \widetilde{\sigma}_{xx}}{1 + \widetilde{\sigma}_{xx}^2 + \widetilde{\sigma}_{xy}^2 + (\frac{\eta}{t} + \frac{t}{\eta})\widetilde{\sigma}_{xx}} e^{it}
\nonumber\\
&&+ 
\frac{1}{\eta} \int_0^\infty \!\!dt \, 
\bigg( 1 + \frac{t^2}{\eta^2} \bigg)
 \frac{\widetilde{\sigma}_{xx}^2 + \widetilde{\sigma}_{xy}^2 + i \frac{t}{\eta} \widetilde{\sigma}_{xx}}{1 + \widetilde{\sigma}_{xx}^2 + \widetilde{\sigma}_{xy}^2 + i (\frac{t}{\eta} - \frac{\eta}{t})\widetilde{\sigma}_{xx}} e^{-t}. 
\nonumber\\
\label{C1}
\ea
\subsection{far-field asymptotics (large $\eta$)}
The threshold at which $\eta$ can be considered asymptotically large depends on the frequency regime one is considering. 
In the high frequency regime, large separations correspond to $\eta \gg (\widetilde{\sigma}_{xx}'')^{-1}$, where we have noted that the threshold is not set by 1 but $(\widetilde{\sigma}_{xx}'')^{-1}$, as the latter can be larger than 1 by a few orders of magnitude. For this regime, using Eqs.~(\ref{C1}) we find 
\ba
&&\Big(\frac{c}{\omega_{10}}\Big)^3 
G_{xx}^R(\rv_0,\rv_0;\omega_{10}) 
\approx 
- \frac{\cos\eta + i \sin \eta}{2\eta}, 
\nonumber\\
&&\Big(\frac{c}{\omega_{10}}\Big)^3 
G_{xy}^R(\rv_0,\rv_0;\omega_{10})
 \approx
- \frac{\widetilde{\sigma}_{xy} (\cos\eta + i \sin \eta)}{\widetilde{\sigma}_{xx} \eta}
, 
\nonumber\\
&&\Big(\frac{c}{\omega_{10}}\Big)^3 
G_{zz}^R(\rv_0,\rv_0;\omega_{10}) 
\approx
\frac{\widetilde{\sigma}_{xx}^2 + \widetilde{\sigma}_{xy}^2}{\widetilde{\sigma}_{xx} \eta^2}
\big( (\eta + i) e^{i\eta} \big). 
\nonumber\\
\ea
For $\widetilde{\sigma}_{xx}'' \gg \widetilde{\sigma}_{xy}'$ we can approximate the last line by 
\ba
\Big(\frac{c}{\omega_{10}}\Big)^3 
G_{zz}^R(\rv_0,\rv_0;\omega_{10}) 
&\!\!\approx\!\!&
- \frac{\widetilde{\sigma}_{xx}''}{\eta} \sin \eta
+ i \frac{\widetilde{\sigma}_{xx}''}{\eta} \cos \eta. 
\nonumber 
\ea

\subsection{intermediate asymptotics ($\eta$ not large, $\widetilde{\sigma}_{xx}''$ small)}
Next, we consider the high frequency regime, without however assuming that $\eta$ is large. Thus our considerations in this subsection can apply to the range of intermediate emitter-surface separations, e.g., $\eta \sim 10$. For such separations, we saw (in Secs.~IV B and C) that the normalised transition rates for the parallel-aligned and circularly polarised dipole configurations exhibit the shape of sine-integral oscillations. The sine-integral oscillations originate from the contribution dependent on $r_{ss}$ in ${{\rm Im}} \, G_{xx}^R$ (cf. Eq.~(\ref{Gxx})), as we may deduce by the following argument. For the intermediate separations at which the transition rates exhibit the shape of sine-integral oscillations, $\eta^{-1}$ is of the order of $0.1$. On the other hand, $\widetilde{\sigma}_{xx}''$ can be as small as $2.8 \times 10^{-3}$ (for $\hbar\omega/t = 10$) or $2.6 \times 10^{-4}$ (for $\hbar\omega/t = 100$). As $\eta^{-1}$ is not smaller than $\widetilde{\sigma}_{xx}''$, we cannot perform a perturbation analysis based on the smallness of $\eta^{-1}$. Instead, let's consider the leading order expansion in $\widetilde{\sigma}_{xx}''$ of ${{\rm Im}} \, G_{xx}^R$, for values where $\widetilde{\sigma}_{xy}' \ll \widetilde{\sigma}_{xx}'' \ll 1$ (and $\widetilde{\sigma}_{xy}''=\widetilde{\sigma}_{xx}' = 0$): 
\ba
&&\Big(\frac{c}{\omega_{10}}\Big)^3 \, 
{{\rm Im}} \, G_{xx}^R 
\nonumber\\
&\!\!\approx\!\!& 
\frac{\widetilde{\sigma}_{xx}''}{2} \bigg( {{\rm Si}}(\eta) + \frac{3(\eta^2-2)\sin\eta - \eta(\eta^2-6) \cos\eta}{\eta^4} \bigg), 
\nonumber
\ea
where ${{\rm Si}} \equiv \int_0^\eta dt \sin t/t$ denotes the sine integral function, which is present in $r_{ss}$. The sine integral has the asymptotic property that ${{\rm Si}}(\eta) \rightarrow \pi/2$ as $\eta \rightarrow \infty$. 

On the other hand, we also find to linear order in $\widetilde{\sigma}_{xx}''$: 
\ba
\Big(\frac{c}{\omega_{10}}\Big)^3 \, 
{{\rm Im}} \, G_{xy}^R 
&\!\!=\!\!& 
\Big(\frac{c}{\omega_{10}}\Big)^3 \, 
{{\rm Im}} \, G_{yx}^R \approx 0, 
\nonumber\\
\Big(\frac{c}{\omega_{10}}\Big)^3 \, 
{{\rm Im}} \, G_{zz}^R 
&\!\!\approx\!\!& 
\frac{2 \widetilde{\sigma}_{xx}''}{\eta^4} 
\big(  
(\eta^2 - 3) \sin \eta + 3 \eta \cos \eta
\big).
\nonumber
\ea
Thus, for intermediate separations and $\widetilde{\sigma}_{xx}'' \ll \widetilde{\sigma}_{xy}'$, the parallel-aligned dipole's transition rate is 
\be
\frac{R_{10}}{R_{10}^{(0)}} \approx 
1 + \frac{3 \widetilde{\sigma}_{xx}''}{4} \bigg( {{\rm Si}}(\eta) + \frac{3(\eta^2-2)\sin\eta - \eta(\eta^2-6) \cos\eta}{\eta^4} \bigg), 
\ee
whilst the perpendicular dipole's transition rate is 
\be
\frac{R_{10}}{R_{10}^{(0)}} \approx 
1 + \frac{3 \widetilde{\sigma}_{xx}''}{\eta^4} 
\big(  
(\eta^2 - 3) \sin \eta + 3 \eta \cos \eta
\big).
\ee
The transition rate for the perpendicular dipole involves only ${{\rm Im}} \, G_{zz}^R$, which does not depend on $r_{ss}$. 
Thus we see that whilst the transition rates for the parallel-aligned and circularly polarised dipole configurations involve $r_{ss}$ and exhibit sine-integral oscillations, the transition rate for the perpendicularly aligned dipole does not exhibit sine-integral oscillations. 

\subsection{Near-field asymptotics ($\eta \rightarrow 0$)} 

The Green tensor components in Eq.~(D1) each consist of two integrals, one which involves an oscillatory integrand with an upper limit that depends on $\eta$, and the other which involves an exponentially decaying integrand and with an upper limit that is taken to infinity.
As we shall see, the two integrals thus have very different near-field asymptotic behaviors, with the oscillatory integral contribution tending towards a finite value as $\eta \rightarrow 0$, whereas the exponentially decaying contribution diverges according to a power law in the same limit.  

The near-field behavior of the normalised dipole transition rate in the low ($\omega < 2(2t-|u|)/\hbar$) and high ($\omega > 2(2t+|u|)/\hbar$) frequency regimes is qualitatively distinct (i.e., characterised by a different scaling law) from that in the intermediate ($2(2t-|u|)/\hbar < \omega < 2(2t+|u|)/\hbar$) frequency regime, because in the low and high frequency regimes, the imaginary part of the exponentially decaying integrand vanishes and only the oscillatory integral contributes, whilst in the intermediate frequency regime, the exponentially decaying integrand does not vanish and the integral with its power-law divergence actually dominates over the oscillatory integral contribution.  

\subsubsection{low and high frequency regimes}

First we consider the low  and high frequency  regimes, where $\widetilde{\sigma}_{xx} = i\widetilde{\sigma}_{xx}''$ and $\widetilde{\sigma}_{xy} = \widetilde{\sigma}_{xy}'$. For these regimes, the imaginary part of the exponentially decaying integrand vanishes, and we have 
\ba
&&\Big(\frac{c}{\omega_{10}}\Big)^3 
G_{xx}^R(\rv_0,\rv_0;\omega_{10}) 
\nonumber\\
&\!\!=\!\!&
- \frac{i}{2\eta} \int_0^\eta \!\!dt \frac{(1 + (\frac{t}{\eta})^2)
\big[ (\widetilde{\sigma}_{xy}')^2 - (\widetilde{\sigma}_{xx}'')^2 \big] 
+ i \big[ \frac{\eta}{t} + (\frac{t}{\eta})^3 \big] \widetilde{\sigma}_{xx}''}{1 + (\widetilde{\sigma}_{xy}')^2 - (\widetilde{\sigma}_{xx}'')^2 + i (\frac{\eta}{t} + \frac{t}{\eta})\widetilde{\sigma}_{xx}''} e^{it}, 
\nonumber\\
&&\Big(\frac{c}{\omega_{10}}\Big)^3 
G_{xy}^R(\rv_0,\rv_0;\omega_{10}) 
\nonumber\\
&\!\!=\!\!&
- \frac{i}{\eta^2} \int_0^\eta \!\!dt \, t \frac{\widetilde{\sigma}_{xy}'}{1 + (\widetilde{\sigma}_{xy}')^2 - (\widetilde{\sigma}_{xx}'')^2 + i (\frac{\eta}{t} + \frac{t}{\eta})\widetilde{\sigma}_{xx}''} e^{it}, 
\nonumber\\
&&\Big(\frac{c}{\omega_{10}}\Big)^3 
G_{zz}^R(\rv_0,\rv_0;\omega_{10}) 
\nonumber\\
&\!\!=\!\!&
\frac{i}{\eta} \int_0^\eta \!\!dt \, 
\bigg( 1 - \frac{t^2}{\eta^2} \bigg)
 \frac{(\widetilde{\sigma}_{xy}')^2 - (\widetilde{\sigma}_{xx}'')^2 + i \frac{t}{\eta} \widetilde{\sigma}_{xx}''}{1 + (\widetilde{\sigma}_{xy}')^2 - (\widetilde{\sigma}_{xx}'')^2 + i (\frac{\eta}{t} + \frac{t}{\eta})\widetilde{\sigma}_{xx}''} e^{it}. 
\nonumber\\
\ea
As $\eta \rightarrow 0$, $t$ also tends to $0$, which means we can replace $t/\eta$ by 1 and $e^{it}$ by 1 in the integrand for $G_{xx}^R$. In the integrand for $G_{zz}^R$ we  make a similar replacement except for $(1 - t^2/\eta^2)$, as we are interested in the scaling behavior with $\eta$ as $\eta \rightarrow 0$. 
\ba
&&\Big(\frac{c}{\omega_{10}}\Big)^3 
G_{xx}^R(\rv_0,\rv_0;\omega_{10}) 
\nonumber\\
&\!\!\approx\!\!&
- \frac{i}{\eta} \int_0^\eta \!\!dt \frac{
(\widetilde{\sigma}_{xy}')^2 - (\widetilde{\sigma}_{xx}'')^2 + i \widetilde{\sigma}_{xx}''}{1 + (\widetilde{\sigma}_{xy}')^2 - (\widetilde{\sigma}_{xx}'')^2 + 2 i \widetilde{\sigma}_{xx}''}
\nonumber\\
&\!\!=\!\!&
- i \frac{
(\widetilde{\sigma}_{xy}')^2 - (\widetilde{\sigma}_{xx}'')^2 + i \widetilde{\sigma}_{xx}''}{1 + (\widetilde{\sigma}_{xy}')^2 - (\widetilde{\sigma}_{xx}'')^2 + 2 i \widetilde{\sigma}_{xx}''}, 
\nonumber\\
&&\Big(\frac{c}{\omega_{10}}\Big)^3 
G_{xy}^R(\rv_0,\rv_0;\omega_{10}) 
\nonumber\\
&\!\!\approx\!\!&
- \frac{i}{\eta^2} \int_0^\eta \!\!dt \, t \frac{\widetilde{\sigma}_{xy}'}{1 + (\widetilde{\sigma}_{xy}')^2 - (\widetilde{\sigma}_{xx}'')^2  + 2 i \widetilde{\sigma}_{xx}''} 
\nonumber\\
&\!\!=\!\!&
- \frac{i}{2} \frac{\widetilde{\sigma}_{xy}'}{1 + (\widetilde{\sigma}_{xy}')^2 - (\widetilde{\sigma}_{xx}'')^2  + 2 i \widetilde{\sigma}_{xx}''}, 
\nonumber\\
&&\Big(\frac{c}{\omega_{10}}\Big)^3 
G_{zz}^R(\rv_0,\rv_0;\omega_{10}) 
\nonumber\\
&\!\!\approx\!\!&
\frac{i}{\eta} \int_0^\eta \!\!dt \, 
\bigg( 1 - \frac{t^2}{\eta^2} \bigg)
 \frac{(\widetilde{\sigma}_{xy}')^2 - (\widetilde{\sigma}_{xx}'')^2 + i \widetilde{\sigma}_{xx}''}{1 + (\widetilde{\sigma}_{xy}')^2 - (\widetilde{\sigma}_{xx}'')^2 + 2 i \widetilde{\sigma}_{xx}''}
\nonumber\\
&\!\!=\!\!&
\frac{2i}{3} 
 \frac{(\widetilde{\sigma}_{xy}')^2 - (\widetilde{\sigma}_{xx}'')^2 + i \widetilde{\sigma}_{xx}''}{1 + (\widetilde{\sigma}_{xy}')^2 - (\widetilde{\sigma}_{xx}'')^2 + 2 i \widetilde{\sigma}_{xx}''}. 
\ea
Taking the imaginary part of $G_{xx}^R$ and $G_{zz}^R$, and the real part of $G_{xy}^R$,  we obtain 
\ba
&&\Big(\frac{c}{\omega_{10}}\Big)^3 \, {{\rm Im}} \, G_{xx}^R(\rv_0,\rv_0;\omega_{10}) 
\nonumber\\
&\!\!=\!\!&
- \frac{1}{2} 
\bigg[ 2 
- \frac{1}{1+(\widetilde{\sigma}_{xx}'' - \widetilde{\sigma}_{xy}')^2}
- \frac{1}{1+(\widetilde{\sigma}_{xx}'' + \widetilde{\sigma}_{xy}')^2}
\bigg], 
\nonumber\\
&&\Big(\frac{c}{\omega_{10}}\Big)^3 
{{\rm Re}} \, G_{xy}^R(\rv_0,\rv_0;\omega_{10})
\nonumber\\
&\!\!=\!\!&
- \frac{\widetilde{\sigma}_{xy}' \widetilde{\sigma}_{xx}''}{\big[ 1 - (\widetilde{\sigma}_{xx}'')^2 +(\widetilde{\sigma}_{xy}')^2 \big]^2 + 4(\widetilde{\sigma}_{xx}'')^2},
\nonumber\\
&&\Big(\frac{c}{\omega_{10}}\Big)^3 \, {{\rm Im}} \, G_{zz}^R(\rv_0,\rv_0;\omega_{10}) 
\nonumber\\
&\!\!=\!\!&
\frac{1}{3} 
\bigg[ 2 
- \frac{1}{1+(\widetilde{\sigma}_{xx}'' - \widetilde{\sigma}_{xy}')^2}
- \frac{1}{1+(\widetilde{\sigma}_{xx}'' + \widetilde{\sigma}_{xy}')^2}
\bigg].
\ea
The transition rate for the parallel-aligned dipole thus behaves as 
\be
\frac{R_{10}}{R_{10}^{(0)}} 
\rightarrow  
1 
- \frac{3}{4} 
\bigg[ 2 
- \frac{1}{1+(\widetilde{\sigma}_{xx}'' - \widetilde{\sigma}_{xy}')^2}
- \frac{1}{1+(\widetilde{\sigma}_{xx}'' + \widetilde{\sigma}_{xy}')^2}
\bigg], 
\ee
whilst that for the perpendicularly aligned dipole behaves as 
\be
\frac{R_{10}}{R_{10}^{(0)}} \rightarrow 
2 
- \frac{1}{2 \big[ 1 + (\widetilde{\sigma}_{xx}'' - \widetilde{\sigma}_{xy}')^2 \big]}
- \frac{1}{2 \big[ 1 + (\widetilde{\sigma}_{xx}'' + \widetilde{\sigma}_{xy}')^2 \big]}.
\ee
The right circularly-polarised dipole transition rate behaves as 
\ba
&&\frac{R_{10}}{R_{10}^{(0)}} 
\\
&\!\!\rightarrow\!\!&
1 - 
\frac{3}{2} \frac{\widetilde{\sigma}_{xy}' \widetilde{\sigma}_{xx}''}{\big[ 1 - (\widetilde{\sigma}_{xx}'')^2 +(\widetilde{\sigma}_{xy}')^2 \big]^2 + 4(\widetilde{\sigma}_{xx}'')^2} 
\nonumber\\
&&-
 \frac{3}{4} 
\left[ 2 
- \frac{1}{2 \big[ 1 + (\widetilde{\sigma}_{xx}'' - \widetilde{\sigma}_{xy}')^2 \big]}
- \frac{1}{2 \big[ 1 + (\widetilde{\sigma}_{xx}'' + \widetilde{\sigma}_{xy}')^2 \big]} 
\right].
\nonumber
\ea
For the low frequency regime, we take the test frequencies $\widetilde{\omega} = 1, 1.9$.
For the case $C = 1$ and $\widetilde{\omega} = 1$, $\widetilde{\sigma}_{xx}'' \approx -0.0055$ and $\widetilde{\sigma}_{xy}' \approx 0.0092$. 
Correspondingly, 
for the perpendicularly aligned dipole, $R_{10}/R_{10}^{(0)} \approx 1.0001$; 
for the parallel aligned dipole, $R_{10}/R_{10}^{(0)} \approx 0.9998$; 
and for the right circularly polarised dipole, $R_{10}/R_{10}^{(0)} \approx 0.9999$.
For $C = 1$ and $\widetilde{\omega} = 1.9$, $\widetilde{\sigma}_{xx}'' \approx -0.049$ and $\widetilde{\sigma}_{xy}' \approx 0.049$. 
Correspondingly, for the perpendicularly aligned dipole, $R_{10}/R_{10}^{(0)} \approx 1.0048$;  
for the parallel aligned dipole, $R_{10}/R_{10}^{(0)} \approx 0.9928$; 
and for the right circularly polarised dipole, $R_{10}/R_{10}^{(0)} \approx 0.996$.  

For the case $C = -1$ and $\widetilde{\omega} = 1$, $\widetilde{\sigma}_{xx}'' \approx -0.0055$ and $\widetilde{\sigma}_{xy}' \approx -0.0092$. 
The corresponding near-field transition rate for a right-circularly polarised dipole is $R_{10}/R_{10}^{(0)} \approx 0.9998$. 
For $C = -1$ and $\widetilde{\omega} = 1.9$, $\widetilde{\sigma}_{xx}'' \approx -0.049$ and $\widetilde{\sigma}_{xy}' \approx -0.049$.
The corresponding near-field transition rate for a right-circularly polarised dipole is $R_{10}/R_{10}^{(0)} \approx 0.989$. 

\subsubsection{intermediate frequency regime} 

Let us now consider the near-field asymptotic behavior of $G_{xx}^R$, $G_{xy}^R$ and $G_{zz}^R$ in the intermediate frequency regime. 
For this regime, $\widetilde{\sigma}_{xx}', \widetilde{\sigma}_{xy}'' \neq 0$, so the second, exponentially decaying integrand contribution to the Green tensor component acquires an imaginary part. The first, oscillatory integrand contribution still tends to a finite limiting value as $\eta \rightarrow 0$, but the imaginary part of the exponentially decaying integrand contribution diverges and dominates the near-field behavior of the transition rate. We thus focus on the second integral contribution:
\ba
&&\Big(\frac{c}{\omega_{10}}\Big)^3 
G_{xx}^R(\rv_0,\rv_0;\omega_{10}) 
\nonumber\\
&\!\!\approx\!\!&
- 
\frac{1}{2\eta} \int_0^\infty \!\!dt \frac{(1 - (\frac{t}{\eta})^2)(\widetilde{\sigma}_{xx}^2 + \widetilde{\sigma}_{xy}^2) - i ((\frac{t}{\eta})^3 + \frac{\eta}{t})\widetilde{\sigma}_{xx}}{1 + \widetilde{\sigma}_{xx}^2 + \widetilde{\sigma}_{xy}^2 + i (\frac{t}{\eta} - \frac{\eta}{t})\widetilde{\sigma}_{xx}} e^{-t}, 
\nonumber\\
&&\Big(\frac{c}{\omega_{10}}\Big)^3 
G_{xy}^R(\rv_0,\rv_0;\omega_{10}) 
\approx
- \frac{1}{\eta} \int_0^\infty \!\!dt \, e^{-t}
\frac{\widetilde{\sigma}_{xy}}{\widetilde{\sigma}_{xx}}, 
\nonumber\\
&&\Big(\frac{c}{\omega_{10}}\Big)^3 
G_{zz}^R(\rv_0,\rv_0;\omega_{10}) 
\nonumber\\
&\!\!\approx\!\!&
\frac{1}{\eta} \int_0^\infty \!\!dt \, 
\bigg( 1 + \frac{t^2}{\eta^2} \bigg)
 \frac{\widetilde{\sigma}_{xx}^2 + \widetilde{\sigma}_{xy}^2 + i \frac{t}{\eta} \widetilde{\sigma}_{xx}}{1 + \widetilde{\sigma}_{xx}^2 + \widetilde{\sigma}_{xy}^2 + i (\frac{t}{\eta} - \frac{\eta}{t})\widetilde{\sigma}_{xx}} e^{-t}. 
\nonumber\\
\ea
By performing a series expansion in powers of $1/\eta$, we find the leading order contributions to the imaginary part of $G_{xx}^R$ and $G_{zz}^R$ and the real part of $G_{xy}^R$ are given by  
\ba
&&\Big(\frac{c}{\omega_{10}}\Big)^3 
{{\rm Im}} \, 
G_{xx}^R(\rv_0,\rv_0;\omega_{10})  
\approx 
\frac{\widetilde{\sigma}_{xx}'}{2 \big( (\widetilde{\sigma}_{xx}')^2 + (\widetilde{\sigma}_{xx}'')^2 \big) \eta^2}, 
\nonumber\\
&&\Big(\frac{c}{\omega_{10}}\Big)^3 
{{\rm Re}} \, G_{xy}^R(\rv_0,\rv_0;\omega_{10}) 
\approx
- \frac{\widetilde{\sigma}_{xy}' \widetilde{\sigma}_{xx}' + \widetilde{\sigma}_{xy}'' \widetilde{\sigma}_{xx}''}{\big( (\widetilde{\sigma}_{xx}')^2 + (\widetilde{\sigma}_{xx}'')^2 \big)\eta},  
\nonumber\\
&&\Big(\frac{c}{\omega_{10}}\Big)^3 
{{\rm Im}} \, G_{zz}^R(\rv_0,\rv_0;\omega_{10}) 
\approx 
\frac{\widetilde{\sigma}_{xx}'}{\big( (\widetilde{\sigma}_{xx}')^2 + (\widetilde{\sigma}_{xx}'')^2 \big) \eta^2}. 
\nonumber\\
\ea
As $\eta \rightarrow 0$, the transition rate for the horizontal dipole is dominated by the divergence in ${{\rm Im}} \, G_{xx}^R$: 
\be
\frac{R_{10}}{R_{10}^{(0)}} \rightarrow \frac{\widetilde{3\sigma}_{xx}'}{4 \big( (\widetilde{\sigma}_{xx}')^2 + (\widetilde{\sigma}_{xx}'')^2 \big) \eta^2}. 
\ee
Similarly, the asymptotic behavior for the transition rate of the vertical dipole is given by 
\be
\frac{R_{10}}{R_{10}^{(0)}} 
\rightarrow 
\frac{3\widetilde{\sigma}_{xx}'}{2\big( (\widetilde{\sigma}_{xx}')^2 + (\widetilde{\sigma}_{xx}'')^2 \big) \eta^2}. 
\ee
As $\widetilde{\sigma}_{xx}' > 0$, we see that the above normalised transition rates diverge to positive infinity as $\eta \rightarrow 0$. 
As ${{\rm Im}} \, G_{xx}^R$ diverges as $\eta^{-2}$ in the near-field limit, which is stronger than the $\eta^{-1}$ near-field divergence of ${{\rm Re}} \, G_{xy}^R$, the leading-order near-field asymptotic behavior of the right circularly-polarised dipole transition rate coincides with that for the horizontal dipole, and is the same for both $C = 1$ and $C = -1$. 

\subsection{near a van Hove singularity with $|u| = t$ and $\omega = 2t/\hbar$}
\label{app:asymp-vhs}

For the case where $|u| = t$ and the frequency approaches $\omega = 2t/\hbar$ (the value associated with the effectively one-dimensional van Hove singularities described in the previous section), 
both $\sigma_{xx}$ and $\sigma_{xy}$ become divergent. 
Thus, in the reflection coefficients in Eqs.~(\ref{r-coeffs}) the terms $\widetilde{\sigma}_{xx}^2$ and $\widetilde{\sigma}_{xy}^2$ are much larger than $\widetilde{\sigma}_{xx}$, and we can approximate the reflection coefficients to leading order by 
\begin{subequations}
\label{r-coeffs-VHS}
\ba
r_{ss} &\!\!=\!\!& - r_{pp} \approx -1, 
\\
r_{ps} &\!\!=\!\!& r_{sp} \approx 0.  
\ea
\end{subequations}
Physically, these coefficient values correspond to the case of a perfectly conducting mirror. 
Correspondingly, the imaginary parts of Green tensor components relevant to our transition rate calculations are given by 
\ba
&&{{\rm Im}} \, G_{xx}^R(\rv_0,\rv_0;\omega_{10}) 
\approx 
- \Big(\frac{\omega_{10}}{c}\Big)^3 
\frac{\eta \cos \eta + (\eta^2 - 1) \sin \eta}{\eta^3}, 
\nonumber\\
&&{{\rm Im}} \, G_{xy}^R(\rv_0,\rv_0;\omega_{10}) 
= 
-
{{\rm Im}} \, G_{yx}^R(\rv_0,\rv_0;\omega_{10})
 \approx 0, 
\nonumber\\
&&{{\rm Im}} \, G_{zz}^R(\rv_0,\rv_0;\omega_{10}) 
\approx
- 2 \Big(\frac{\omega_{10}}{c}\Big)^3  \frac{\eta \cos \eta - \sin \eta}{\eta^3}. 
\ea
The normalised transition rate for the perpendicularly aligned dipole becomes 
\be
\frac{R_{10}}{R_{10}^{(0)}} \approx 1 - 3 \left( \frac{\eta \cos \eta - \sin \eta}{\eta^3} \right), 
\ee
and for the parallel aligned and circularly polarised dipoles it becomes 
\be
\frac{R_{10}}{R_{10}^{(0)}} \approx 1 - \frac{3}{2} \left( \frac{\eta \cos \eta + (\eta^2 - 1) \sin \eta}{\eta^3} \right).  
\ee


\begin{thebibliography}{99}

\bibitem{weng2015}
H. Weng, R. Yu, X. Hu, X. Dai, and Z. Fang, 
``Quantum anomalous Hall effect and related topological electronic states." 
{\em Adv. Phys.}~{\bf 64}, 227 (2015). 

\bibitem{liu2016}
C.-X. Liu, S.-C. Zhang, and X.-L. Qi, 
``The quantum anomalous Hall effect: theory and experiment." 
{\em Annu. Rev. Condens. Matter Phys.}~{\bf 7}, 301 (2016). 

\bibitem{cayssol2013}
J. Cayssol, 
``Introduction to Dirac materials and topological insulators."
{\em Comptes Rendus Physique}~{\bf 14}, 760 (2013). 

\bibitem{bernevig2013}
B. A. Bernevig and T. L. Hughes, {\em Topological insulators and topological superconductors} (Princeton University Press, Princeton, 2013). 

\bibitem{ren2016}
Y. Ren, Z. Qiao, and Q. Niu, 
``Topological phases in two-dimensional materials: a review." 
{\em Rep. Prog. Phys.}~{\bf 79}, 066501 (2016). 

\bibitem{zhang2016} 
J. Zhang, B. Zhao, T. Zhou, and Z. Yang, 
``Quantum anomalous Hall effect in real materials." 
{\em Chin. Phys. B}~{\bf 25}, 117308 (2016). 

\bibitem{lifshitz1955} 
E. M. Lifshitz, ``The theory of molecular attractive forces between solids."
{\em Zh. Eksp. Teor. Fiz.}~{\bf 29}, 94--110 (1955); English translation {\em Sov. Phys. JETP}~{\bf 2}, 73--83 (1956).

\bibitem{grushin2011}
A. G. Grushin, P. Rodriguez-Lopez, and A. Cortijo, 
``Effect of finite temperature and uniaxial anisotropy on the Casimir effect with three-dimensional topological insulators." 
{\em Phys. Rev. B}~{\bf 84}, 045119 (2011).

\bibitem{pablo2014}
P. Rodriguez-Lopez, A. G. Grushin, 
``Repulsive Casimir Effect with Chern insulators."
{\em Phys. Rev. Lett.}~{\bf 112}, 056804 (2014). 

\bibitem{hoye2018}
J. S. H\o ye and I. Brevik, 
``Repulsive Casimir force." 
{\em Phys. Rev. A}~{\bf 98}, 022503 (2018).

\bibitem{song2014} 
G. Song, J.-P. Xu, and Y.-P. Yang, 
``Spontaneous emission of a two-level system near the interface of topological insulators." 
{\em EPL}~{\bf 105}, 64001 (2014).

\bibitem{fuchs2017}
S. Fuchs, J. A. Crosse, and S. Y. Buhmann, ``Casimir-Polder shift and decay rate in the presence of nonreciprocal media." 
{\em Phys. Rev. A}~{\bf 95}, 023805 (2017). 

\bibitem{fang2015} 
W. Fang, Z.-X. Yang, and G.-X. Li, 
``Quantum properties of an atom in a cavity constructed by topological insulators." 
{\em J. Phys. B: At. Mol. Opt. Phys.}~{\bf 48}, 245504 (2015). 

\bibitem{zeng2019} 
R. Zeng, et al., 
``Spontaneous emission interference in topological multilayers." 
{\em J. Opt. Soc. Amer. B}~{\bf 36}, 1890 (2019).    

\bibitem{wilczek1987}
F. Wilczek, 
``Two applications of axion electrodynamics." 
{\em Phys. Rev. Lett.}~{\bf 58}, 1799 (1987).

\bibitem{qi2008}
X.-L. Qi, T. L. Hughes, and S.-C. Zhang, 
``Topological field theory of time-reversal invariant insulators." 
{\em Phys. Rev. B}~{\bf 78}, 195424 (2008).  

\bibitem{lu2018}
B.-S. Lu, ``van der Waals torque and force between anisotropic topological insulator slabs." 
{\em Phys. Rev. B}~{\bf 97}, 045427 (2018). 

\bibitem{ziegler2013} 
A. Hill, A. Sinner, and K. Ziegler, 
``Optical Hall conductivity of systems with gapped spectral nodes."
{\em Eur. Phys. J. B}~{\bf 86}, 53 (2013).    












\bibitem{qwz2006} 
X.-L. Qi, Y.-S. Wu, and S.-C. Zhang, 
``Topological quantization of the spin Hall effect in two-dimensional paramagnetic semiconductors." 
{\em Phys. Rev. B}~{\bf 74}, 085308 (2006). 

\bibitem{asboth2016}
J. K. Asboth, L. Oroszl\'{a}ny, and A. P\'{a}lyi, 
{\em A Short Course on Topological Insulators} 
(Springer, Heidelberg, 2016), Ch. 6.

\bibitem{eykholt1986}
R. Eykholt, ``Extension of the Kubo formula for the electrical-conductivity tensor to arbitrary polarisations of the electric field." 
{\em Phys. Rev. B}~{\bf 34}, 6669 (1986). 

\bibitem{bassani1975} 
F. Bassani and G. P. Parravicini, {\em Electronic States and Optical Transitions in Solids} (Pergamon Press, Oxford, 1975). 

\bibitem{dressel2002}
M. Dressel and G. Gr\"{u}ner, \emph{Electrodynamics of Solids: Optical Properties of Electrons in Matter} (Cambridge University Press, Cambridge, UK, 2002). 

\bibitem{gonzalez-tudela2019}
A. Gonz\'{a}lez-Tudela and F. Galve, 
``Anisotropic Quantum Emitter Interactions in Two-Dimensional Photonic-Crystal Baths." 
{\em ACS Photonics}~{\bf 6}, 221 (2019). 

\bibitem{LL5}
L. D. Landau and E. M. Lifshitz, 
{\em Statistical Physics, Third Edition, Part 1} 
(Butterworth-Heinemann, 1980). 

\bibitem{fain1969}
V. M. Fain and Ya. I. Khanin, 
{\em Quantum Electronics, Volume 1: Basic Theory}
(Pergamon Press, Oxford, 1969). 

\bibitem{tomas1}
M. S. Toma\v{s}, 
``Green function for multilayers: light scattering in planar cavities." 
{\em Phys. Rev. A}~{\bf 51}, 2545 (1995). 

\bibitem{footnote1}
Our reflection coefficients agree with the ones in Ref.~\cite{pablo2014}, though it may appear that our sign for $r_{ps}$ and $r_{sp}$ is opposite to theirs. 
The difference arises from different choices of coordinate frame. 
In our case, we chose the incident wave to propagate in the $-\hat{z}$ direction, whereas in Ref.~\cite{pablo2014} the incident wave was chosen to propagate in the $+\hat{z}$ direction. Thus, if the magnetization direction (or orientation vector of the Hall current) of the Chern insulator is chosen to be opposite to the normally propagating direction of the incident wave, then the magnetization/Hall current vector in our chosen coordinate frame is positive, whereas that in the coordinate frame of Ref.~\cite{pablo2014} is negative. Thus a positive magnetization/Hall current vector in our coordinate frame gives rise to a positive Chern number $C > 0$, whereas it gives rise to a negative Chern number $C < 0$ in the coordinate frame of Ref.~\cite{pablo2014}. 

\bibitem{czycholl2017}
G. Czycholl, {\em Theoretische Festk\"{o}rperphysik Band 2} (4. Auflage, Springer Verlag, Berlin, 2017).

\bibitem{wylie-sipe1}
J. M. Wylie and J. E. Sipe, ``Quantum electrodynamics near an interface." 
{\em Phys. Rev. A}~{\bf 30}, 1185 (1984)

\bibitem{wylie-sipe2}
J. M. Wylie and J. E. Sipe, ``Quantum electrodynamics near an interface. II" 
{\em Phys. Rev. A}~{\bf 32}, 2030 (1985)

\bibitem{sakurai1967}
J. J. Sakurai, {\em Advanced Quantum Mechanics} 
(Addison-Wesley, 1967).     

\bibitem{grynberg2010} 
G. Grynberg, A. Aspect, and C. Fabre, 
{\em Introduction to Quantum Optics} 
(Cambridge University Press, Cambridge, UK, 2010).

\bibitem{milonni2019}
P. W. Milonni, {\em An Introduction to Quantum Optics and Quantum Fluctuations} 
(Oxford University Press, Oxford, UK, 2019).

\bibitem{footnote2}
We thank D. Wilkowski for his comment on this. 










\bibitem{xue2018}
Y. Xue, J. Y. Zhang, B. Zhao, X. Y. Wei and Z. Q. Yang, 
``Non-Dirac Chern insulators with large band gaps and spin-polarised edge states." 
{\em Nanoscale}~{\bf 10}, 8569 (2018).

\bibitem{grushin2012}
A. G. Grushin, T. Neupert, C. Chamon, and C. Mudry, 
``Enhancing the stability of a fractional Chern insulator against competing phases." 
{\em Phys. Rev. B}~{\bf 86}, 205125 (2012). 



\end{thebibliography}
\end{document}